\documentclass[conference]{IEEEtran}
\IEEEoverridecommandlockouts

\usepackage{cite}
\usepackage[hidelinks]{hyperref}
\usepackage{amsmath,amssymb,amsfonts}
\usepackage[labelformat=simple]{subcaption}

\usepackage{algorithm}
\usepackage{algpseudocode}
\usepackage{graphicx}
\usepackage{textcomp}
\usepackage{cuted}
\usepackage{xcolor}
\usepackage{comment}
\usepackage{float}
\usepackage{array}
\usepackage{tabularx}
\usepackage{makecell}
\usepackage{booktabs}  
\usepackage{multirow} 
\usepackage{lipsum} 
\usepackage{quantikz}

\usepackage{pifont}
\usepackage[T1]{fontenc}
\usepackage{tikz}
\usepackage{mathrsfs}
\usepackage{eufrak}
\usetikzlibrary{shapes.geometric, arrows, positioning}
\usepackage[normalem]{ulem}
\usepackage{enumitem}

\usepackage{amsthm}

\newtheorem{theorem}{Theorem}
\newtheorem{corollary}{Corollary}

\tikzstyle{startstop} = [rectangle, rounded corners, minimum width=3cm, minimum height=1cm, text centered, draw=black, fill=red!30]
\tikzstyle{process} = [rectangle, minimum width=3cm, minimum height=1cm, text centered, draw=black, fill=orange!30]
\tikzstyle{decision} = [diamond, aspect=2, text centered, draw=black, fill=green!30]
\tikzstyle{arrow} = [thick,->,>=stealth]

\def\BibTeX{{\rm B\kern-.05em{\sc i\kern-.025em b}\kern-.08em
    T\kern-.1667em\lower.7ex\hbox{E}\kern-.125emX}}
    
\begin{document}

\title{Markov Chain Model of Entanglement Setup and Utilization in Noisy Dynamic LEO Satellite Networks
}

\author{\IEEEauthorblockN{Yifan Gao
~~~~Alvin Valera
~~~~Winston K.G. Seah}
\IEEEauthorblockA{\textit{School of Engineering and Computer Science} \\
\textit{Victoria University of Wellington}\\
Wellington, New Zealand \\
yifan.gao@vuw.ac.nz, \{alvin.valera,winston.seah\}@ecs.vuw.ac.nz}}

\maketitle

\begin{abstract}
Quantum entanglement distribution in dynamic Low Earth Orbit (LEO) satellite networks faces significant challenges from dynamic topology, limited quantum resources and strict coherence time constraints. 
To our best knowledge, existing works treat entanglement Einstein-Podolsky-Rosen (EPR) pairs as immediately available resources without modeling real EPR pair parameters, neglecting propagation delay and probabilistic link failure in the physical free-space quantum channel.
This paper considers both stochastic error and systematic error in LEO free-space quantum channels, addresses both gaps through mathematical methods, presenting a comprehensive Markov chain model for a one-hop LEO satellite quantum link, with a state space defined by link storage age and physical distance. The model captures the full three-stage entangled link lifecycle under two strategies (pre-generation and on-demand) and derives expressions for request satisfaction rate (RSR), average waiting time (AWT), link utilization efficiency (LUE) and average consumed link fidelity (ACF) of each strategy. Validation against Monte Carlo baselines confirms agreement within 0.5\% across all metrics. Moreover, numerical evaluation under realistic parameters establishes a maximum one-hop transmission distance of 50km, a cutoff time bound of 0.3s and confirms that polarization rotation has a negligible impact on maximum transmission distance within \(\mathbf{40-50}\)km, providing theoretical foundations for designing and optimizing quantum entanglement distribution strategies in satellite networks. 

\end{abstract}

\begin{IEEEkeywords}
LEO satellite, Free Space, Markov Chains, Decoherence
\end{IEEEkeywords}

\section{Introduction}\label{sec:In}

Quantum networks exploit entanglement to enable applications such as quantum key distribution (QKD), quantum-enhanced sensing and distributed quantum computing over geographically separated nodes. Scaling these applications globally requires distributing entanglement over long distances, which terrestrial optical fiber can not support beyond approximately 100--200 km due to exponential photon loss~\cite{pittaluga2021600}. LEO satellites provide an alternative through free-space optical channels with lower photon loss~\cite{gonzalez2024satellite}. However, the performance of a satellite quantum network depends on its constituent links, each comprising both a quantum and classical channel, as entanglement must be established and maintained at the link level before network-level operations can proceed. Modeling how a single link is generated, how its fidelity degrades under physical impairments, and how it is consumed by quantum protocols is therefore a necessary foundation for satellite quantum network design~\cite{meister2025simulation}. To the best of our knowledge, existing literature~\cite{liao2017satellite,pan2023free,sisodia2024fso,liao2017long} has yet to address these combined aspects. 

Most existing analytical models for quantum entanglement distribution were developed for terrestrial fiber networks and rely on assumptions that do not hold in free-space satellite environments. Specifically, they treat entanglement links as immediately available resources, ignoring the physical parameters governing Einstein-Podolsky-Rosen (EPR)\footnote{An EPR pair is a maximally entangled two-qubit state, typically of the form \(|\Phi \rangle = \frac{1}{\sqrt{2}} (|00\rangle+|11\rangle)\), in which the measurement outcomes of the two qubits are perfectly correlated regardless of the physical distance separating them. EPR pairs serve as the fundamental resource for quantum communication protocols, including quantum teleportation and entanglement-based QKD.} pair generation, photon transmission or fidelity evolution under free-space channel conditions. Consequently, distance-dependent free-space effects, including Gaussian beam diffraction, pointing errors, polarization rotation and transmission loss are absent from existing models~\cite{wallnofer2022simulating,fittipaldi2024entanglement}. Propagation delay and probabilistic link failure, which directly affect entanglement attempt success, are also consistently neglected. Furthermore, existing Markov-chain models assume static topologies with fixed inter-node distances and derive steady-state results over time scales much longer than the 5-minute contact windows typical of LEO satellites~\cite{mollakhani2026inter}. As a result, current analytical models do not accurately capture the behavior of entangled links in realistic LEO satellite networks.

To address these limitations, we develop a distance-aware Markov-chain model for entanglement generation and utilization over a one-hop LEO quantum inter-satellite link (QISL). The model captures the complete lifecycle of an entangled link, from probabilistic generation and storage in quantum memory to eventual consumption by a quantum communication protocol. We analyze two entanglement strategies:\textit{ pre-generation}, where links are proactively generated and stored, and \textit{on-demand generation}, where links are created only after a request arrives. From the resulting Markov models, we derive closed-form expressions for request satisfaction rate (RSR), average waiting time (AWT), link utilization efficiency (LUE) and average consumed fidelity (ACF). Unlike existing Markov-chain models developed for static terrestrial fiber networks~\cite{inesta2023optimal,haldar2024fast,dai2022capacity,li2023dynamic,panigrahy2023capacity}, the proposed framework is typically for the dynamic free-space characteristics of LEO satellite networks.

The analytical framework enables a quantitative comparison between pre-generation and on-demand entanglement strategies. The results show that pre-generation can improve request responsiveness at the expense of storage-induced fidelity degradation whereas on-demand generation preserves fidelity but incurs higher waiting time. Numerical evaluation under realistic LEO satellite parameters further indicates a maximum one-hop transmission distance of approximately 40--50km, a link lifetime below 0.3s and negligible polarization-rotation effects within this operating range. These findings provide practical design guidelines for entanglement management and quantum resource allocation in satellite quantum networks.

The rest of this paper is structured as follows. Section~\ref{sec:RW} surveys related work on Markov chains for quantum networks and free-space quantum communications. Section~\ref{sec:SM} presents our entanglement link generation model. Section~\ref{sec: MCMC} presents our Markov chain constructions, transition dynamics and analytical performance metrics. Section~\ref{sec:E} evaluates the model with realistic LEO satellite parameters. Finally, section~\ref{sec:Conclusion} concludes this paper.

\section{Related Work}\label{sec:RW}

This section reviews existing research on Markov chain models for quantum networks and free-space quantum communication systems, highlighting the gaps our work addresses. While prior work has analyzed either terrestrial quantum networks or classical satellite communications, no existing model captures the unique challenges of LEO satellite quantum networks operating in dynamic free-space environments.

\subsection{Markov Chains}

Markov chain models provide a mathematical framework for analyzing stochastic entanglement distribution processes and deriving key performance measures in quantum networks. Vinay et al.~\cite{vinay2019statistical} established the foundation by modeling probabilistic entanglement establishment and demonstrating significant improvements in secret-key-rate estimation. Building on this work, Brand et al.~\cite{brand2020efficient} developed efficient polynomial-time techniques for analyzing large repeater chains with purification, while Zang et al.~\cite{zang2024analytical} derived closed-form expressions for throughput and latency in continuously operating quantum repeater networks.
Despite these advances, existing Markov chain models were developed for terrestrial fiber-based quantum networks and assume static topologies with fixed inter-node distances. Consequently, they do not capture the time-varying topology, distance-dependent free-space channel effects and short contact durations that characterize LEO satellite quantum networks.

\begin{table*}[bt]
\centering
\caption{Comparison of Existing Models}
\label{tab:comparison}
\footnotesize
\renewcommand{\arraystretch}{1.2}
\setlength{\tabcolsep}{2.5pt}
\begin{tabular}{|l|c|c|c|c|c|c|c|c|c|c|c|}
\hline
\multirow{2}{*}{\textbf{Categories}} & \multicolumn{3}{c|}{\textbf{Markov Chains}} & \multicolumn{7}{c|}{\textbf{Free space}} & \multirow{2}{*}{\textbf{Our work}} \\
\cline{2-11}
& \cite{vinay2019statistical} & \cite{brand2020efficient} & \cite{zang2024analytical} & \cite{lee2004part} & \cite{puryear2009optical,puryear2011experimental,puryear2011time} & \cite{liao2017long,liao2017satellite} & \cite{vasylyev2016atmospheric,vasylyev2018theory} & \cite{pan2020experimental,pan2023free} & \cite{klen2023numerical} & \cite{sisodia2024fso} & \\
\hline
\textbf{Free-space} & $\boldsymbol{\times}$ & $\boldsymbol{\times}$ & $\boldsymbol{\times}$ & \checkmark & $\boldsymbol{\times}$ & \checkmark & \checkmark & \checkmark & \checkmark & \checkmark & \checkmark \\
\hline
\textbf{Dynamic topology} & $\boldsymbol{\times}$ & $\boldsymbol{\times}$ & $\boldsymbol{\times}$ & $\boldsymbol{\times}$ & $\boldsymbol{\times}$ & $\boldsymbol{\times}$ & $\boldsymbol{\times}$ & $\boldsymbol{\times}$ & $\boldsymbol{\times}$ & $\boldsymbol{\times}$ & \checkmark \\
\hline
\textbf{Variable distance} & $\boldsymbol{\times}$ & $\boldsymbol{\times}$ & $\boldsymbol{\times}$ & $\boldsymbol{\times}$ & $\boldsymbol{\times}$ & $\boldsymbol{\times}$ & $\boldsymbol{\times}$ & $\boldsymbol{\times}$ & $\boldsymbol{\times}$ & $\boldsymbol{\times}$ & \checkmark \\
\hline
\textbf{Decoherence} & \checkmark & \checkmark & \checkmark & N/A & \checkmark & N/A & \checkmark & N/A & \checkmark & \checkmark & \checkmark \\
\hline
\textbf{Polarization Rotation} & $\boldsymbol{\times}$ & $\boldsymbol{\times}$ & $\boldsymbol{\times}$ & \checkmark & \checkmark & \checkmark & \checkmark & \checkmark & \checkmark & \checkmark & \checkmark \\
\hline
\textbf{Large networks} & $\boldsymbol{\times}$ & \checkmark & \checkmark & $\boldsymbol{\times}$ & $\boldsymbol{\times}$ & $\boldsymbol{\times}$ & $\boldsymbol{\times}$ & $\boldsymbol{\times}$ & $\boldsymbol{\times}$ & $\boldsymbol{\times}$ & \checkmark \\
\hline
\end{tabular}
\end{table*}

\subsection{Entanglement Generation in Free Space}

To overcome the distance limitations of terrestrial fiber-based quantum networks, significant research has explored free-space quantum communication via satellites, particularly in LEO constellations. Experimental studies have demonstrated the feasibility of satellite-based quantum communication, including satellite-to-ground QKD from the Micius satellite~\cite{liao2017satellite,liao2017long,liao2018satellite}, free-space quantum secure direct communication~\cite{pan2020experimental,pan2023free}, and comparative evaluations of Free Space Optics (FSO)-QKD protocols under realistic channel impairments~\cite{sisodia2024fso}. These works show that free-space channels can support quantum communication over distances beyond the reach of optical fibers, but remain susceptible to beam spreading, pointing errors, polarization rotation, and atmospheric effects.

Many of these challenges have previously been studied in classical FSO systems. Lee and Chan~\cite{lee2004part} showed that spatial diversity can mitigate atmospheric turbulence, while Puryear et al.~\cite{puryear2011experimental,puryear2011time,puryear2009optical} demonstrated that turbulence-induced channel dynamics can be modeled using Markov processes. Although these studies focus on classical communications, they provide useful insights into modeling time-varying free-space channels.
In the quantum domain, atmospheric channels are commonly modeled through probability distributions of transmittance (PDT)~\cite{vasylyev2016atmospheric,vasylyev2018theory,klen2023numerical}. Vasylyev et al.~\cite{vasylyev2016atmospheric,vasylyev2018theory} proposed models that capture beam wandering, beam broadening and turbulence-induced distortions, while Klen and Semenov~\cite{klen2023numerical} proposed a Beta-distribution-based approximation for PDT obtained through numerical simulations. However, these models focus primarily on transmittance fluctuations and do not explicitly consider polarization rotation, which can significantly reduce the initial fidelity of entangled pairs.

\subsection{Classical Communication in Quantum Communication}

Existing studies on classical communication in quantum networks primarily focus on the entanglement establishment phase. Research on entanglement routing, resource allocation and quantum network protocols extensively models the classical communication required for entanglement generation, Bell-state measurements, heralding, entanglement swapping, purification and network coordination~\cite{shi2020concurrent,pant2019routing,vasan2025entanglement,kumar2025zbr,li2021effective,yang2024asynchronous,hu2024high,halder2024optimal,abane2025entanglement}. However, these works generally assume that once an end-to-end (E2E) EPR pair has been established and stored in quantum memory, it can be utilized instantaneously. Consequently, they neglect the additional classical communication required to consume the entangled link for applications such as quantum teleportation and QKD basis reconciliation.

\subsection{Limitations of Existing Models}

Table~\ref{tab:comparison} summarizes the capabilities and limitations of existing analytical models.
Markov chain models provide powerful tools for analyzing stochastic entanglement distribution, but have been developed primarily for static, terrestrial fiber networks with fixed link distances. Conversely, free-space quantum communication studies have focused on channel characterization and experimental demonstrations, without incorporating these physical-layer effects into analytical models of entanglement generation, storage and utilization. As a result, key characteristics of LEO satellite quantum networks, including time-varying link distances, distance-dependent generation success, transmission and storage decoherence, and the limited duration of satellite contact windows, remain largely absent from existing analytical frameworks.

Furthermore, while prior work models the classical communication required to establish entanglement, the additional coordination required to utilize an established entangled link has received little attention. Consequently, the combined impact of free-space channel impairments, quantum-memory dynamics and classical communication delays on satellite quantum link performance remains insufficiently understood.

To address these challenges, we develop a distance-aware Markov-chain model for one-hop LEO satellite quantum links. The proposed framework explicitly incorporates distance-dependent entanglement generation, transmission and storage decoherence, and classical communication delays during both link establishment and utilization. Using this framework, we analyze both pre-generation and on-demand entanglement management strategies and derive closed-form expressions for request satisfaction rate, average waiting time, link utilization efficiency and average consumed fidelity.

\section{System Model}\label{sec:SM}

We consider a one-hop quantum link between two LEO satellites, denoted as $A$ (source) and $B$ (destination), as the foundation for larger network analysis. The physical distance between them at time \(t\) is \(d^{t} = || \textbf{v}_{A}^{t} - \textbf{v}_{B}^{t} ||\), where \(\textbf{v}_{A}^{t}\) and \(\textbf{v}_{B}^{t}\) are the three-dimensional position vectors of $A$ and $B$, respectively. An entangled link between two satellites can be established only when \(d^t \leq d_{\max}\), where \(d_{\max}\) is the maximum transmission distance determined by the free-space channel constraints. The system operates in discrete time slots of duration \(\Delta t\) and quantum communication requests arrive sequentially with at most one request per time slot.

The lifecycle of an entangled link consists of three stages: (\textit{i}) \textit{generation}, (\textit{ii}) \textit{storage} and (\textit{iii}) \textit{utilization}. In the generation stage, $A$ produces an EPR pair and transmits one photon to $B$ through the free-space optical channel. In the storage stage, the successfully established link is held in quantum memory and experiences fidelity degradation due to decoherence. In the utilization stage, the link is consumed by a quantum communication protocol to satisfy a request. The system does not consider any link whose fidelity falls below the minimum acceptable threshold \(F_\text{th}\), corresponding to a maximum storage age of \(K(d)\) time slots.

The following assumptions are adopted throughout this work. Over the analysis interval $[0,T]$, the inter-satellite distance is assumed to vary slowly relative to the timescale of entanglement generation and utilization, allowing distance-dependent quantities to be treated as constant within the interval. Atmospheric absorption, scattering and turbulence are neglected because the communication path lies predominantly above the atmosphere, such that transmittance variations are dominated by diffraction and pointing errors. Each satellite is assumed to store at most one entangled link, enabling a tractable link-level analysis. Request arrivals follow a Bernoulli process with probability $\lambda$ per time slot. Since the system does not maintain a request queue, only the occurrence of at least one arrival within a time slot affects the state evolution\footnote{Modeling multiple simultaneous requests would require introducing queueing states in addition to the quantum-link states, substantially increasing the Markov-chain state space. Since the objective of this work is to characterize link-level entanglement dynamics rather than request scheduling, queueing effects are left for future work.}. Quantum-memory decoherence is modeled using an amplitude-damping channel.
Finally, a successfully generated link becomes available for utilization in the subsequent time slot, yielding a minimum storage age of \(\Delta t\). The key symbols used throughout this paper are summarized in Table~\ref{tab: symbols}.  Detailed derivation is provided in Appendix ~\ref{app: SM}.

\begin{table}[bt]
\centering
\caption{Key Symbols}
\label{tab: symbols}
\begin{tabular}{ll}
\hline
\textbf{Symbol} & \textbf{Description} \\
\hline
$d$ & Transmission distance between satellites \\
$d_{\max}$ & Maximum allowable transmission distance \\
$\Delta t$ & Duration of one time slot \\
$\lambda$ & Request arrival probability per time slot \\
$p$ & EPR pair generation probability \\
$q_\text{error}(d)$ & Photon capture probability at satellite \textit{B} \\
&with pointing error effects \\
$p'(d)$ & Overall link generation success probability \\
$\eta(d)$ & Free-space channel transmittance \\
$\sigma_\delta$ & Pointing error standard deviation \\
$\varepsilon$ & Quantum bit error rate (QBER) \\
$\sigma_\text{rot}$ & Rotational jitter standard deviation \\
$F'_0 (d)$ & Initial link fidelity after transmission \\
$F(d,i)$ & Link fidelity at storage age $i \cdot \Delta t$ of distance \(d\) \\
$F_\text{th}$ & Minimum acceptable fidelity threshold \\
$\Gamma$ & Amplitude damping rate of quantum memory \\
$K(d)$ & Maximum storage age in time slots \\
$T_\text{cutoff}(d)$ & Maximum link lifetime \\
$G_{\max}$ & Maximum number of generation attempts \\
$C(d)$ & Time slots per generation attempt \\
$L(d)$ & Time slots per utilization attempt \\
$p_\text{util}(d)$ & Utilization success probability \\
$T$ & Total observation time window \\
\hline
\end{tabular}
\end{table}

\subsection{Link Generation}

Satellite \textit{A} generates an EPR pair with success probability \(p\) locally and transmits one photon to satellite \textit{B}. The overall link generation success probability is given by
\begin{equation}
    p'(d) = p \cdot q_\text{error}(d) \cdot p_\text{suc}(d)
    \label{eq: link total success establish rate}
\end{equation}
where \(q_\text{error}(d)\) is the photon capture probability with pointing error effects and \(p_\text{suc}(d)\) is the probability that the link fidelity meets the minimum threshold. 

\subsubsection{Gaussian Beam Diffraction}

For inter-satellite distances much greater than the Rayleigh length \(d_\mathcal{R}=\frac{\pi W_0^2}{\omega}\), the spot radius grows linearly with distance. Integrating the Gaussian intensity distribution over the receiving aperture of radius \(R_{\rm ap}\) yields the ideal transmittance as
\begin{equation}
    \eta(d) 
    = 1 - e^{-\frac{2\pi^2 W_0^2 R_{\text{ap}}^2}{\omega^2 d^2}}
    \label{eq: ideal transmittance}
\end{equation}
where \(\pi\) is the circumference ratio, \(W_0\) is the transmitter beam waist radius and \(\omega\) is the wavelength of a single photon.

\subsubsection{Pointing Error Effects}

Due to satellite attitude control limitations, a random pointing deviation $\delta$ (radians) shifts the beam center by distance \(d_{\text{dev}}=d\cdot \delta\), reducing the received optical power and consequently the probability of successful photon capture $q_{\text{error}}$. Assuming that the pointing error components follow independent Gaussian distributions with standard deviation $\sigma_{\delta}$, then
\begin{equation}
q_\text{error}(d)=\left(1-e^{-\frac{2R_{\text{ap}}^2}{\theta^2 d^2}}\right)/\left(1+\frac{4\sigma_\delta^2}{\theta^2}\right)
\label{eq: q-error}
\end{equation}
where \(\theta= \omega/(\pi W_0)\) is the beam divergence half angle.

\subsubsection{Polarization Rotation}
The moving mirrors within each satellite's pointing and tracking system introduce time-dependent polarization rotation \(\vartheta_{\text{rot}}(t)=\vartheta_{\text{sys}}(t)+\vartheta_{\text{jit}}(t)\), where the jitter component follows \(\vartheta_{\text{jit}}(t) \sim \mathcal{N}(0, \sigma_{\text{rot}}^{2})\)~\cite{tayebi2023dynamics}, \(\sigma_{\text{rot}}^{2}\) is the rotational jitter variance and \(t\) is the time duration of photon propagation. Taking the expectation over the jitter, the expected fidelity of a transmitted EPR pair under polarization rotation is given by
\begin{equation}
\begin{split}
    \mathbb{E}[F_{\text{rot}}(t)]
= \frac{1}{2}\Bigl[
1 + \cos\!\left(2\vartheta_{\text{sys}}(t)\right)   e^{-2\sigma_{\text{rot}}^{2}}
\Bigr]
\label{eq:rotation fidelity}
\end{split}
\end{equation}
where \(\vartheta_{\text{sys}}(t)=\frac{v_{\text{sat}}}{R_{\rm E}+ h_{\text{sat}}} \cdot t\) ~\cite{toyoshima2011polarization}, with \(R_{\rm E}=6371\) km being the radius of Earth, \(h_{\text{sat}}\) is the height of the satellite orbit and \(v_{\text{sat}}\) is the orbital velocity of a LEO satellite.

\subsubsection{Fidelity after Transmission} 
When the transmittance \(\eta(d)\) decreases, background noise and dark counts contribute more significantly to fidelity degradation. Combining the polarization rotation effect and the transmission loss, the initial link fidelity $F_0'(d,t)$ after successful photon capture is given by~\cite{nielsen2010quantum}
\begin{equation}
    F_0' (d,t) = \mathbb{E}[F_{\text{rot}}(t)] \times \frac{\eta(d)+[1-\eta(d)]  \varepsilon}{1+3[1-\eta(d)] \varepsilon}
\label{eq:initial fidelity}
\end{equation}
where \(\varepsilon\) is the quantum bit error rate (QBER). Since the polarization rotation accumulates only while the photon propagates from \textit{A} to \textit{B}, we evaluate~\eqref{eq:initial fidelity} at \(t = \frac{d}{c}\), so \(F_0' (d,t) = F_0' (d, \frac{d}{c}) \) for the resulting initial link fidelity, which depends on \(d\) only.

\subsubsection{Fidelity Qualified Rate}

Since \(F_0'(d)\) is monotonically decreasing with distance,
\begin{equation}
p_{\text{suc}}(d)
=
\begin{cases}
1, & d \leq d_{\text{max}},\\
0, & \text{otherwise}.
\end{cases}
\end{equation}

\subsubsection{Maximum Transmission Distance}

The maximum transmission distance \(d_{\text{max}} \) is the distance when the initial link fidelity equals to the fidelity threshold \(F_{\text{th}}\). For the short ISL distances considered in this work (40--50 km), the polarization rotation effect is negligible since \(e^{-2\sigma_{\text{rot}}^2}\approx 1\). Under this approximation, \(d_{\text{max}}\) can be expressed as
\begin{equation}
    d_{\max}
= \sqrt{\frac{-2R_{\text{ap}}^2}{\theta^2 \ln\!\left[ \frac{1 - F_{\text{th}}}{1 + \varepsilon(3F_{\text{th}} - 1)} \right]}}
\label{eq: simple dmax}
\end{equation}
which provides useful design insight into the factors limiting the achievable transmission distance of one-hop LEO quantum links while avoiding the complexity of the full expression.

\subsubsection{Link Generation Attempt Duration}

Each link generation attempt involves EPR pair generation, quantum channel transmission and classical communication acknowledgment~\cite{mahafza2017introduction,kurose2019computer}. The number of time slots required per attempt of the LEO free space environment is
\begin{equation}
    C(d) =  \left\lceil \frac{1}{\Delta t} \left( \frac{1}{R_\text{src}}+ \frac{3d}{c} + t^{(\text{Cls})}_\text{proc} \right)  \right\rceil 
    \label{eq: waiting time}
\end{equation}
where \(R_\text{src}\) is the initial EPR pair generation rate, \(t^\text{(Cls)}_\text{proc} = t_\text{mod}+t_\text{demod}+t_\text{pkt}\), where \(t_\text{mod}\), \(t_\text{demod}\) and \(t_\text{pkt}\) are the modulation, demodulation and processing times, respectively, of the protocol and packet, and $\lceil x \rceil$ denotes the ceiling of $x$.

\subsection{Link Storage}

After distribution, the entangled link is stored in quantum memory until it is consumed by a quantum communication protocol. During storage, both qubits of the EPR pair undergo independent amplitude damping due to interactions with the surrounding environment~\cite{nielsen2010quantum}. As a result, the fidelity of the link decays exponentially with storage time according to~\cite{li2023dynamic}
\begin{equation}
F(d, i) 
= F_0'(d) e^{-2\Gamma  i\Delta t}
\label{eq: storing fidelity}
\end{equation}
where \(F_0'(d)\) is the initial fidelity after transmission, \(i\) is the storage time in slots and \(\Gamma\) is the damping rate characterizing the coupling between the quantum memory and its environment.

An entangled link remains usable only while its fidelity exceeds the threshold \(F_{\text{th}}\). Combining (\ref{eq: storing fidelity}) with this requirement yields the maximum allowable storage duration, expressed in time slots as
\begin{equation}
K(d) = \left\lfloor \frac{T_{\text{cutoff}}(d)}{\Delta t} \right\rfloor 
= \left\lfloor \frac{1}{\Delta t} \left( \frac{d}{c}- \frac{1}{2\Gamma} \ln \frac{F_{\text{th}}}{F_0'(d)} \right) \right\rfloor
\label{eq: max age}
\end{equation}
where \(c\) is the speed of light and $\lfloor x \rfloor$ denotes the floor of $x$.

\subsection{Link Utilization}

Once an entangled link has been established, it can be consumed by a quantum communication application. In this work, we consider the E91 QKD protocol~\cite{E91} as a representative utilization process. E91 consists of four phases: quantum measurement, basis sifting, error estimation and privacy amplification~\cite{ekert1991quantum}. Together, these phases require a sequence of one-way classical transmissions between the two satellites.

\subsubsection{Utilization Process Duration}

The total utilization duration comprises the measurement time and the classical communication delays across all four phases, yielding
\begin{equation}
    L(d) = \left\lceil \frac{1}{\Delta t} \left( \frac{5d}{c} + t_{\mathrm{util}}^{(\text{proc})} \right) \right\rceil 
    \label{eq: link utilization time}
\end{equation}
where \(t_{\mathrm{util}}^{(\text{proc})}\) is the total processing overhead across all phases. 

\subsubsection{Utilization Success Probability}

A utilization attempt succeeds only if all classical message exchanges are completed successfully, both single-photon detectors remain operational, and synchronization between the satellites is maintained. Since the protocol requires five one-way classical transmissions (see Appendix A
), the overall utilization success probability is
\begin{equation}
    p_\text{util} = \left( p^\text{(Cls)}_{\text{single}}  \right) ^5 \cdot p_\text{det}^{(A)} \cdot p_\text{det}^{(B)} \cdot p_\text{sync}
    \label{eq: utilization success rate}
\end{equation}
where \(p^\text{(Cls)}_{\text{single}}\) is the success probability of a single classical transmission, \(p_{\mathrm{det}}^{(A)}\) and \(p_{\mathrm{det}}^{(B)}\) are the detector availability probabilities at satellites \(A\) and \(B\), respectively. \(p_{\mathrm{sync}}\) is the probability of maintaining synchronization throughout the utilization process.

\section{Markov Chain Model Construction}\label{sec: MCMC}

To analyze the impact of entanglement management on satellite quantum-link performance, we develop Markov chain models for two representative strategies: \textit{pre-generation}~\cite{inesta2023performance} and \textit{on-demand} generation~\cite{kolar2022adaptive}. In the pre-generation strategy, entangled links are proactively generated and stored in quantum memory before request arrival. In contrast, in the on-demand strategy, links are generated only after a request is received. These approaches represent a trade-off between request responsiveness and resource efficiency.

\subsection{Overview}

A request arriving in time slot \(t\) is represented by the source-destination pair \((\mathfrak{s}^{t},\mathfrak{d}^{t})\), wherein the nodes are separated by distance \(d\). For both strategies, entanglement generation succeeds with probability \(p'(d)\) given by Eq.~\eqref{eq: link total success establish rate}. A maximum of \(G_{\max}\) generation attempts is permitted with each attempt requiring \(C(d)\Delta t\) time slots according to Eq.~\eqref{eq: waiting time}.

Once an entangled link is available, the utilization process occupies \(L(d)\Delta t\) time slots and succeeds with probability \(p_{\mathrm{util}}\) given by Eq.~\eqref{eq: utilization success rate}. Regardless of the outcome, the system returns to the initial state after \(L(d)\Delta t\) time slots. A failed utilization attempt permanently consumes the link because the Bell state measurements (BSM) collapse the entangled states.

For both strategies, a newly generated link must be stored for at least one time slot before utilization, yielding a minimum link age of \(\Delta t\) and a corresponding fidelity of
\(
F(d,1)=F'_0 (d)\cdot e^{-2\Gamma\Delta t}.
\)
In the pre-generation strategy, a link may remain in memory for up to \(K(d)\) time slots according to Eq.~(\ref{eq: max age}) before being discarded due to decoherence. In contrast, the on-demand strategy generates links only after request arrival and therefore does not require a storage-age cutoff.

\subsection{Markovian Property}

The proposed models satisfy the Markov property because the system state at time slot \(t\) contains all information required to determine its evolution in the next slot. Specifically, the state records both the current operational phase and, when applicable, the storage age \(i\) of the entangled link. State transitions are driven by independent request arrivals with probability \(\lambda\), entanglement-generation outcomes with probability \(p'(d)\), and utilization outcomes with probability \(p_{\mathrm{util}}\), all of which depend only on the current state. Consequently, the conditional distribution of the next state is independent of the system history once the current state is known~\cite{norris1998markov}.

Although the inter-satellite distance \(d^t\) evolves continuously as the satellites move, we assumed it remains approximately constant within the analysis interval \([0,T]\) corresponding to a single satellite communication window. Under this quasi-static assumption, all distance-dependent quantities, including \(p'(d)\), \(F_0'(d)\), \(K(d)\), \(C(d)\) and \(L(d)\), remain constant throughout the interval and therefore do not introduce additional temporal dependencies. As a result, the complete system state forms a finite discrete-time Markov chain whose one-step transition probabilities are fully determined by the current state.

\subsection{Preliminaries}

The proposed analytical framework models the evolution of a single-hop QISL as a discrete-time Markov chain 
\[
\mathcal{M}=(\mathcal{S},P),
\]
where \(\mathcal{S}\) denotes the state space and
\(P=[p_{ij}]\) is the one-step transition probability matrix, whose entries are
\[
p_{ij}
=
\Pr(S_{k+1}=j \mid S_k=i),
\qquad
i,j\in\mathcal S,
\]
where \(S_k\) denotes the system state in slot \(k\).
The system state is represented by the tuple
\[
S_k=(X_k,Y_k),
\]
where \(X_k\) identifies the current operational phase and \(Y_k\) denotes the corresponding phase-dependent index. The exact interpretation of \(Y_k\) depends on the operational phase and is specified separately for the pre-generation and on-demand strategies. Specifically,
\begin{itemize}
    \item \(X_k=0\) denotes the generation phase. Here, \(Y_k=0\) represents a state with no active request, while \(Y_k=g\), \(1\leq g\leq G_{\max}\), corresponds to the \(g\)-th entanglement-generation attempt.
    \item \(X_k=1\) denotes the storage phase. Here, \(Y_k=i\), \(1\leq i\leq K(d)\), represents a stored entangled link with age \(i\Delta t\). This is not present in the on-demand strategy.
    \item \(X_k=2\) denotes the utilization phase. Here, \(Y_k=i\), \(1\leq i\leq K(d)\), represents execution of the utilization protocol using a link of age \(i\Delta t\).
\end{itemize}

\subsection{Pre-Generation Strategy}

\begin{figure}[t]
  \centering
\includegraphics[width=0.5\textwidth]{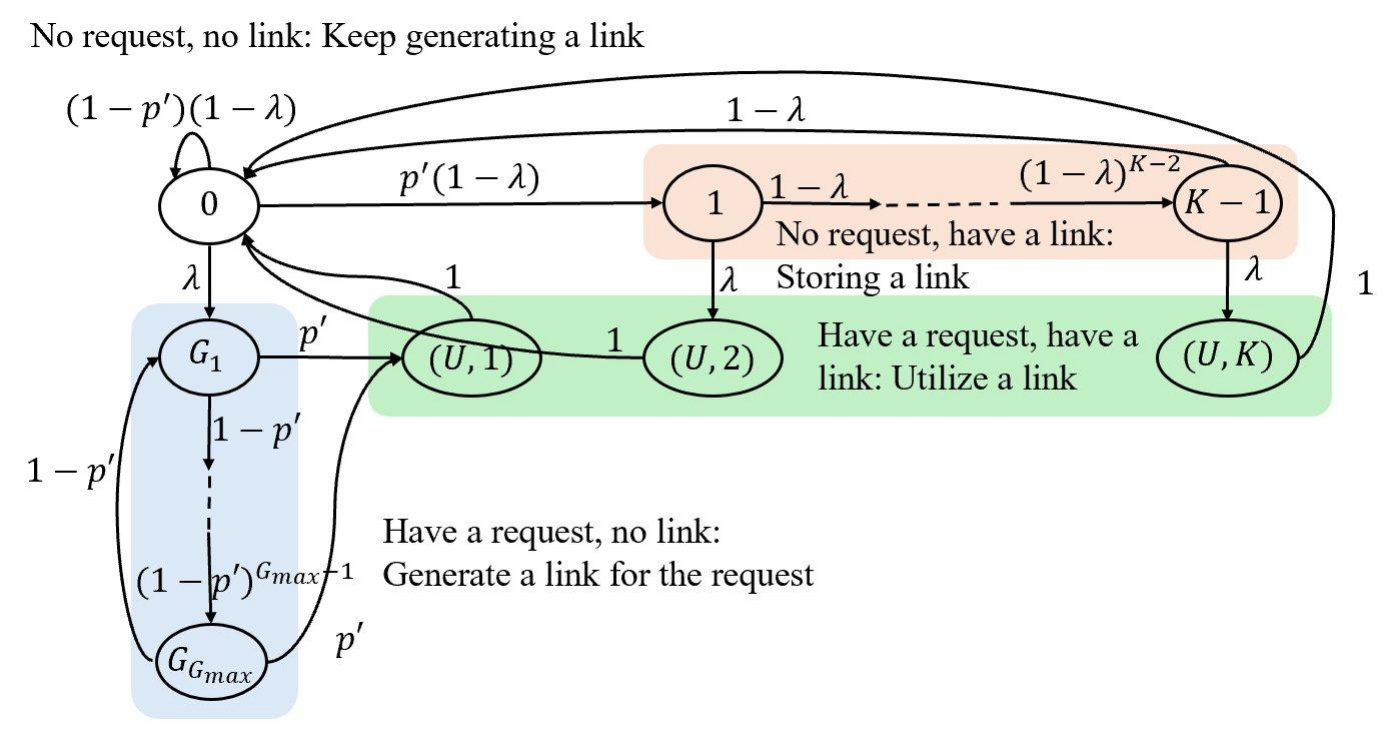}
  \caption{Pre-generation link strategy Markov chain}
  \label{fig:Pre-generation Link Mode}
\end{figure}

Fig.~\ref{fig:Pre-generation Link Mode} shows the state transition diagram of the pre-generation Markov chain. In this strategy, entangled links are proactively generated and stored in quantum memory before requests arrive. This reduces request latency by maintaining readily available links but may lead to fidelity degradation due to storage decoherence and resource wastage when stored links expire before use.

\subsubsection{Transitions from state \((0,0)\)}

At state $(0,0)$, the system continuously attempts proactive link generation while monitoring for incoming requests. A request arrival, occurring with probability $\lambda$, causes an immediate transition to generation state $(0,1)$. In the absence of a request, which occurs with probability $1-\lambda$, a successful generation attempt with probability $p'(d)$ produces a stored link of age 1 and transitions the system to state $(1,1)$; otherwise, the system remains in state $(0,0)$. The resulting transition probabilities are
 
\begin{equation}
p^\text{(p)}_{(0,0),s}
=
\begin{cases}
\lambda, & s=(0,1),\\
(1-\lambda)p'(d), & s=(1,1),\\
(1-\lambda)(1-p'(d)), & s=(0,0),\\
0, & \text{otherwise}.
\label{eq: pre-gen trans prob from (0,0)}
\end{cases}
\end{equation}

\subsubsection{Transitions from stored-link state \((1,i), 1\leq i<K(d)-1\)}
At stored-link states \((1,1),(1,2),\ldots,(1,K(d)-2)\), the system holds a pre-generated link of age \(i\Delta t\) and fidelity \(F(d,i)=F'_0 (d)e^{-2\Gamma i\Delta t}\). A request arrival with probability \(\lambda\) initiates the utilization process, whereas the absence of a request with probability \(1-\lambda\) causes the link to remain in quantum memory and age by one time slot, transitioning to state \((1,i+1)\). During this transition, the link fidelity decreases from \(F(i)\) to \(F(i+1)\) due to storage decoherence. This yields
\begin{equation}
p^\text{(p)}_{(1,i),s}
=
\begin{cases}
\lambda, & s=(2,i+1),\\
1-\lambda, & s=(1,i+1),\\
0, & \text{otherwise}.
\end{cases}
\label{eq: pre-gen trans prob from (1,i)}
\end{equation}

\subsubsection{Transitions from maximum age stored-link state \((1,K(d)-1)\)}
At state \((1,K(d)-1)\), the system holds a pre-generated link of age \((K(d)-1)\Delta t\) and fidelity \(F(d,K(d)-1)=F'_0(d) e^{-2\Gamma (K(d)-1)\Delta t}\), which is close to the threshold \(F_{\mathrm{th}}\). A request arrival with probability \(\lambda\) initiates utilization of the link, whereas the absence of a request with probability \(1-\lambda\) causes the link to expire, after which it is discarded and the system transitions to state \((0,0)\). The resulting transition probabilities are
\begin{equation}
p^\text{(p)}_{(1,K(d)-1),s}
=
\begin{cases}
\lambda, & s=(2,K(d)),\\
1-\lambda, & s=(0,0),\\
0, & \text{otherwise}.
\label{eq: pre-gen trans prob from (1,K-1)}
\end{cases}
\end{equation}

\subsubsection{Transitions from generating states $(0,g)$}
At generating state \((0,g)\), the system is performing the \(g\)-th entanglement-generation attempt for the current request. A successful attempt with probability \(p'(d)\) establishes an entangled link and initiates the utilization process in state \((2,1)\). Otherwise, with probability \(1-p'(d)\), the generation attempt fails. If \(g<G_{\max}\), the system proceeds to the next generation state \((0,g+1)\); otherwise, the request times out and the system returns to state \((0,0)\). The resulting transition probabilities are

\begin{equation}
p^\text{(p)}_{(0,g),s}
=
\begin{cases}
p'(d), & s=(2,1),\\
1-p'(d), & s=(0,g+1), \quad g<G_{\max},\\
1-p'(d), & s=(0,0), \quad g=G_{\max},\\
0, & \text{otherwise}.
\label{eq: pre-gen trans prob from (0,g)}
\end{cases}
\end{equation}

\subsubsection{Transitions from utilization states \((2,i),1 \leq i \leq K(d)\)}
At utilization state \((2,i)\), the system is executing the quantum communication protocol using a link of storage age \(i \Delta t\). The utilization process occupies the system for a fixed duration, after which the link is consumed and the system returns to state \((0,0)\), regardless of whether the utilization succeeds or fails. Consequently, the resulting transition probabilities are
\begin{equation}
p^\text{(p)}_{(2,i),s}
=
\begin{cases}
1, & s=(0,0),\\
0, & \text{otherwise}.
\label{eq: pre-gen trans prob from (2,i)}
\end{cases}
\end{equation}
The utilization success probability \(p_{\mathrm{util}}\) does not appear explicitly in the state transition probabilities because both successful and unsuccessful utilization attempts lead to the same next state. Instead, \(p_{\mathrm{util}}\) influences the performance metrics by determining whether a request entering the utilization state is ultimately satisfied or unsuccessful.

\subsection{On-Demand Strategy}

\begin{figure}[t]
  \centering
\includegraphics[width=0.5\textwidth]{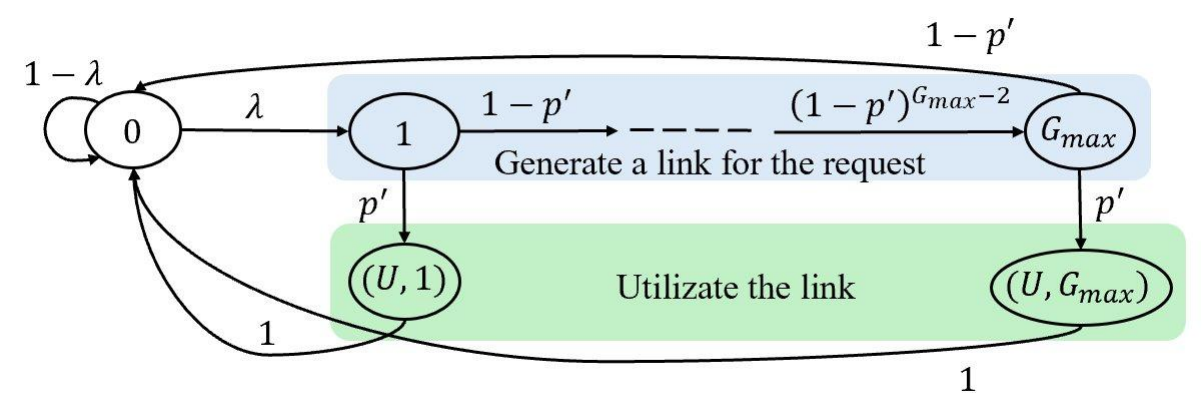}
  \caption{On-demand link strategy Markov chain}
  \label{fig:On-demand Mode}
\end{figure}

Fig.~\ref{fig:On-demand Mode} shows the state transition diagram of the on-demand generation Markov chain. In this strategy, entanglement generation is triggered by request arrival rather than performed proactively. The system follows a locked service policy whereby, upon accepting a request, it dedicates all resources to that request until the generation and utilization processes complete. Consequently, subsequent requests arriving during this interval are rejected. Under the assumed single-link quantum memory capacity, this strategy avoids storage decoherence and link wastage due to expiration but incurs additional waiting time because entangled links must be generated on demand.

\subsubsection{Transitions from state \((0,0)\)}
At state $(0,0)$, the system waits for a request. A request arrival with probability $\lambda$ initiates the first generation attempt and transitions the system to state $(0,1)$, whereas the absence of a request with probability $1-\lambda$ leaves the system in state $(0,0)$. The resulting transition probabilities are
\begin{equation}
p^\text{(o)}_{(0,0),s}
=
\begin{cases}
1-\lambda, & s=(0,0),\\
\lambda, & s=(0,1),\\
0, & \text{otherwise}.
\label{eq: on-de trans prob from (0,0)}
\end{cases}
\end{equation}

\subsubsection{Transitions from generation states}
At generation state \((0,g)\), the system is performing the \(g\)-th entanglement-generation attempt for the current request. A successful attempt with probability \(p'(d)\) establishes an entangled link and initiates the utilization process in state \((2,g)\). Otherwise, with probability \(1-p'(d)\), the generation attempt fails. If \(g<G_{\max}\), the system proceeds to state \((0,g+1)\) for the next generation attempt. If \(g=G_{\max}\), the request exceeds the attempt limit and the system returns to state \((0,0)\) to await new requests. The resulting transition probabilities are

\begin{equation}
p^\text{(o)}_{(0,g),s}
=
\begin{cases}
p'(d), & s=(2,g),\\
1-p'(d), & s=(0,g+1), \quad g<G_{\max},\\
1-p'(d), & s=(0,0), \quad g=G_{\max},\\
0, & \text{otherwise}.
\end{cases}
\label{eq: on-de trans prob from (0,g)}
\end{equation}

\subsubsection{Transitions from utilization state \((2,g),1 \leq g < G_{\max}\)}
For intermediate utilization states \((2,g)\), the system is executing the quantum communication protocol with a link generated at the \(g\)-th attempt. The utilization process takes a fixed duration of \(L \cdot \Delta t\) time slots. After completion, the system returns to the state \((0,0)\) with probability \(1\). Therefore, the transition probabilities are

\begin{equation}
p^\text{(o)}_{(2,g),s}
=
\begin{cases}
1, & s=(0,0),\\
0, & \text{otherwise}.
\end{cases}
\label{eq: on-de trans prob from (2,g)}
\end{equation}

\subsection{Existence of Steady-State Solution}

\begin{theorem}[Existence and uniqueness of the stationary distribution]
For fixed distance \(d\) over the analysis interval \([0,T]\), the pre-generation and on-demand models define finite, irreducible and positive recurrent discrete-time Markov chains. Therefore, each model admits a unique stationary distribution.
\end{theorem}

\begin{proof}
Both Markov chains have finite state spaces because \(G_{\max}\) and \(K(d)\) are finite. In each chain, all states can return to the state \((0,0)\): utilization states return to \((0,0)\) after completion, failed generation after \(G_{\max}\) attempts returns to \((0,0)\) and expired stored links in the pre-generation chain are discarded and return to \((0,0)\). Conversely, from \((0,0)\), every state can be reached with non-zero probability through some sequence of request arrivals, successful or failed generation attempts, and, for the pre-generation model, storage-age transitions. Hence, all states communicate and the chains are irreducible.

Since every finite irreducible Markov chain is positive recurrent, both chains are positive recurrent. Therefore, each chain has a unique stationary distribution.
\end{proof}

\begin{corollary}[Ergodicity]
If \(0<\lambda<1\) and \(0<p'(d)<1\), then the pre-generation and on-demand Markov chains are aperiodic and therefore ergodic. Consequently, their limiting distributions exist and are equal to their unique stationary distributions, independent of the initial state.
\label{cor: Ergodicity}
\end{corollary}

\begin{proof}
For the pre-generation model, state \((0,0)\) has a self-loop with probability
\[
p^{\text{(p)}}_{(0,0),(0,0)}
=
(1-\lambda)(1-p'(d))>0.
\]
Likewise, for the on-demand model, state \((0,0)\) has a self-loop with probability
\[
p^{\text{(o)}}_{(0,0),(0,0)}
=
1-\lambda>0.
\]
An irreducible Markov chain with at least one state having a positive self-loop is aperiodic. Therefore, both chains are finite, irreducible, positive recurrent and aperiodic, and hence ergodic.
\end{proof}

\section{Performance Analysis}

Having established the existence of a unique stationary distribution for both Markov chains in the preceding section, we now derive analytical expressions for their long-run performance. 
We derive four key performance metrics:\textit{ request satisfaction rate} (\(\mathcal{R}\)), \textit{average waiting time} (\(\mathcal{W}\)), \textit{link utilization efficiency} (\(\mathcal{U}\)) and \textit{average consumed fidelity} (\(\mathcal{F}\)). These metrics characterize the reliability, responsiveness, resource efficiency and fidelity performance of a single-hop QISL under the pre-generation and on-demand strategies. Derivation details are provided in Appendix ~\ref{app:PMD}.

\subsection{Request Satisfaction Rate}

We consider two kinds of request satisfaction rate, 1) the request satisfaction rate for the long-run fraction of accepted requests that are ultimately served successfully \(\mathcal{R}_\text{ac}\) and 2) the request satisfaction rate of all arrival requests \(\mathcal{R}_\text{total}\), including the ones that are blocked out. A request is considered successful if an entangled link is established within \(G_{\max}\) generation attempts and the subsequent utilization process succeeds with probability \(p_{\mathrm{util}}\).

For the pre-generation strategy, a request is accepted either when a stored entangled link is immediately available or when the system is in state \((0,0)\). In the former case, satisfaction depends only on successful utilization, whereas in the latter, it additionally requires at least one successful generation attempt within \(G_{\max}\) trials. Consequently, the two kinds of request satisfaction rates, denoted by \(\mathcal{R}_\text{ac}^\text{(p)}(\lambda,d)\) and \(\mathcal{R}_\text{total}^\text{(p)}(\lambda,d)\), is
\begin{equation}
\begin{split}
\mathcal{R}_\text{ac}^\text{(p)} (\lambda,d)
& = p_{\mathrm{util}} \, \left[ 1 + \frac{\lambda (\gamma (d)- 1)}{\lambda + \iota(d)}     \right]
\label{eq:R_p_ac}
\end{split}
\end{equation}
\begin{equation}
\begin{split}
\mathcal{R}&^{\text{(p)}}_\text{total} (\lambda,d) \\
 &= \frac{p'(d) p_\text{util} [ \lambda\gamma (d)+ \iota(d)]}{p' (d)\lambda + \lambda^2 \gamma (d)+ \lambda^2 \gamma(d) p'(d) + p'(d) \iota(d) (1+\lambda) }
    \label{eq:R_p_total}
\end{split}
\end{equation}
where 
\[
\gamma (d)= 1-[1-p'(d)]^{G_{\max}}, 
\]
\[
\iota (\lambda,d)= p'(d) \left[ 1-\lambda - (1-\lambda)^{K(d)} \right].
\]

For the on-demand strategy, requests are accepted only when the system is in the state (0,0), after which every accepted request follows the same generation and utilization procedures. Therefore, the two kinds of satisfaction rates, denoted by $\mathcal{R}^\text{(o)}_\text{ac}(d)$ and $\mathcal{R}^\text{(o)}_\text{total}(\lambda,d)$ are given by
\begin{equation}
\mathcal{R}^\text{(o)}_\text{ac}(d)
=
\gamma (d) p_{\mathrm{util}}
\label{eq:R_o_ac}
\end{equation}
\begin{equation}
\mathcal{R}_\text{total}^{\text{(o)}}(\lambda,d) 
= \frac{\gamma (d)p'(d) p_\text{util}}{p'(d) + \gamma(d)\lambda + \gamma(d)\lambda p'(d)}
\label{eq:R_o_total}
\end{equation}
where $\mathcal{R}^\text{(o)}_\text{ac}(d)$ is independent of the request arrival instant but $\mathcal{R}^\text{(o)}_\text{total}(\lambda,d)$ does.

\subsection{Average Waiting Time}

The average waiting time is the \textit{expected delay from request acceptance to successful satisfaction}, conditioned on the request being satisfied. 

For the pre-generation strategy, a satisfied request is served either by an already stored link or by a newly generated link from state \((0,0)\). If a stored link is available, the number of waiting time slots is \(1+L(d)\). If the request is accepted in state \((0,0)\), additional delay is incurred by the generation process. 
Its average waiting time \(\mathcal{W}^{\text{(p)}} (\lambda,d)\) is given by
\begin{equation}
\begin{split}
\mathcal{W}^{\text{(p)}}(\lambda,d) 
& = 2 + L (d)+ \frac{\lambda \kappa (d)- \iota(d)}{\lambda \gamma(d) + \iota (d)}
\end{split}
\label{eq:W_p}
\end{equation}
where 
\[
\kappa (d)= \frac{C(d) [\gamma (d)- (1-\gamma(d))p'(d)G_{\max}]}{p'(d)}.
\]

For the on-demand strategy, every accepted request starts from state \((0,0)\) and follows the same generation process. If generation succeeds on the \(g\)-th attempt, the request waits for the generation duration and the subsequent utilization duration. Thus, its average waiting time slots, denoted by $\mathcal{W}^{\text{(o)}}(d)$, is independent of the stationary state probabilities and is 
\begin{equation}
\begin{split}
\mathcal{W}^{\text{(o)}} (d)
& = 2 + L(d) + \frac{\kappa(d)}{\gamma(d)}.
\end{split}
\label{eq:W_o}
\end{equation}

\subsection{Link Utilization Efficiency}

The link utilization efficiency is the long-run \textit{fraction of generated entangled links that are consumed by a request before expiring due to decoherence}. A proactively generated link may either remain in quantum memories until consumed by a request or expire before use. In contrast, links generated upon request arrival are utilized immediately whenever generation succeeds. Therefore, the link utilization efficiency of the pre-generation strategy, denoted by $\mathcal{U}^{\text{(p)}}(\lambda,d)$ is
\begin{equation}
\mathcal{U}^{\text{(p)}}(\lambda,d) 
= \frac{\iota (d)+ \lambda \gamma(d)}{p'(d) - \lambda p'(d) + \lambda \gamma(d)}.
\label{eq:U_p}
\end{equation}

For the on-demand strategy, every successfully generated entangled link is immediately used in a utilization attempt. Since no link is stored for future requests, no generated link can expire due to storage decoherence. Hence, no matter the distance

\begin{equation}
\mathcal{U}^{\text{(o)}} = 1.
\label{eq:U_o}
\end{equation}

\subsection{Average Consumed Fidelity}

The average consumed fidelity is the \textit{expected fidelity of an entangled link at the instant it enters the utilization phase}. Since quantum measurements immediately collapse the entangled state, subsequent classical communication introduces no additional decoherence. Consequently, this metric depends only on the storage age of the link and is independent of the utilization success probability \(p_{\mathrm{util}}\).

For the pre-generation strategy, links entering utilization originate either from successful on-demand generation in state \((0,0)\) with fidelity \(F(d,1)\), or from stored-link states \((1,i)\), \(1\le i\le K(d)-1\), with fidelity \(F(d,i+1)\). Therefore, the average consumed fidelity of this strategy, denoted by $\mathcal{F}^{\text{(p)}}(\lambda,d)$, is the weighted average over all links entering the utilization phase,
\begin{equation}
\begin{split}
\mathcal{F}^{\text{(p)}}(\lambda,d)
& = \frac{\nu \lambda F'_0 (d)[\gamma (d)(1-\nu) + p'(d)(\nu - \nu^{K(d)})]}{(1-\lambda)(1-\nu)[\gamma (d) \lambda + \iota(d)]}.
\end{split}
\label{eq:F_p}
\end{equation}
where 
\[
\nu (\lambda) = (1-\lambda)e^{-2\Gamma \Delta t}.
\]

For the on-demand strategy, every consumed link is generated immediately before utilization and therefore has the minimum storage age of one time slot. Consequently, all consumed links have identical fidelity, yielding

\begin{equation}
\mathcal{F}^{\text{(o)}}(d)
=
F(d,1)
=
F_0' (d) e^{-2\Gamma\Delta t}.
\label{eq:F_o}
\end{equation}

\section{Evaluation}\label{sec:E}

This section evaluates the proposed single-hop QISL model. We first examine the photon capture probability with pointing error effects \(q_\text{error}\), EPR-pair initial fidelity \(F'_0(d)\), fidelity degradation during quantum-memory storage \(F(d,i)\), cutoff time \(K(d)\) and maximum transmission distance \(d_{\max}\). We then evaluate the pre-generation and on-demand strategies in terms of request satisfaction rate \(\mathcal{R}\), average waiting time \(\mathcal{W}\), link utilization efficiency \(\mathcal{U}\) and link consumed fidelity \(\mathcal{F}\). Table~\ref{tab:symbols_values} summarizes the parameter values used in the evaluation, together with their corresponding references. We assume \(F_{\rm th} \geq 0.5\), representing the minimum fidelity required for entanglement purification to produce higher-fidelity pairs~\cite{bennett1996purification}.

\begin{table}[tbp]
\centering
\caption{Parameters of Evaluation}
\label{tab:symbols_values}
\begin{tabular}{lll}
\toprule
\textbf{Symbol} & \textbf{Value} & \textbf{Facility / Note} \\
\midrule
$\theta$ & $5~\mu\text{rad}$ & Micius~\cite{liao2017satellite,liao2018satellite}, Jinan-1~\cite{li2025microsatellite} \\
$R_{ap}$ & $100$--$150$ mm & Micius~\cite{liao2017satellite}, Jinan-1~\cite{li2025microsatellite} \\
$\sigma_{\delta}$ & $0.5$--$1.0~\mu$rad & Micius~\cite{zhang2020design} \\
$\varepsilon$ & $1\%$--$3\%$ & Micius~\cite{gozzard2021vulnerability} \\
$\Gamma$ & $0.5$--$1~\text{s}^{-1}$ & Trapped ion~\cite{tubio2024satellite} \\
$F_{\mathrm{th}}$ & $0.5$ & Min.\ $F_{\mathrm{purify}}$~\cite{bennett1996purification} \\
$d$ & $40$--$150~\text{km}$ & Starlink~\cite{cakaj2021parameters} \\
$v_{\mathrm{sat}}$ & $7.59~\text{km/s}$ & Starlink~\cite{cakaj2021parameters} \\
$h_{\mathrm{sat}}$ & $550~\text{km}$ & Starlink~\cite{cakaj2021parameters} \\
$\sigma_{\mathrm{rot}}$ & $0.5$--$1.0~\mu$rad & GOES-16~\cite{dennehy2019survey} \\
\bottomrule
\end{tabular}
\end{table}

\subsection{Entangled Link}

\begin{figure}[htbp]
    \centering
    \captionsetup{font=footnotesize}
    \begin{subfigure}{0.47\linewidth}
        \centering
        \includegraphics[width=\linewidth]{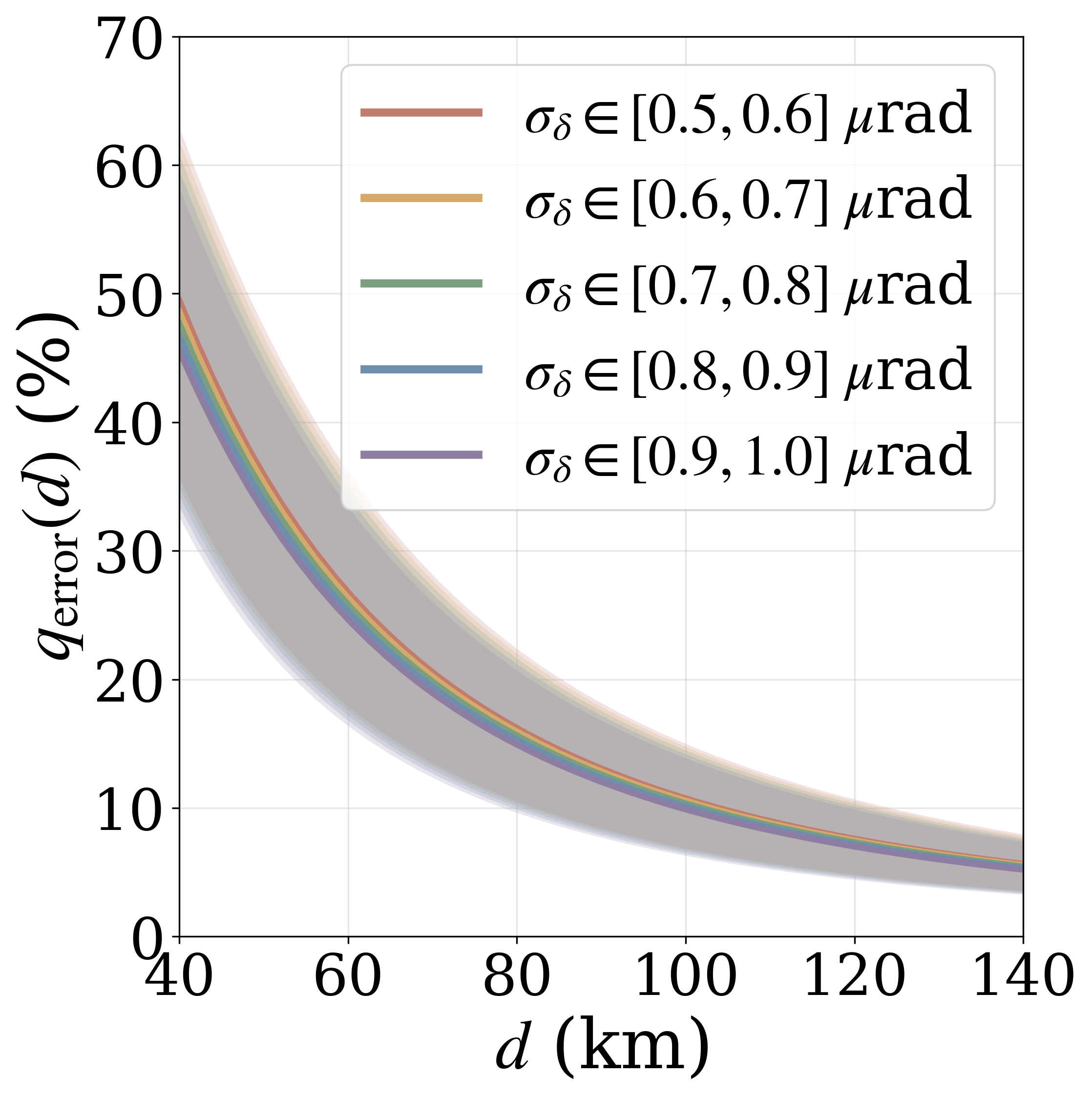}
        \caption{Photon Capture Probability}
        \label{fig:q-error range}
    \end{subfigure}
    \hfill
    \begin{subfigure}{0.48\linewidth}
        \centering
        \includegraphics[width=\linewidth]{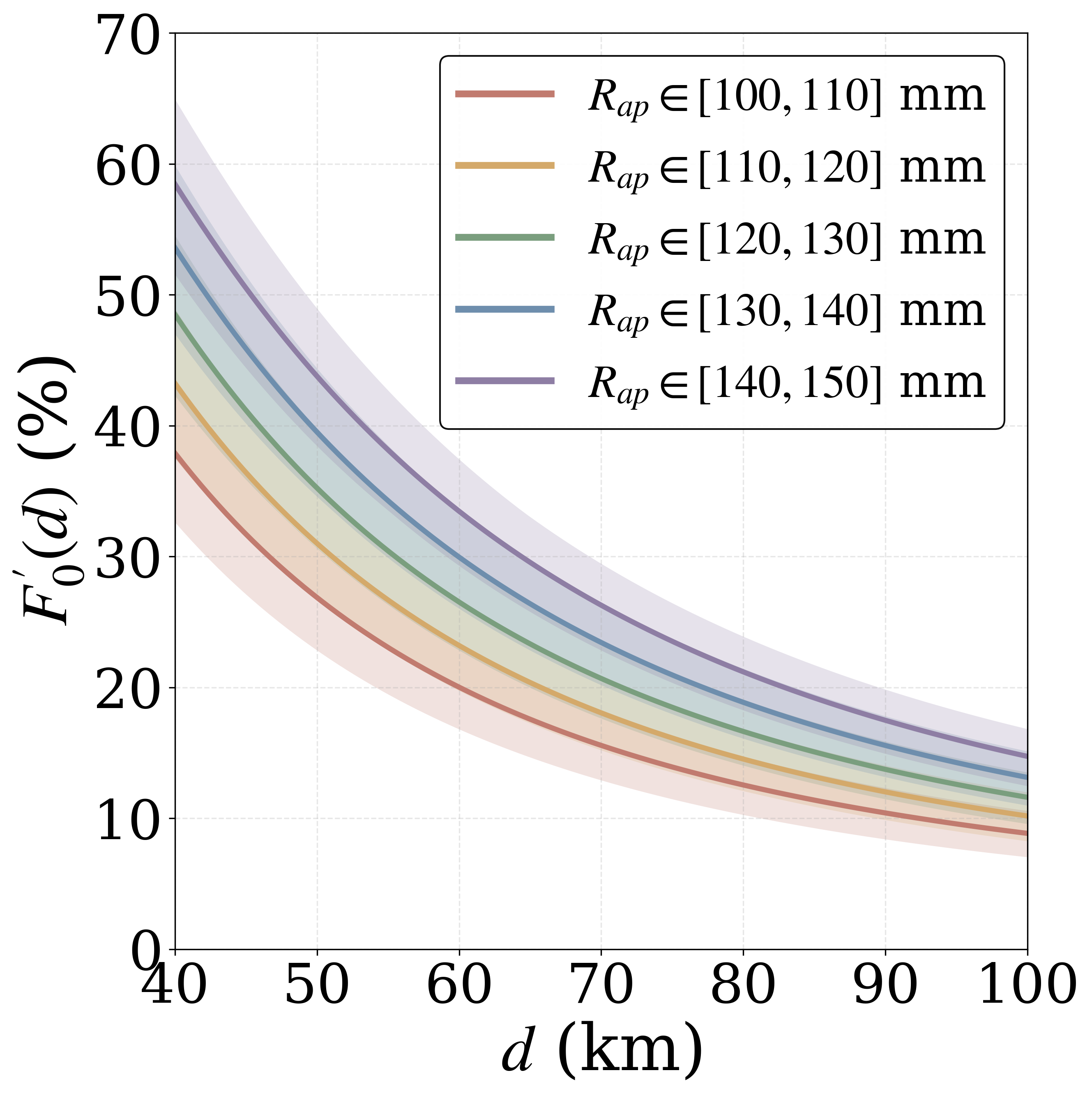}
        \caption{Initial Fidelity}
        \label{fig:Initial Fidelity}
    \end{subfigure}
    \caption{Generate an EPR Pair}
    \label{fig:LUE}
\end{figure}

Fig.~\ref{fig:q-error range} shows the trend of photon capture probability with pointing error effects \(q_{\rm error}(d)\) (Eq.~\eqref{eq: q-error}) in shaded area. 
It decreases monotonically with both link distance \(d\) and pointing error \(\sigma_\delta\), from approximately \(57\)--\(63\%\) at \(d=40\) km to \(7\)--\(8\%\) at \(d=140\) km. The substantial overlap among the five bands indicates relatively low sensitivity to \(\sigma_\delta\) over this range. At \(d=40\) km, increasing \(\sigma_\delta\) from \(0.5\) to \(1.0~\mu\mathrm{rad}\) reduces the mean \(q_{\rm error}(d)\) by approximately \(6\%\), whereas the difference narrows to about \(0.8\%\) at \(d=140\) km. Thus, the absolute impact of pointing error diminishes with distance as diffraction loss becomes dominant.

Fig.~\ref{fig:Initial Fidelity} shows the initial fidelity \(F_0'(d)\) over \(d=40\)--\(100\) km for five receiver aperture-radius intervals. Fidelity decreases rapidly with distance due to increasing diffraction loss and transmission time. At \(d=40\) km, the mean \(F_0'(d)\) ranges from \(55.86\%\) for \(R_{\rm ap}\in[100,110]\) mm to \(64.95\%\) for \(R_{\rm ap}\in[140,150]\) mm, showing that a larger aperture improves entangled-link fidelity. At \(d=100\) km, the corresponding mean fidelities decrease to \(8.84\%\) for \(R_{\rm ap}\in[100,110]\) mm and \(15.12\%\) for \(R_{\rm ap}\in[140,150]\) mm, respectively. The results showing that \(F_0'(d)<50\%\) when \(d>50\) km, indicating that \(d_\text{max} \leq 50\)km.

\begin{figure}[htbp]
    \centering
    \captionsetup{font=footnotesize}
    \begin{subfigure}{0.40\linewidth}
        \centering
        \includegraphics[width=\linewidth]{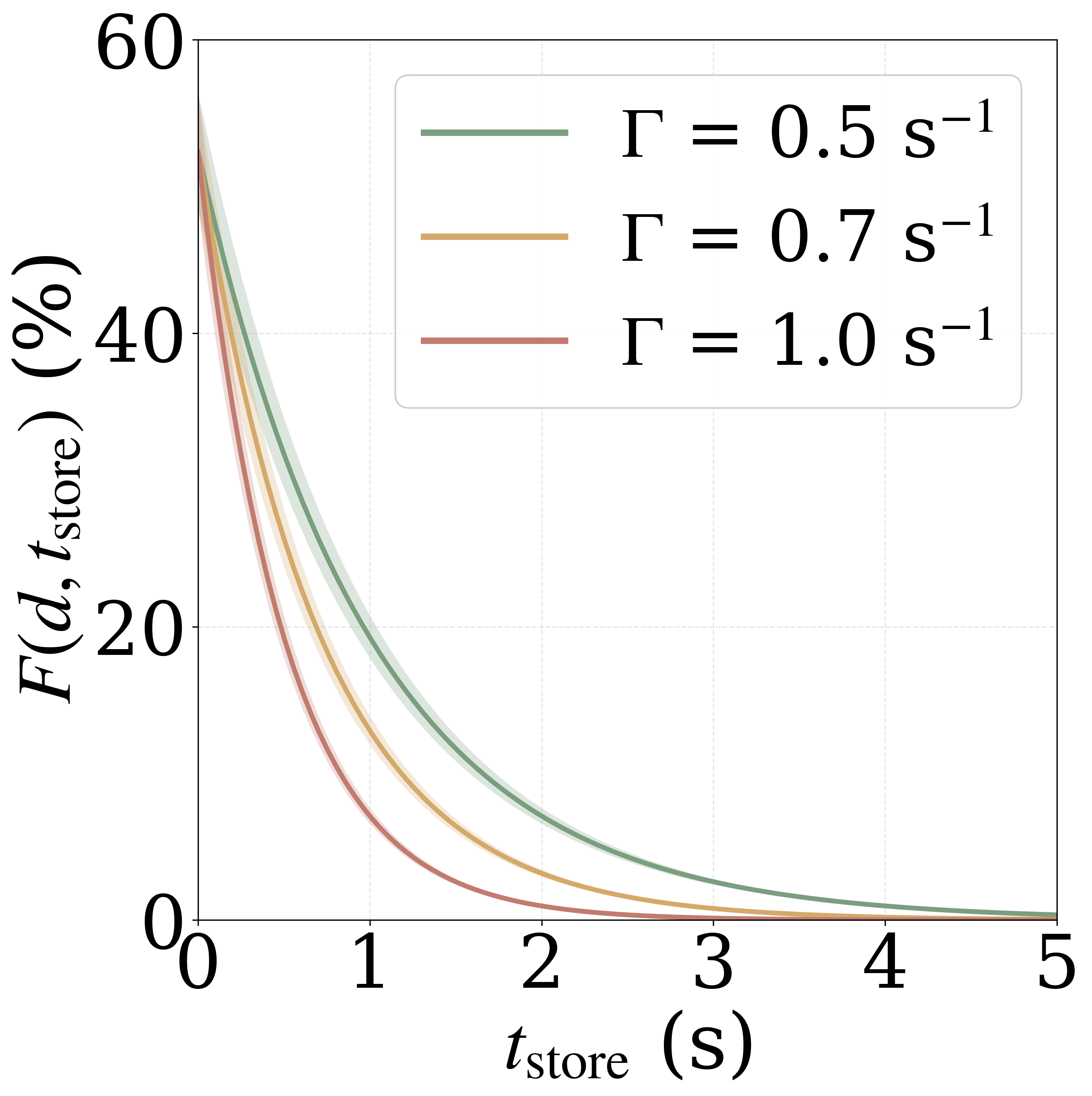}
        \caption{Storing Link Decoherence}
        \label{fig:Fidelity Evolution Range}
    \end{subfigure}
    \hfill
    \begin{subfigure}{0.55\linewidth}
        \centering
        \includegraphics[width=\linewidth]{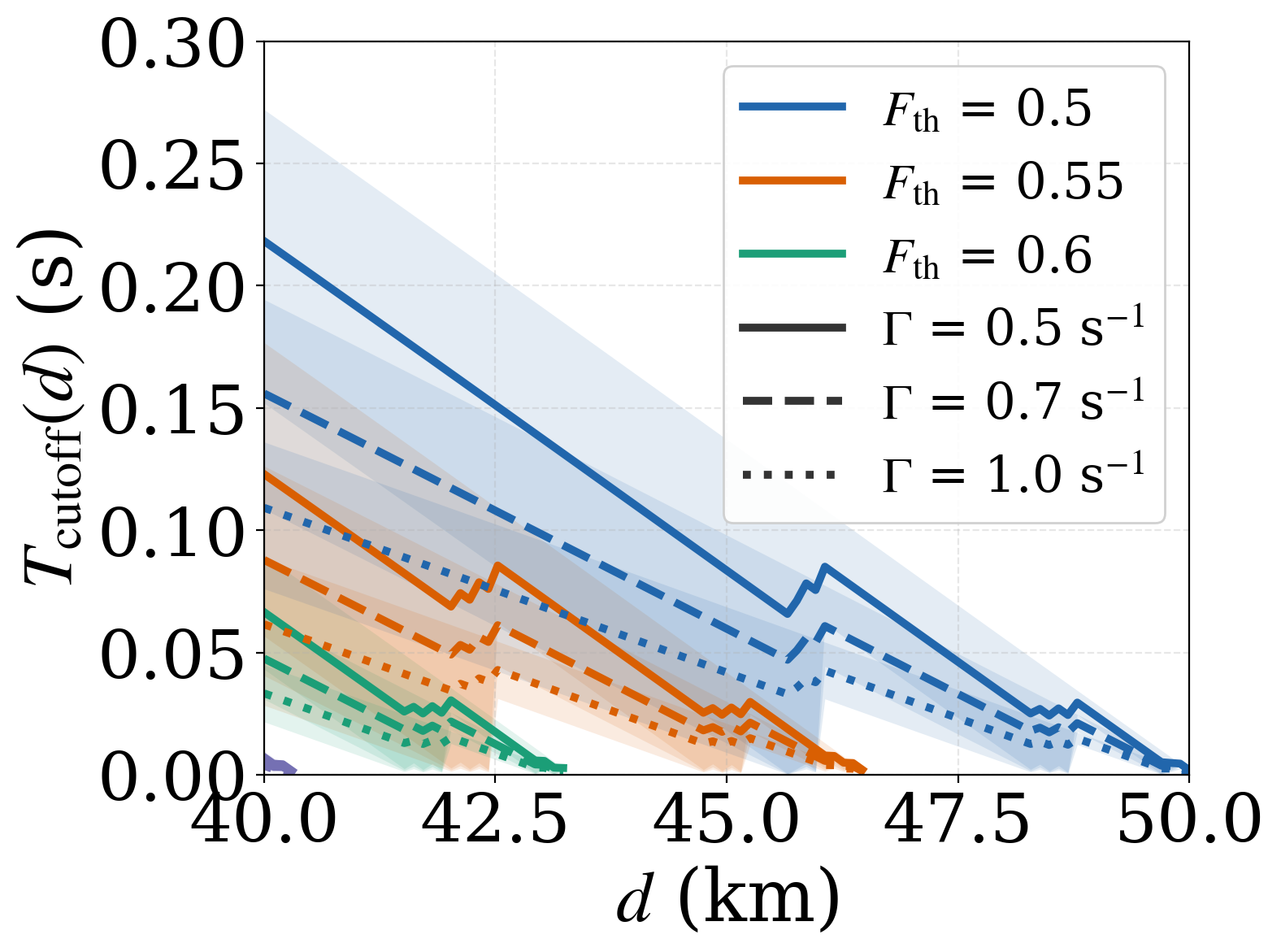}
        \caption{Cutoff Time}
        \label{fig:Cutoff Time}
    \end{subfigure}
    \caption{Store an EPR Pair}
    \label{fig:LUE}
\end{figure}

Fig.~\ref{fig:Fidelity Evolution Range} shows the evolution of stored EPR-pair fidelity \(F(d,t_{\rm store})\) (Eq.~\eqref{eq: storing fidelity}) over \(t_{\rm store} \in (0,5]\) s, with the shaded band representing the range across all parameter combinations. 
The green area shows the fidelity evolution across the full range of distances \(d \in [40,45]\) km.
At \(t_{\rm store}=0\), link fidelity ranges from \(48.54\%\) to \(54.95\%\), reflecting variations in the initial fidelity \(F'_0(d)\) due to the free-space channel parameters. Fidelity then decays exponentially due to amplitude damping and approaches zero by \(t_{\rm store}=5\) s. The results indicate an entanglement half-life well below \(0.5\) s under the considered decoherence rates, highlighting the importance of utilizing an EPR pair as quickly as possible in practical satellite quantum communication. The results show \(F(d,t_\text{store})<50\%\) when \(t>0.5\) s, indicating that \(T_\text{cutoff}(d)<0.5\) s.

Fig.~\ref{fig:Cutoff Time} shows the cutoff time \(T_{\rm cutoff}(d)\) (Eq.~\eqref{eq: max age}) over \(d=40\)--\(50\) km, with the shaded bands representing variations across the considered different \(F_{\rm th}\). In general, \(T_{\rm cutoff}\) decreases with increasing \(d\), \(F_{\rm th}\), \(\Gamma\) and \(\sigma_\delta\). For \(\varepsilon=3\%\), the maximum \(T_{\rm cutoff}(d)\) is \(0.26\)s at \(d=40\) km, with \(F_{\rm th}=0.5\), \(\Gamma=0.5~\mathrm{s}^{-1}\) and \(\sigma_\delta=0.5~\mu\mathrm{rad}\). At the same distance and decoherence rate, the mean \(T_{\rm cutoff}(d)\) decreases from \(0.20\) s at \(F_{\rm th}=0.5\) to \(0.11\) s at \(F_{\rm th}=0.55\) and \(0.05\) s at \(F_{\rm th}=0.6\).

\begin{figure}[t]
  \centering
    \includegraphics[width=0.33\textwidth]{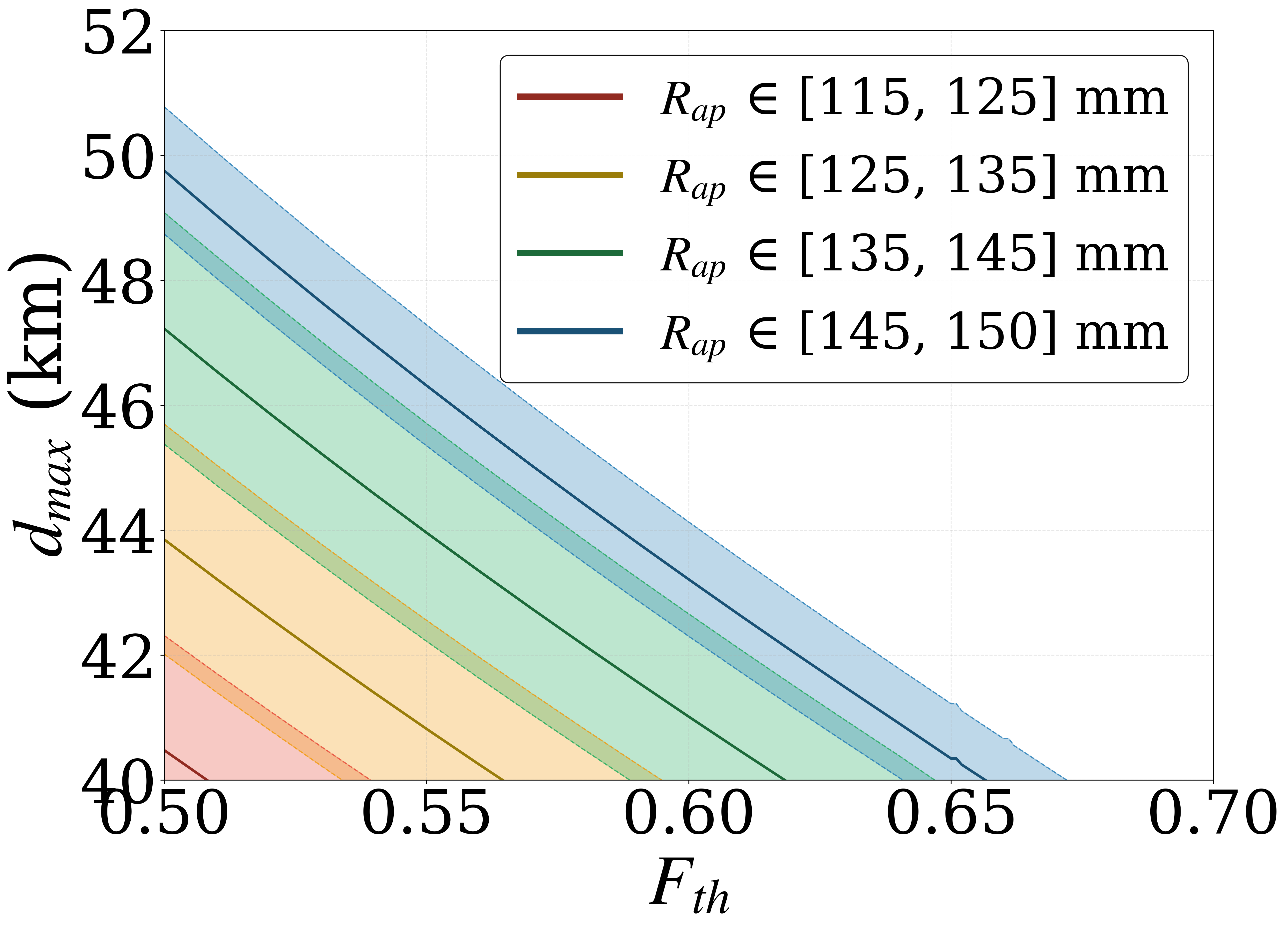}
    \captionsetup{font=footnotesize}
    \caption{Maximum Transmission Distance}
    \label{fig:Maximum Transmission Distance with Rotation}
\end{figure}

Fig.~\ref{fig:Maximum Transmission Distance with Rotation} shows the dependence of the maximum one-hop transmission distance \(d_{\max}\) on the fidelity threshold \(F_{\rm th}\) and receiver aperture radius \(R_{\rm ap}\), with each shaded band representing variations across QBER \(\varepsilon\) and rotational jitter \(\sigma_{\rm rot}\). In general, \(d_{\max}\) increases with \(R_{\rm ap}\) but decreases with increasing \(F_{\rm th}\) and \(\varepsilon\). For the minimum considered link distance of \(40\) km, the highest achievable threshold is approximately \(F_{\rm th}=0.66\). The maximum \(d_{\max}=50.78\) km occurs at \(F_{\rm th}=0.5\), \(\varepsilon=1\%\) and \(R_{\rm ap}=150\) mm. For \(F_{\rm th}=0.5\), an aperture radius of at least \(115\) mm is required to achieve \(d_{\max}\geq40\) km with the effect of polarization rotation, while \(R_{\rm ap}\geq150\) mm provides greater robustness under less favorable conditions.

For \(R_{\rm ap}\in[100,110]\) mm, polarization rotation combined with high \(\varepsilon\) can reduce the initial fidelity \(F'_0(d)\) below \(F_{\rm th}=0.5\), resulting in \(p'(d)=0\) and hence no feasible \(d_{\max}\), explaining the absence of results in this range. Table~\ref{tab:D_max} further compares \(d_{\max}\) with and without polarization rotation and shows that, whenever a feasible link exists, its effect on \(d_{\max}\) is small. Thus, polarization rotation can reasonably be neglected when analyzing the maximum one-hop transmission distance over the considered \(40\)--\(50\) km range with \(R_{\rm ap} \geq 115\) mm.

\begin{table}[tbp]
\centering
\caption{\(d_{max}\) Comparison}
\label{tab:D_max}
\begin{tabular}{c c c c}
\toprule
\(\varepsilon\) & Mean Diff. (km) & Max Diff. (km) & Relative Diff. (\%) \\
\midrule
1\% & \(5.3\times10^{-9}\)  & \(2.48\times10^{-7}\) & $1.06 \times 10^{-7}$ \\
2\% & \(2.07\times10^{-8}\) & \(2.41\times10^{-7}\) & $1.12 \times 10^{-7}$ \\
3\% & \(1.11\times10^{-8}\) & \(2.31\times10^{-7}\) & $1.09 \times 10^{-7}$ \\
\bottomrule
\end{tabular}
\end{table}

\subsection{Markov Chain Models}\label{sec:Markov Chain Models}

We evaluate the pre-generation and on-demand strategies using four performance metrics: request satisfaction rate \(\mathcal{R}\), average waiting time \(\mathcal{W}\), link utilization efficiency \(\mathcal{U}\) and average consumed fidelity \(\mathcal{F}\). The analytical models are cross-validated against four Monte Carlo simulation baselines, two for each strategy.

\begin{figure}[htbp]
    \centering
    \captionsetup{font=footnotesize}
    \begin{subfigure}{0.48\linewidth}
        \centering
        \includegraphics[width=\linewidth]{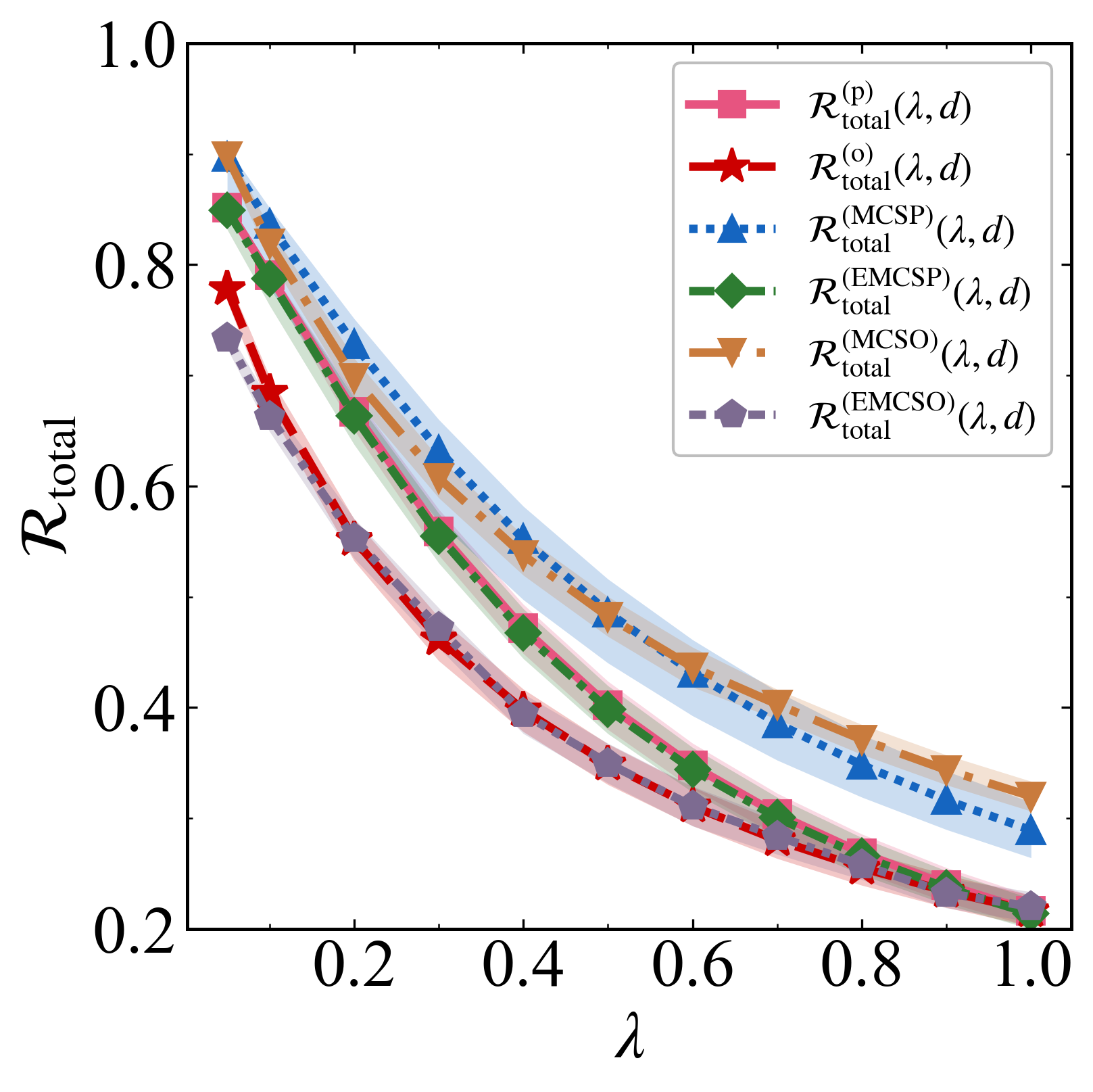}
        \caption{Total Requests}
        \label{fig: total Request Satisfaction Rate}
    \end{subfigure}
    \hfill
    \begin{subfigure}{0.48\linewidth}
        \centering
        \includegraphics[width=\linewidth]{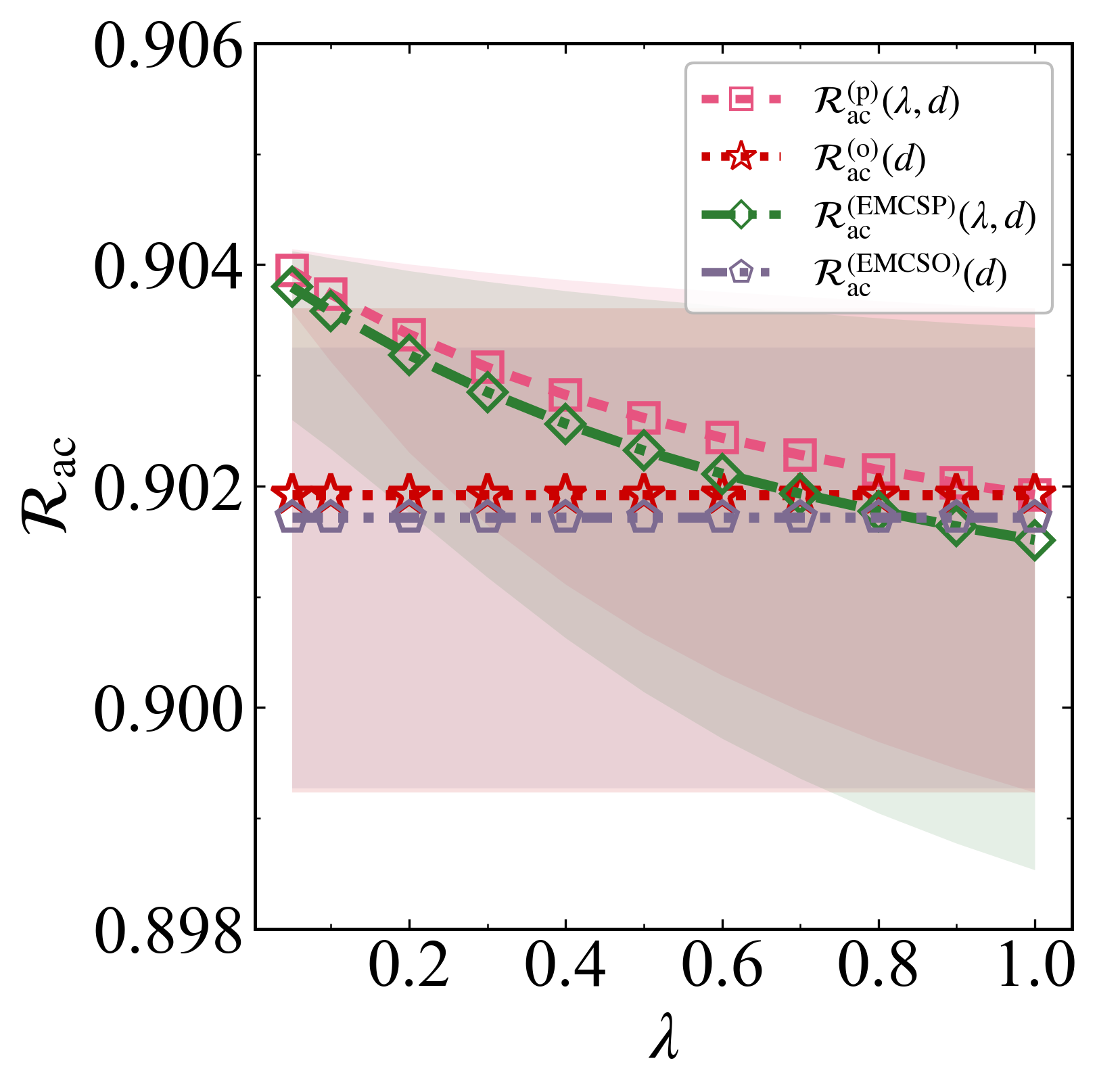}
        \caption{Accepted Requests}
        \label{fig: accepted Request Satisfaction Rate}
    \end{subfigure}
    \caption{Request Satisfaction Rate}
    \label{fig:RSR}
\end{figure}

\begin{figure*}[htbp]
    \centering
    \captionsetup{font=footnotesize}

    \begin{minipage}[t]{0.74\textwidth}
        \vspace{0pt}  
        \centering
        \begin{subfigure}[t]{0.32\linewidth}
            \centering
            \includegraphics[width=\linewidth]{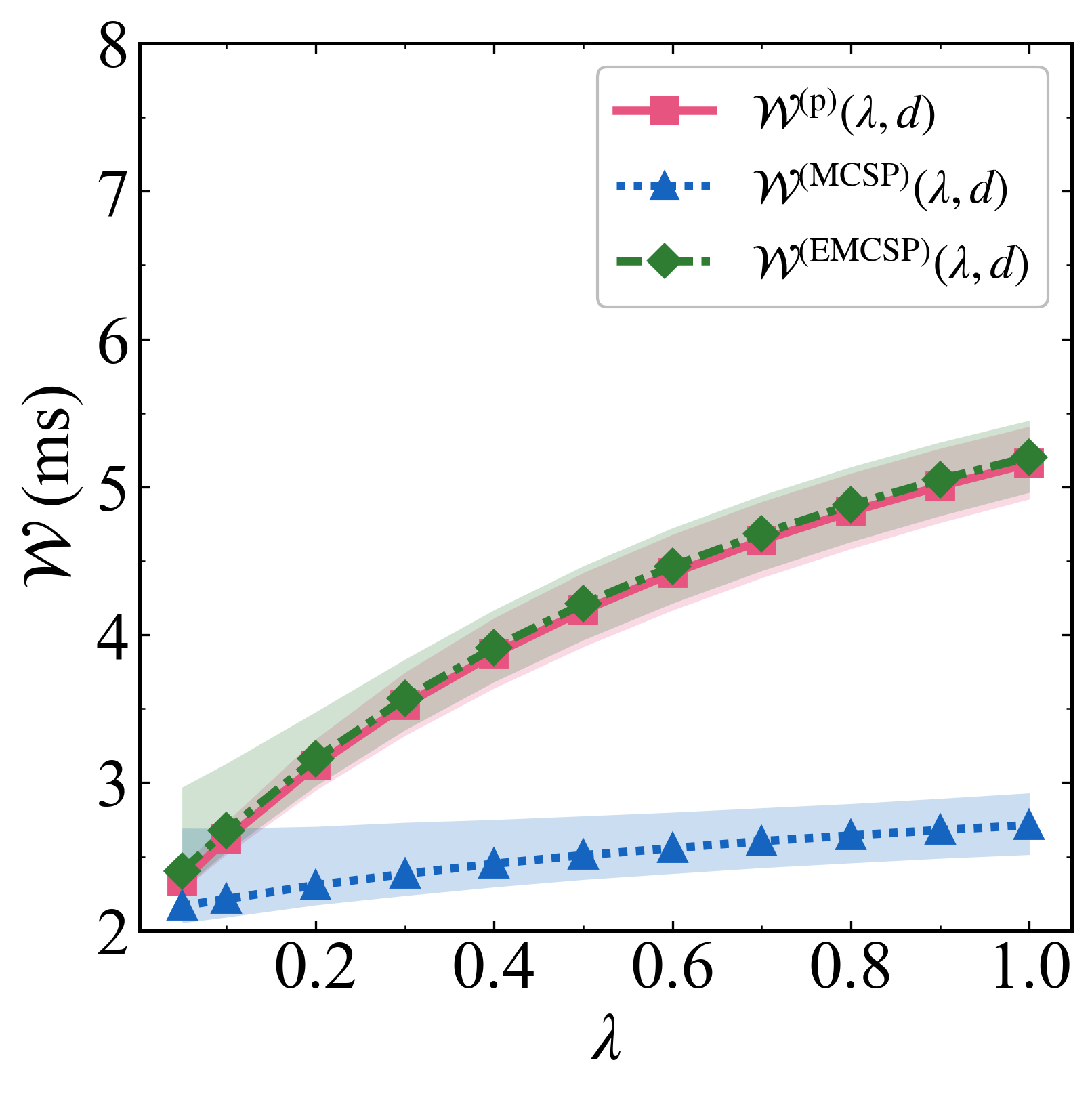}
            \caption{Pre-generation Models}
            \label{fig:pre-gen average waiting time}
        \end{subfigure}
        \hfill
        \begin{subfigure}[t]{0.32\linewidth}
            \centering
            \includegraphics[width=\linewidth]{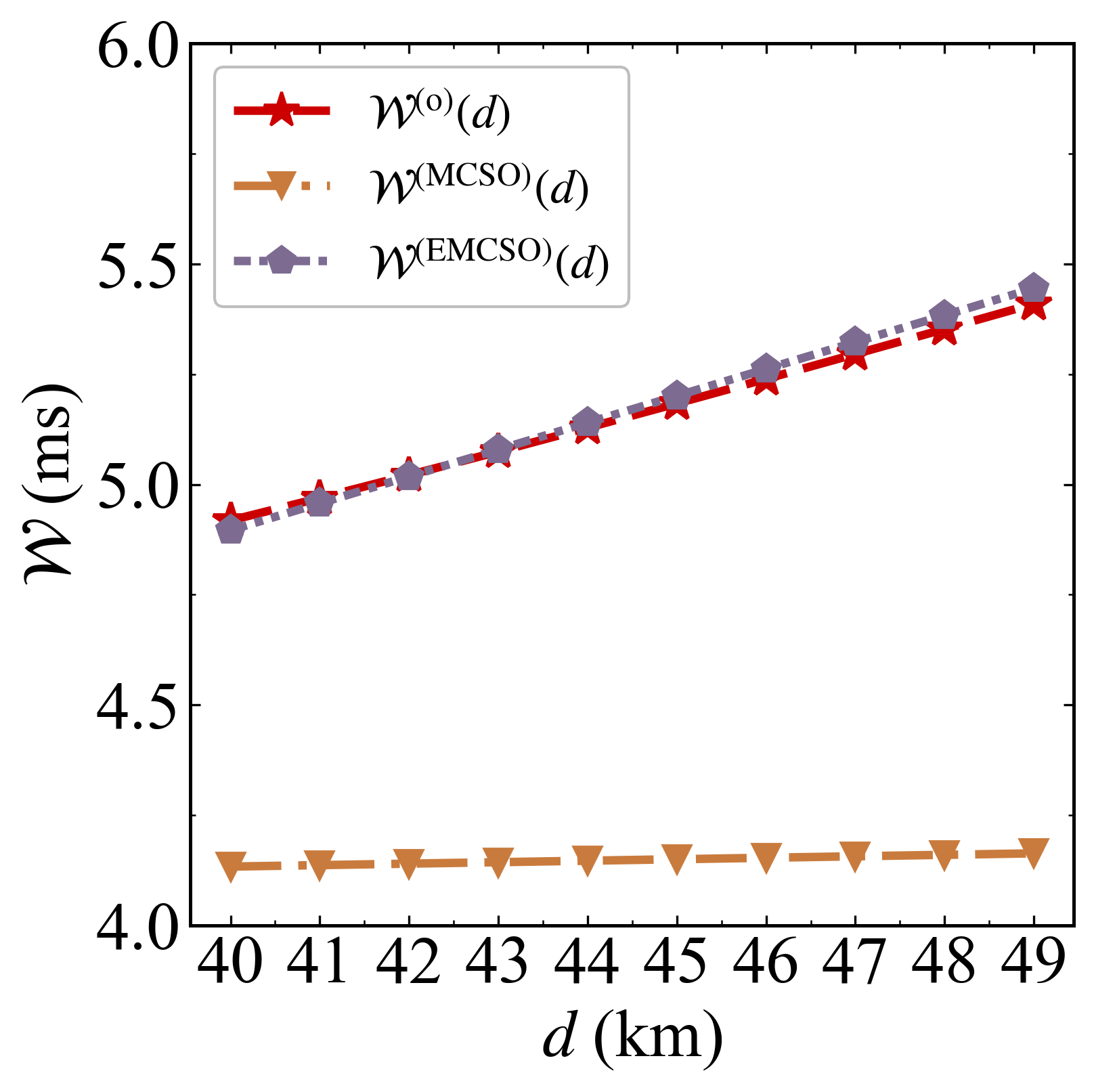}
            \caption{On-demand vs. \(d\) (\(\lambda=0.5\))}
            \label{fig:on-demand average waiting time of distance}
        \end{subfigure}
        \hfill
        \begin{subfigure}[t]{0.32\linewidth}
            \centering
            \includegraphics[width=\linewidth]{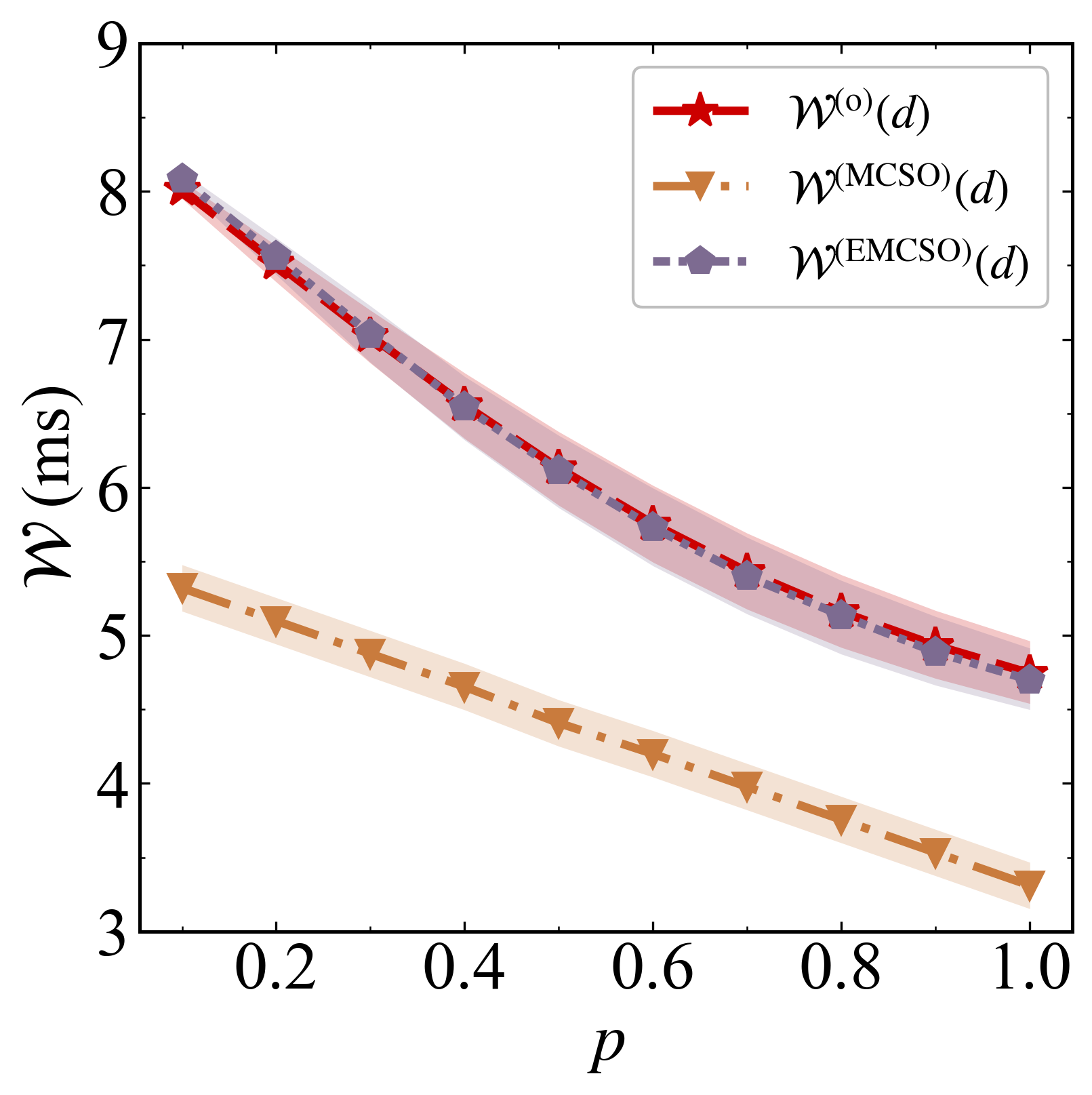}
            \caption{On-demand vs. \(p\) (\(\lambda=0.5\))}
            \label{fig:on-demand average waiting time of EPR pair generation rate}
        \end{subfigure}
        \caption{Average Waiting Time}
        \label{fig:AWT}
    \end{minipage}
    \hfill
    \begin{minipage}[t]{0.24\textwidth}
        \vspace{0pt}   
        \centering
        \includegraphics[width=\linewidth]{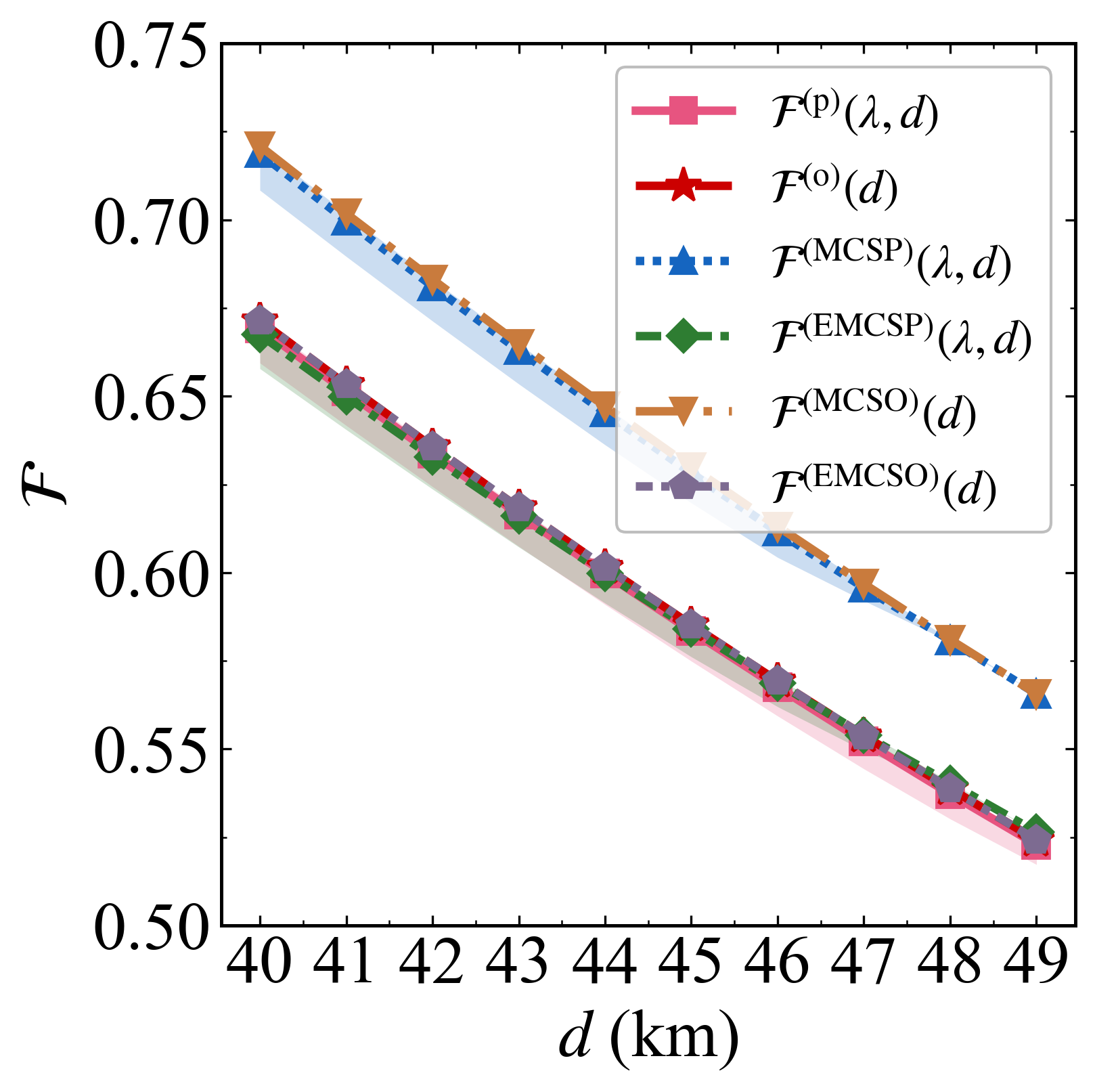}
        \caption{Average Consumed Fidelity}
        \label{fig:average consumed link fidelity}
    \end{minipage}
\end{figure*}

\begin{description}[leftmargin=1em]
\item[\textit{Baseline for Pre-Generation Strategy:}] 
We adopt Walln{\"o}fer's original Monte Carlo simulation (MCSP)~\cite{wallnofer2022simulating} and an enhanced version (denoted by EMCSP). MCSP focuses primarily on entanglement generation for evaluating QKD rates and represents free-space degradation through a damping rate, without explicitly modeling photon loss, beam transmittance, polarization rotation, storage cutoff time or utilization delay. EMCSP extends MCSP to include these effects, together with the fidelity loss and delay associated with E91 utilization, and therefore closely corresponds to our analytical pre-generation model. We use EMCSP to validate the analytical results of the pre-generation strategy.
\item[\textit{Baseline for On-Demand Strategy:}] For the on-demand strategy, no existing free-space simulation is directly applicable. We therefore adapt a Monte Carlo simulation originally developed for a fiber-based entanglement network~\cite{chakraborty2019distributed}, replacing its fiber channel with the free-space model of MCSP to obtain MCSO. Like MCSP, MCSO does not explicitly model photon loss, beam transmittance, polarization rotation, storage cutoff time, or utilization delay. We extend MCSO with the same effects incorporated into EMCSP, together with link consumption, to obtain EMCSO. We use EMCSO to validate the analytical results of the on-demand strategy. 
\end{description}
All simulations use \(1000\) Markov steps and the LEO ISL parameters are in Table~\ref{tab:symbols_values}. Shaded areas in Fig.~(\ref{fig:RSR})(\ref{fig:LUE})(\ref{fig:AWT}) and~(\ref{fig:average consumed link fidelity}) show the range over \(d \in [40, 49]\) km and solid lines with markers show the mean values. Unless otherwise stated or varies in a figure, we consider request arrival rates \(\lambda\in(0,1]\), \(\Delta t = 1\)ms, \(G_\text{max} = 10, R_\text{ap}=150\) mm, \(\sigma_\delta=0.5~\mu\)rad, \(\varepsilon=1\% , \Gamma=0.5~\mathrm{s}^{-1},\sigma_\text{rot}=0.7~\mu\)rad, \(p^\text{(Cls)}_{\text{single}} = p_\text{sync} = 0.99, p_\text{det}^{(A)} = p_\text{det}^{(B)} = 0.98, p=0.8\) and link distances \(d=\{40,45,49\}\) km corresponding to \(K(d)=\{295,158,48\}\) time slots, respectively. The following subsections examine each performance metric separately.

\begin{figure}[htbp]
    \centering
    \captionsetup{font=footnotesize}
    \begin{subfigure}{0.48\linewidth}
        \centering
        \includegraphics[width=\linewidth]{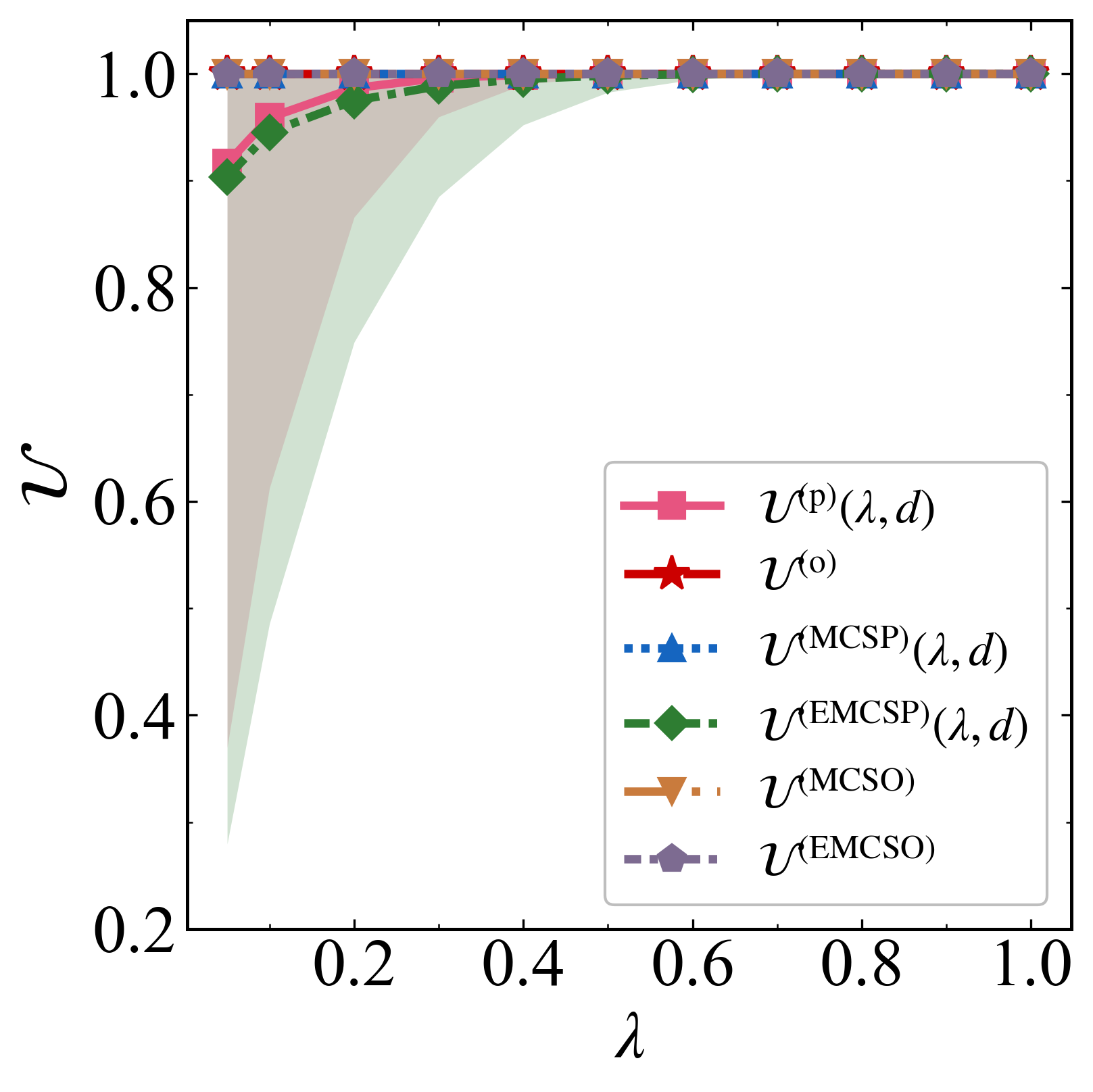}
        \caption{Varying Request Arrival Rate}
        \label{fig:link utilization efficiency of lambda}
    \end{subfigure}
    \hfill
    \begin{subfigure}{0.48\linewidth}
        \centering
        \includegraphics[width=\linewidth]{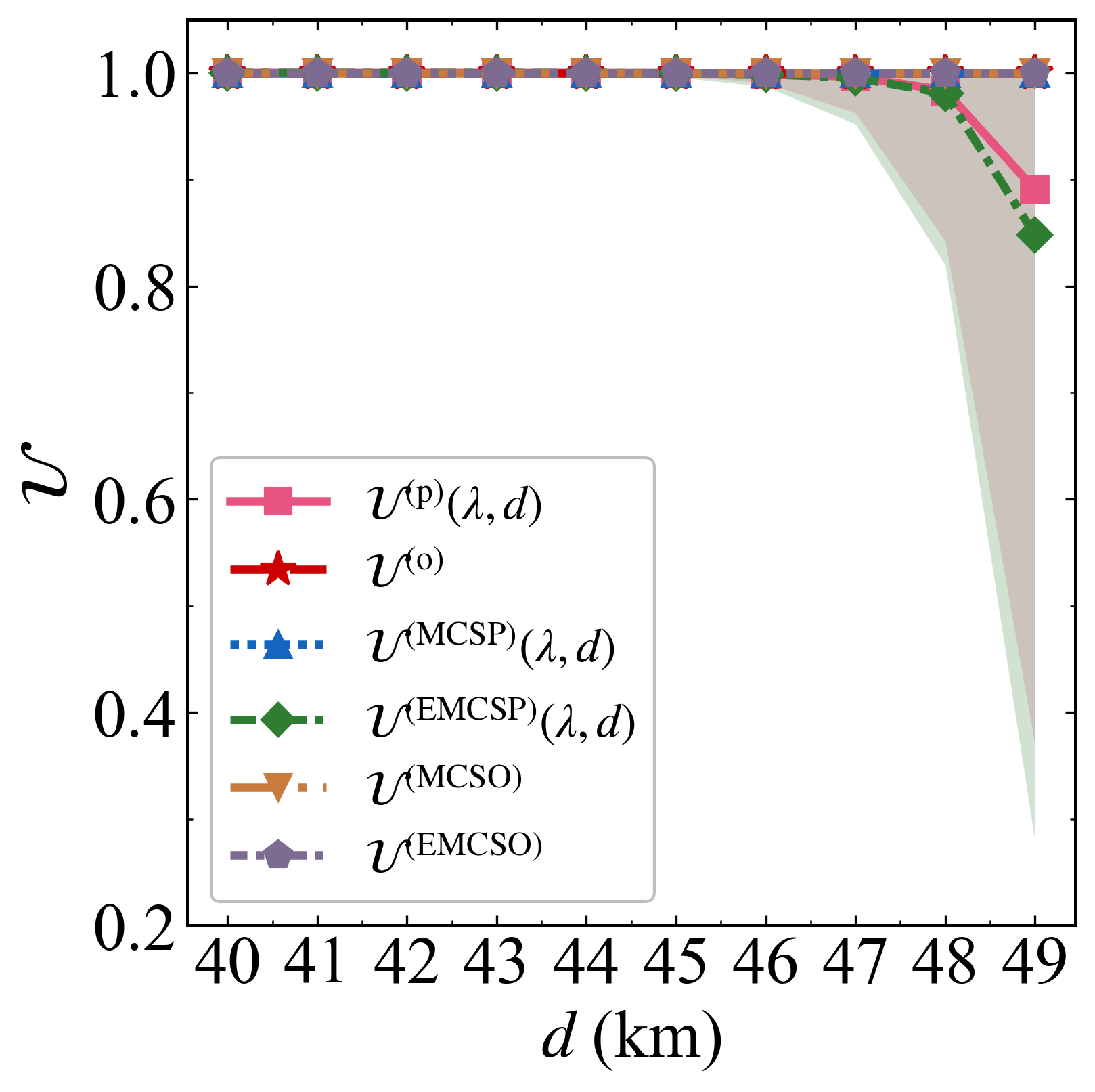}
        \caption{Varying Distance}
        \label{fig:link utilization efficiency of distance}
    \end{subfigure}
    \caption{Link Utilization Efficiency}
    \label{fig:LUE}
\end{figure}

\subsubsection{Request Satisfaction Rate}

As shown in Fig.~\ref{fig: total Request Satisfaction Rate}, the analytical results for both strategies closely agree with their enhanced simulations. For pre-generation, \(\mathcal{R}_{\rm total}^{(\mathrm{p})}(\lambda,d)\) (Eq.~\eqref{eq:R_p_total}) and \(\mathcal{R}_{\rm total}^{(\mathrm{EMCSP})}(\lambda,d)\) differ by at most \(0.005\) for \(d\leq48\) km, increasing to \(0.02\) at \(d=49\) km, where the shorter cutoff \(K(d)=48\) increases sensitivity to stored-link availability. For on-demand, \(\mathcal{R}_{\rm total}^{(\mathrm{o})}(\lambda,d)\) (Eq.~\eqref{eq:R_o_total}) and \(\mathcal{R}_{\rm total}^{(\mathrm{EMCSO})}(\lambda,d)\) differ by at most \(0.012\) for \(\lambda\geq0.2\) at all distances. Larger differences occur only at low arrival rates: \(0.043\)--\(0.046\) at \(\lambda=0.05\) and \(0.019\)--\(0.023\) at \(\lambda=0.1\). Overall, the close agreement validates both analytical models over the evaluated range.

Both \(\mathcal{R}_{\rm total}^{(\mathrm{p})}(\lambda,d)\) and \(\mathcal{R}_{\rm total}^{(\mathrm{o})}(\lambda,d)\) decrease with \(\lambda\), from approximately \(0.85\) and \(0.79\), respectively, at \(\lambda=0.05\) to \(0.20\)--\(0.23\) at \(\lambda=1\). At \(\lambda=1\), the two strategies converge because a request arrives every slot, preventing the pre-generation strategy from entering a storage state. For all \(\lambda<1\), \(\mathcal{R}_{\rm total}^{(\mathrm{p})}(\lambda,d)>\mathcal{R}_{\rm total}^{(\mathrm{o})}(\lambda,d)\), since pre-generated links can serve arriving requests immediately, whereas the on-demand strategy must first generate a link and blocks subsequent requests during generation.
 
In contrast, Fig.~\ref{fig: accepted Request Satisfaction Rate} shows that \(\mathcal{R}_{\rm ac}\) is nearly insensitive to \(\lambda\), with both analytical models closely matching their corresponding simulation models. Across all evaluated \(\lambda\) and \(d\), \(\mathcal{R}_{\rm ac}^{(\mathrm{p})}(\lambda,d)\) (Eq.~\eqref{eq:R_p_ac}) differs from \(\mathcal{R}_{\rm ac}^{(\mathrm{EMCSP})}(\lambda,d)\) by less than \(0.1\%\), while \(\mathcal{R}_{\rm ac}^{(\mathrm{o})}(d)\) (Eq.~\eqref{eq:R_o_ac}) differs from \(\mathcal{R}_{\rm ac}^{(\mathrm{EMCSO})}(d)\) by less than \(0.05\%\). Both strategies remain within \(0.899\)--\(0.904\). From Eq.~\eqref{eq:R_p_ac}, \(\mathcal{R}_{\rm ac}^{(\mathrm{p})}(\lambda,d)\) ranges from \(p_{\rm util}\) as \(\lambda\rightarrow0\), when most accepted requests use stored links, to \(\gamma (d) p_{\rm util}\) as \(\lambda \rightarrow 1\), when stored links become unavailable. With \(p_{\rm util}=0.90
\) and \(\gamma (d) \geq 0.99
\), these bounds differ by less than \(0.005\). In comparison, Eq.~\eqref{eq:R_o_ac} gives \(\mathcal{R}_{\rm ac}^{(\mathrm{o})}(d)=\gamma (d) p_{\rm util}\), independent of \(\lambda\). Thus, both strategies approach the same accepted-request service ceiling at high arrival rates.

\subsubsection{Link Utilization Efficiency (LUE)}

Figs.~\ref{fig:link utilization efficiency of lambda} and~\ref{fig:link utilization efficiency of distance} show link utilization efficiency for different arrival rates ($\lambda$) and distances ($d$), 
respectively. For on-demand, \(\mathcal{U}^{(\mathrm{o})}\), \(\mathcal{U}^{(\mathrm{MCSO})}\), and \(\mathcal{U}^{(\mathrm{EMCSO})}\) remain at \(1\) across all evaluated 
arrival rates and distances. Since links are generated only after a request arrives, every successfully generated link is immediately consumed and none expires unused.

For pre-generation, over the evaluated range, \(\mathcal{U}^{(\mathrm{p})}(\lambda,d)\) and \(\mathcal{U}^{(\mathrm{EMCSP})}(\lambda,d)\) are very close, differing by at most \(0.004\) for \(d \leq 46\) km. They degrades as \(d \rightarrow 49\) km (Fig.~\ref{fig:link utilization efficiency of distance}) and increases as \(\lambda \rightarrow 1\) (Figs.~\ref{fig:link utilization efficiency of lambda}).
\(\mathcal{U}^{(\mathrm{p})}(\lambda,d)\) and \(\mathcal{U}^{(\mathrm{EMCSP})}(\lambda,d)\) remain above 0.99 for \(\lambda \in [0.6,1]\) at \(d=40\)--\(45\) km, so almost every pre-generated link in this range is consumed before it expires. At \(\lambda = 0.05\), \(\mathcal{U}^{(\mathrm{p})}(\lambda,d)\) falls to 0.84 at \(d=48\) km and to 0.37 at \(d=49\) km, while \(\mathcal{U}^{(\mathrm{EMCSP})}(\lambda,d)\) falls to 0.82 and 0.28 at the same two distances. At \(d=49\) km, \(\mathcal{U}^{(\mathrm{p})}(\lambda,d) < 0.87\) and \(\mathcal{U}^{(\mathrm{EMCSP})}(\lambda,d) < 0.75\) for \(\lambda \leq 0.2\) and they goes up to \(> 0.98\) for \(\lambda \geq 0.5\). 
At longer distances, the shorter cutoff time reduces the interval in which a pre-generated link can wait for a request, so the probability that a stored link expires before any request arrives grows as \(\lambda\) falls. At \(d=49\) km, the cutoff is so short that a stored link can meet fewer than 1 request within its lifetime, so most pre-generated links are discarded.

\subsubsection{Average Waiting Time}

For pre-generation (Fig.~\ref{fig:pre-gen average waiting time}), \(\mathcal{W}^{(\mathrm{p})}(\lambda,d)\) increases with \(\lambda\) but varies little with \(d\). At \(d=40\) km, \(\mathcal{W}^{(\mathrm{p})}(\lambda,d)\) and \(\mathcal{W}^{(\mathrm{EMCSP})}(\lambda,d)\) increase from \(2.27\) and \(2.28\) ms at \(\lambda=0.05\) to \(4.92\) and \(4.96\) ms at \(\lambda=1\), respectively. The analytical and simulated results differ by less than \(3\%\) for \(d\leq48\) km, except at \(d=40\) km and \(\lambda\leq0.2\). 
Under these conditions, the long cutoff allows more links to remain stored, making the average waiting time more sensitive to the fraction of requests served directly from storage.

Unlike pre-generation, the on-demand strategy generates a link only after accepting a request. Consequently, Eq.~\eqref{eq:W_o} shows that \(\mathcal{W}^{(\mathrm{o})}(d)\) is independent of \(\lambda\), since every accepted request follows the same generation-then-utilization sequence. We therefore evaluate \(\mathcal{W}^{(\mathrm{o})}(d)\) against the parameters that affect it directly, viz., transmission distance \(d\) and EPR-pair generation probability \(p\).
Fig.~\ref{fig:on-demand average waiting time of distance} shows the average waiting time results of the on-demand strategy for various \(d\) for \(p=0.8\) and \(\lambda=0.5\). As \(d\) increases from \(40\) to \(49\) km, \(\mathcal{W}^{(\mathrm{o})}(d)\) increases from \(4.92\) to \(5.41\) ms, while \(\mathcal{W}^{(\mathrm{EMCSO})}(d)\) increases from \(4.87\) to \(5.37\) ms, with less than \(1\%\) difference between them at all distances. This trend follows Eq.~\eqref{eq:W_o}, where increasing \(d\) increases both the generation duration \(C(d)\) (Eq.~\eqref{eq: waiting time}) and utilization duration \(L(d)\) (Eq.~\eqref{eq: link utilization time}), while reducing the overall generation success probability \(p'(d)\) (Eq.~\eqref{eq: link total success establish rate}) through lower photon capture probability. Consequently, more attempts are typically required to establish a link. In contrast, MCSO underestimates its average waiting time by approximately \(25\)--\(30\%\) because it omits the utilization delay and part of the classical communication overhead.

Fig.~\ref{fig:on-demand average waiting time of EPR pair generation rate} shows the average waiting time results of the on-demand strategy of different \(p\) for \(\lambda=0.5\). As \(p\) increases from \(0.1\) to \(1\), \(\mathcal{W}^{(\mathrm{o})}\) decreases from \(7.95\) to \(4.54\) ms at \(d=40\) km and from \(8.07\) to \(4.96\) ms at \(d=49\) km. A higher \(p\) increases \(p'(d)\), reducing the expected number of generation attempts \(\bar g (d)\) before success. EMCSO agrees with the analytical results within \(1\%\) across all \(p\), while MCSO again underestimates the waiting time because it omits the utilization delay and part of the classical communication overhead.

Comparing the two strategies, \(\mathcal{W}^{(\mathrm{p})} (\lambda,d) \leq \mathcal{W}^{(\mathrm{o})}(d)\) across all \(\lambda,d\) and \(p\), but the gap is decreasing and \(\mathcal{W}^{(\mathrm{p})} (\lambda,d) = \mathcal{W}^{(\mathrm{o})}(d)\) when \(\lambda = 1\), when the pre-generation strategy no longer enters a storage state, which acts the same as the on-demand strategy. For \(\lambda<1\), pre-generation can serve requests from stored links, whereas on-demand requests must wait for link generation. This waiting time has a fidelity cost, that on-demand links are consumed at the minimum storage age \(1 \cdot \Delta t\), whereas pre-generated links may remain stored for up to \(K(d)-1\) additional slots before consumption. Consequently, \(\mathcal{F}^{(\mathrm{p})}(\lambda,d) \leq \mathcal{F}^{(\mathrm{o})}(d)\), despite \(\mathcal{W}^{(\mathrm{p})}(\lambda,d) \leq \mathcal{W}^{(\mathrm{o})}(d)\).

\subsubsection{Averaged Consumed Fidelity}

Fig.~\ref{fig:average consumed link fidelity} shows the analytical results for both strategies agree with their corresponding simulations within \(0.01\) across all evaluated distances. For the pre-generation strategy, \(\mathcal{F}^{(\mathrm{p})}(\lambda,d)\) decreases with \(d\), from \(0.67\) at \(d=40\) km to \(0.52\) at \(d=49\) km, compared with \(0.67\) and \(0.53\) for EMCSP, respectively. At a fixed distance, \(\mathcal{F}^{(\mathrm{p})}(\lambda,d)\) increases slightly with \(\lambda\) by less than \(0.03\), as more frequent requests reduce the time stored links can age before consumption.
For the on-demand strategy, \(\mathcal{F}^{(\mathrm{o})}(d)\) is independent of \(\lambda\) but decreases with \(d\), as Eq.~\eqref{eq:F_o}, from \(0.68\) at \(d=40\) km to \(0.53\) at \(d=49\) km.

In summary, the analytical models closely match the enhanced Monte Carlo baselines (EMCSP/EMCSO), whereas the raw baselines (MCSP/MCSO) deviate because they omit free-space and utilization effects. Pre-generation strategies achieve a higher request satisfaction rate and lower waiting time, while on-demand strategies achieve full link utilization and higher consumed fidelity.

\section{Conclusion}\label{sec:Conclusion}

This paper presents Markov chain models for dynamic one-hop LEO satellite quantum links under pre-generation and on-demand strategies. The models capture link generation, storage, utilization, distance-dependent physical effects and yield closed-form expressions for four performance metrics, which are request satisfaction rate \(\mathcal{R}\), average waiting time \(\mathcal{W}\), link utilization efficiency \(\mathcal{U}\), and average consumed fidelity \(\mathcal{F}\). The analytical results are validated against Monte Carlo simulations for both strategies.

The results reveal a fundamental trade-off between responsiveness and fidelity. Pre-generation achieves lower \(\mathcal{W}\) and higher \(\mathcal{R}_{\rm total}\) by allowing requests to use stored links. Still, these links may expire unused or lose fidelity through decoherence during storage. In contrast, on-demand generation eliminates unused-link expiry and minimizes storage-induced fidelity loss, but requires every accepted request to wait for link generation. Thus, neither strategy is universally superior, pre-generation favors responsiveness and request satisfaction, whereas on-demand favors link utilization and consumed fidelity. In practice, the two strategies may be combined or selected dynamically according to the request arrival rate \(\lambda\), link distance \(d\) and fidelity threshold \(F_{\rm th}\).

Beyond the strategy-level trade-offs, the analysis provides practical design guidelines for LEO quantum networks. Photon capture probability and initial fidelity degrade rapidly with distance, limiting the feasible one-hop QISL range to approximately \(50\) km under the considered error conditions. The corresponding coherence time is at most \(0.26\) s, highlighting the importance of minimizing link storage and utilization delays in quantum applications. For \(F_{\rm th}=0.5\), a receiver aperture radius of at least \(115\) mm is required under the evaluated conditions, while \(150\) mm extends the maximum distance to \(50.78\) km. Polarization rotation has negligible effect on this maximum distance for the short links considered, so photon loss, pointing error, aperture size and memory decoherence are the more important physical design considerations.

Together, the closed-form performance expressions and these physical design guidelines provide a basis for selecting an entanglement-management strategy and configuring one-hop LEO quantum links. Our future work will develop adaptive policies that dynamically select between pre-generation and on-demand operations according to network conditions and extend the present free-space framework to multi-hop satellite quantum networks.

\bibliographystyle{IEEEtran}
\bibliography{IEEEabrv,references}

\clearpage
\appendices

\section{Derivation of the System Model}\label{app: SM}

To improve the readability of the main text, the detailed derivations of the single-hop QISL model are presented in this appendix. The derivations include the physical channel model, entanglement-generation success probability, storage fidelity evolution, maximum transmission distance, utilization duration and utilization success probability.

\begin{figure}[h]
  \centering
\includegraphics[width=0.475\textwidth]{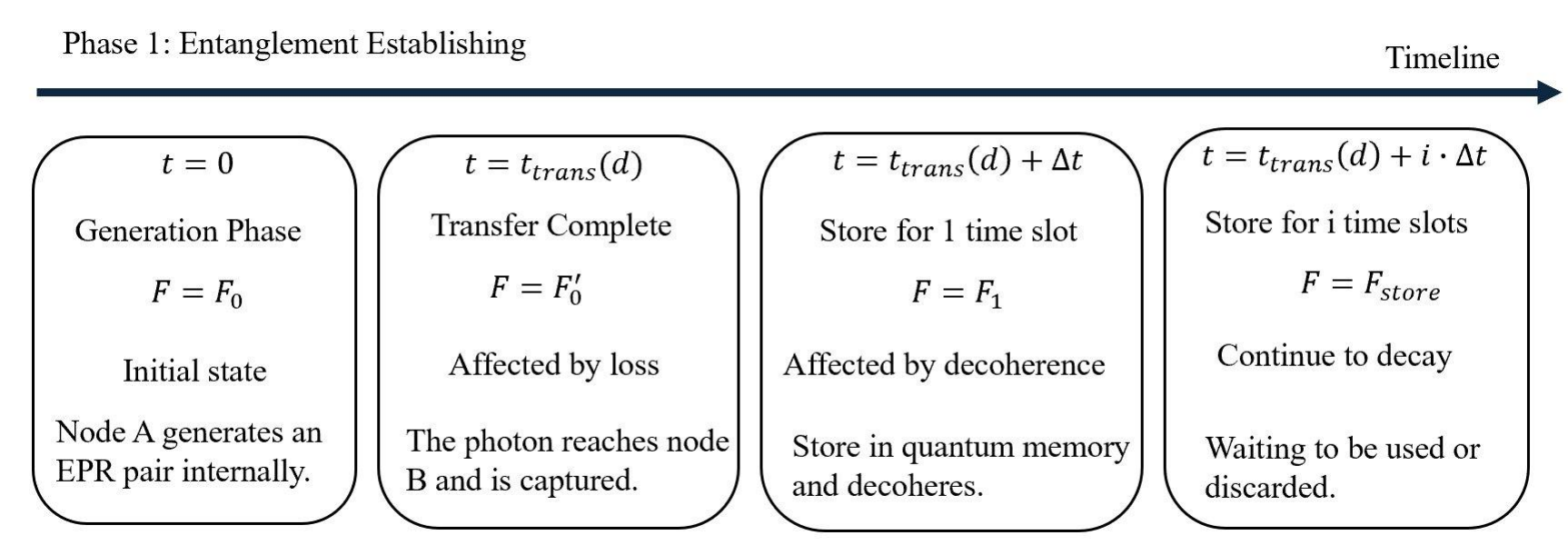}
  \caption{EPR Pair Fidelity Timeline}
  \label{fig:epr_timeline}
\end{figure}

\subsection{Link Generation in Free Space}

The free-space entangled link generation model considers three factors:
\begin{itemize}
    \item photon transmission success probability \(q(d)\);
    \item polarization rotation \(\sigma_{\mathrm{rot}}\); and 
    \item quantum memory decoherence \(\Gamma\).
\end{itemize}

\subsubsection{Gaussian Beam Diffraction}

The photon transmission success probability \(q(d)\) depends on whether the receiver aperture captures the transmitted photon. For ISLs above \(200\) km, atmospheric absorption, scattering and turbulence are neglected. Assuming Gaussian beam propagation~\cite{kogelnik1966laser}, the spot radius at distance \(d\) is
\[
W(d)=W_0\sqrt{1+\left(\frac{\omega d}{\pi W_0^2}\right)^2},
\]
where \(W_0\) is the beam waist radius and \(\omega\) is the photon wavelength. Since LEO ISL distances greatly exceed the Rayleigh length \(d_{\mathcal R}=\pi W_0^2/\omega\),
\[
W(d)\approx\frac{\omega d}{\pi W_0},
\]
yielding the divergence angle and beam area as
\[
\theta\approx\frac{\omega}{\pi W_0},\ \ \ \ A_{\rm beam}=\pi W^2(d)=\frac{\omega^2d^2}{\pi W_0^2}.
\]
Let the receiver aperture radius be \(R_{\rm ap}\). Under perfect alignment, the received intensity follows
\[
I(r,d)
=
I_0
\frac{W_0^2}{W^2(d)}
e^{-2r^2/W^2(d)},
\]
where \(r\) is the radial distance from the beam axis. Integrating over the receiver aperture gives the transmittance as
\begin{equation}
\eta(d)
=
\frac{\int_0^{R_{\rm ap}}I(r,d)\,2\pi r\,dr}
{\int_0^\infty I(r,d)\,2\pi r\,dr}
=
1-e^{-\frac{2\pi^2W_0^2R_{\rm ap}^2}{\omega^2d^2}}.
\label{eq:perfect_point_q1}
\end{equation}
Practical LEO satellites experience pointing errors due to finite attitude-control accuracy. Let the pointing error be \(\delta\) (rad), producing a beam offset
\[
d_{\rm dev}=d\delta.
\]
Assuming \(R_{\rm ap}\ll W(d)\), the intensity across the aperture is locally uniform, so
\[
I(d_{\rm dev},d)
=
I_0
\frac{W_0^2}{W^2(d)}
e^{-2d_{\rm dev}^2/W^2(d)}.
\]
Since the collected power is proportional to the local intensity, the pointing loss factor is
\[
\frac{P_{\rm rec}(d_{\rm dev})}{P_{\rm rec}(0)}
=
e^{-2d_{\rm dev}^2/W^2(d)}.
\]
Therefore, the transmittance including pointing error is
\begin{equation}
\eta_{\rm error}
=
\eta
e^{-2d_{\rm dev}^2/W^2(d)},
\label{eq:Single_pointing_error_transmittance1}
\end{equation}
where larger offsets reduce transmittance, while a larger beam radius increases tolerance to pointing errors.

\subsubsection{Pointing Error Effects}

We then account for random pointing errors. Let the pointing error components
\((\delta_x,\delta_y)\) be independent Gaussian random variables with zero mean and variance \(\sigma_\delta^2\). Their joint probability density function is
\[
P(\delta_x,\delta_y)
=
\frac{1}{2\pi\sigma_\delta^2}
e^{-(\delta_x^2+\delta_y^2)/(2\sigma_\delta^2)}.
\]
Since the transmittance depends only on the pointing error magnitude, we transform to polar coordinates,
\[
\delta_x=\rho\cos\nu,\qquad
\delta_y=\rho\sin\nu,
\]
whose Jacobian is \(|J|=\rho\). Integrating over the azimuth angle gives the Rayleigh distribution,
\[
P(\rho)
=
\frac{\rho}{\sigma_\delta^2}
e^{-\rho^2/(2\sigma_\delta^2)}.
\]
Identifying the pointing error magnitude as \(\delta=\rho\),
\[
P(\delta)
=
\frac{\delta}{\sigma_\delta^2}
e^{-\delta^2/(2\sigma_\delta^2)},
\qquad
\delta\ge0.
\]
The average transmittance is obtained by averaging Eq.~\eqref{eq:Single_pointing_error_transmittance1} over the pointing error distribution as
\begin{equation*}
\begin{aligned}
\mathbb{E}[\eta_{\rm error}]
&=
\int_0^\infty
\eta_{\rm error}(\delta)
P(\delta)
\,d\delta \\
&=
\frac{\eta}{\sigma_\delta^2}
\int_0^\infty
\delta
\exp\!\left[
-\delta^2
\left(
\frac{2d^2}{W^2(d)}
+
\frac{1}{2\sigma_\delta^2}
\right)
\right]
d\delta \\
&=
\frac{\eta}
{1+\dfrac{4d^2\sigma_\delta^2}{W^2(d)}}.
\end{aligned}
\end{equation*}
Substituting \(W_0=\omega/(\pi\theta)\) yields the photon transmission success probability. Under perfect alignment,
\begin{equation}
q_{\rm ideal} (d)
=
\eta(d)
=
1-
e^{-2R_{\rm ap}^2/(\theta^2d^2)},
\label{eq:q_ideal}
\end{equation}
whereas with pointing errors,
\begin{equation}
q_{\rm error} (d)
=
\frac{
1-
e^{-2R_{\rm ap}^2/(\theta^2d^2)}
}
{
1+\dfrac{4\sigma_\delta^2}{\theta^2}
}.
\label{eq:q_err}
\end{equation}
which is essentially Eq. (\ref{eq: q-error}).
Here, \(\theta\) is the beam divergence half-angle, \(R_{\rm ap}\) is the receiver aperture radius and \(\sigma_\delta\) is the pointing error standard deviation.

\begin{figure}[t]
  \centering
\includegraphics[width=0.5\textwidth]{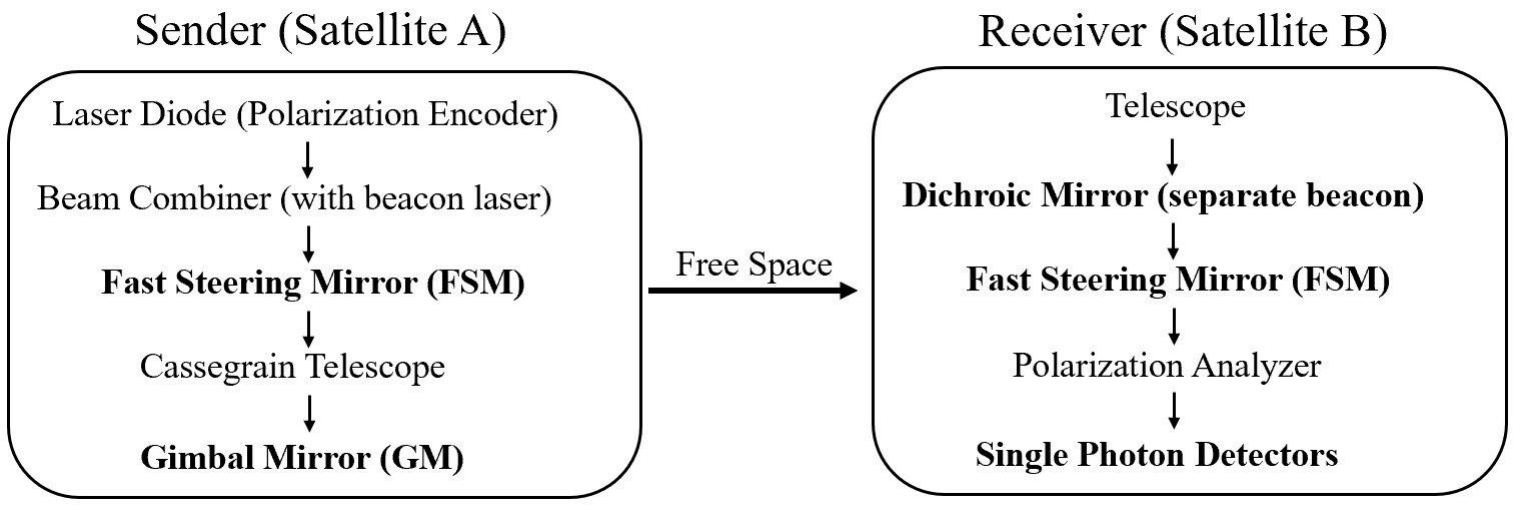}
  \caption{Single Photon Path}
  \label{fig:Light structure}
\end{figure}

\subsubsection{Polarization Rotation}

We then model the polarization transformation experienced by photons during transmission. As shown in Fig.~\ref{fig:Light structure}, Satellite \textit{A} transmits encoded photons through a Cassegrain telescope after coarse pointing by a gimbal mirror and fine pointing by a fast steering mirror (FSM). Satellite \textit{B} collects the photons using a telescope, separates the beacon signal with a dichroic mirror, performs fine tracking using an FSM and measures the quantum states using a polarization analyzer and single-photon detector~\cite{wu2020polarization}. The moving mirrors in the pointing and tracking systems introduce time-dependent polarization rotations, which can degrade entanglement fidelity if uncompensated.

Consider a satellite moving with velocity \(\vec{v}_{\rm sat}\). The FSMs continuously adjust their orientation to maintain optical alignment and each mirror reflection induces a coordinate transformation in the polarization basis. If the mirror normal forms an angle \(\zeta(t)\) with the optical axis and the incidence plane rotates by \(\vartheta(t)\), the photon polarization undergoes a rotation on the Bloch sphere~\cite{bassett2014quantum}. A single-photon polarization state in the computational basis \(\{|H\rangle,|V\rangle\}\) can be written as
\[
|\phi\rangle
=
\cos\!\left(\frac{\chi}{2}\right)|H\rangle
+
e^{i\psi}\sin\!\left(\frac{\chi}{2}\right)|V\rangle,
\]
where \(\chi\in[0,\pi]\) and \(\psi\in[0,2\pi]\) are the Bloch-sphere angles. A rotation by angle \(\vartheta_{\rm rot}\) around axis
\(\hat{\varsigma}=(\varsigma_x,\varsigma_y,\varsigma_z)\) is described by
\[
R_{\rm rot}(\vartheta_{\rm rot},\hat{\varsigma})
=
\exp\!\left(
-i\frac{\vartheta_{\rm rot}}{2}
\hat{\varsigma}\cdot\vec{\sigma}
\right),
\]
where \(\vec{\sigma}=(\sigma_x,\sigma_y,\sigma_z)\) denotes the Pauli matrices.

For the ISL considered here, the dominant rotation is in the linear-polarization basis~\cite{pistoni1995simplified}. Thus, the rotation around the propagation axis is modeled as
\[
R_{\rm prop}(\vartheta_{\rm rot})
=
\begin{pmatrix}
\cos\vartheta_{\rm rot} & -\sin\vartheta_{\rm rot} \\
\sin\vartheta_{\rm rot} & \cos\vartheta_{\rm rot}
\end{pmatrix},
\]
which acts on the polarization amplitudes as
\[
\begin{pmatrix}
\kappa_H\\
\kappa_V
\end{pmatrix}_{\rm out}
=
R_{\rm prop}(\vartheta_{\rm rot})
\begin{pmatrix}
\kappa_H\\
\kappa_V
\end{pmatrix}_{\rm in},
\]
where \(\kappa_H\) and \(\kappa_V\) are the horizontal and vertical polarization amplitudes.
The rotation angle is time-dependent and consists of a deterministic component and random jitter as
\[
\vartheta_{\rm rot}(t)
=
\vartheta_{\rm sys}(t)
+
\vartheta_{\rm jit}(t).
\]
The deterministic component arises from orbital motion and is approximated as~\cite{toyoshima2011polarization}
\begin{equation}
\vartheta_{\rm sys}(t)
=
\frac{v_{\rm sat}}{R_{\rm E}+h_{\rm sat}}t,
\label{eq: vartheta sys}
\end{equation}
where \(R_{\rm E}=6371\,\mathrm{km}\) is the Earth radius, \(h_{\rm sat}\) is the satellite orbital altitude, \(v_{\rm sat}\) is the LEO satellite orbital velocity and \(t\) is the time during photon propagation. The jitter component, caused by platform vibration, attitude-control error and mirror-positioning uncertainty, is modeled as
\(
\vartheta_{\rm jit}(t)\sim \mathcal{N}(0,\sigma_{\rm rot}^2),
\)
where \(\sigma_{\rm rot}^2\) is the rotational jitter variance~\cite{tayebi2023dynamics}.

We now evaluate the effect of polarization rotation on Bell-state fidelity. For the ideal Bell state
\[
|\Phi^+\rangle
=
\frac{|HH\rangle+|VV\rangle}{\sqrt{2}},
\]
assume photon \textit{a} undergoes rotation \(R_{\rm prop}(\vartheta_{\rm rot} (t))\), while photon \textit{b} remains unchanged. The rotated state is
\[
\begin{split}
|\phi_{\rm rot} (t)\rangle
&=
\left(R_{\rm prop}(\vartheta_{\rm rot} (t))\otimes I\right)
|\Phi^+\rangle \\
&=
\frac{1}{\sqrt{2}}
\bigl[
\cos(\vartheta_{\rm rot} (t))|HH\rangle
-\sin(\vartheta_{\rm rot} (t))|HV\rangle\\
& 
+\sin(\vartheta_{\rm rot} (t))|VH\rangle
+\cos(\vartheta_{\rm rot} (t))|VV\rangle
\bigr],
\end{split}
\]
where \(I\) is the identity operator acting on photon \textit{b}. The rotation introduces cross-polarized components \(|HV\rangle\) and \(|VH\rangle\), whose weights increase with \(\vartheta_{\rm rot}(t)\).

The fidelity with respect to the ideal Bell state is therefore
\[
\begin{split}
F_{\rm rot} (t) &
=
|\langle\Phi^+|\phi_{\rm rot} (t)\rangle|^2
=
\cos^2(\vartheta_{\rm rot} (t)) \\
&=
\cos^2(\vartheta_{\rm sys} (t) +\vartheta_{\rm jit} (t)).
\end{split}
\]
Taking the expectation over the Gaussian jitter gives
\[
\begin{split}
\mathbb{E}[F_{\rm rot} (t)] 
&=
\int_{-\infty}^{\infty}
\cos^2(\vartheta_{\rm sys} (t) +\vartheta_{\rm jit} )
p(\vartheta_{\rm jit} )\,d\vartheta_{\rm jit}  \\
&=
\frac{1}{2}
\left[
1+
\cos(2\vartheta_{\rm sys} (t))
e^{-2\sigma_{\rm rot}^2}
\right].
\end{split}
\]

\subsubsection{Decoherence}

We then model the effects of free-space channel and quantum memory decoherence on entanglement fidelity. An entangled link remains usable only for a finite duration before environmental noise causes its fidelity to decay exponentially. We therefore assume that any link whose fidelity falls below a prescribed threshold \(F_{\rm th}\) is discarded, requiring all quantum operations to complete within the corresponding cutoff time.

\paragraph{Fidelity Loss During Transmission}
We first determine the fidelity of an entangled link after free-space transmission from Satellite \textit{A} to Satellite \textit{B}. In general, atmospheric quantum channels transform the input quantum state according to~\cite{klen2023numerical,vasylyev2018theory,vasylyev2016atmospheric}
\[
\Theta_{\rm out}(\phi)
=
\int_0^1
\mathcal{P}(\eta)
\frac{1}{\eta}
\Theta_{\rm in}\!\left(\frac{\phi}{\sqrt{\eta}}\right)
d\eta,
\]
where \(\Theta_{\rm in}\) and \(\Theta_{\rm out}\) denote the Glauber--Sudarshan quasi-probability distributions~\cite{fante2005electromagnetic,fante2005electromagnetics}, \(\mathcal{P}(\eta)\) is the probability distribution of the transmittance (PDT), and \(\eta\in[0,1]\) is the channel transmittance.
For atmospheric free-space links, \(\mathcal{P}(\eta)\) models random transmittance fluctuations, commonly using the Beta distribution~\cite{klen2023numerical}. However, for ISLs above \(200\) km, atmospheric effects are negligible and only the remaining source of transmittance fluctuation is the pointing error, whose effect on photon capture is already accounted for by \(q_{\rm error}\)~\eqref{eq:q_err}. Conditioned on successful photon capture, we therefore evaluate the received fidelity using the diffraction-limited transmittance \(\eta\).

Photon loss reduces the number of detected photons, increasing the relative contribution of background photons and detector dark counts. Following~\cite{nielsen2010quantum}, the Bell-state fidelity after transmission is
\begin{equation*}
F_0(\eta)
=
\frac{\eta+(1-\eta)\varepsilon}
{1+3(1-\eta)\varepsilon},
\label{eq:fidelity_eta}
\end{equation*}
where \(\varepsilon\) is the quantum bit error rate (QBER).

Combining the transmission loss with the expected polarization-rotation fidelity yields the initial fidelity of the generated entangled link as
\begin{equation*}
\begin{split}
F_0' (d,t)
&=
\mathbb{E}[F_{\rm rot}(t)]\,F_0(\eta(d))\\
&=
\frac{1}{2}
\left[
1+
\cos(2\vartheta_{\rm sys}(t))
e^{-2\sigma_{\rm rot}^2}
\right]
\frac{\eta(d)+[1-\eta(d)]\varepsilon}
{1+3[1-\eta(d)]\varepsilon}.
\end{split}
\end{equation*}
Using Eq.~\ref{eq: vartheta sys}, the initial fidelity becomes
\begin{equation}
\begin{split}
F_0'&(d,t) \\
&=
\frac{1}{2}
\left[
1+
\cos\!\left(
\frac{2v_{\rm sat} \cdot t}
{R_{\rm E}+h_{\rm sat}}
\right)
e^{-2\sigma_{\rm rot}^2}
\right]
\frac{\eta(d) + [1-\eta(d)]\varepsilon}
{1+3[1-\eta(d)]\varepsilon}.
\end{split}
\label{eq:initial_fidelity}
\end{equation}

\paragraph{Quantum Memory Decoherence}
We model quantum memory decoherence using the amplitude-damping channel for energy dissipation in quantum memories~\cite{nielsen2010quantum}. For a qubit stored for time \(t_{\rm store}\), the channel is described by the Kraus operators
\[
\mathrm{KO}_0=
\begin{bmatrix}
1 & 0\\
0 & \sqrt{1-\alpha(t_{\rm store})}
\end{bmatrix}, \
\mathrm{KO}_1=
\begin{bmatrix}
0 & \sqrt{\alpha(t_{\rm store})}\\
0 & 0
\end{bmatrix},
\]
where
\[
\alpha(t_{\rm store})
=
1-e^{-\Gamma t_{\rm store}},
\]
and \(\Gamma\) (\(\mathrm{s}^{-1}\)) is the damping rate that characterizes the coupling between the quantum memory and its environment.
Consider an EPR pair initially prepared in the Bell state
\[
|\Phi^+\rangle
=
\frac{|00\rangle+|11\rangle}{\sqrt{2}}
\]
with initial fidelity \(F_0'(d)\). As both qubits are stored independently, amplitude damping causes population decay and coherence loss. The probability that both qubits remain undecayed after storage time \(t_{\rm store}\) is
\[
P_{\rm survive} (t_\text{store})
=
(1-\alpha(t_\text{store}))^2
=
e^{-2\Gamma t_{\rm store}}.
\]
Accordingly, the fidelity of the stored EPR pair evolves as~\cite{li2023dynamic}
\begin{equation}
F(d,t_{\rm store})
=
F_0'(d) P_{\rm survive}(t_\text{store})
=
F_0'(d) e^{-2\Gamma t_{\rm store}},
\end{equation}
where \(F_0'(d) = F_0'(d, \frac{d}{c})\) is the post-transmission fidelity given by Eq.~\eqref{eq:initial_fidelity} at \(t = t_{\rm trans} = \frac{d}{c}\).
Given a minimum fidelity threshold \(F_{\rm th}\), the link expires once
\(F(d, T_{\rm cutoff}-t_{\rm trans}(d))=F_{\rm th}\), yielding the cutoff time as
\[
T_{\rm cutoff} (d)
=
\frac{d}{c}
-
\frac{1}{2\Gamma}
\ln\!\left(
\frac{F_{\rm th}}{F_0' (d)}
\right),
\]
where \(c\) is the speed of light. The corresponding maximum storage age in discrete time slots is
\begin{equation}
K(d)
=
\left\lfloor
\frac{T_{\rm cutoff}(d)}{\Delta t}
\right\rfloor
=
\left\lfloor
\frac{1}{\Delta t} \left(
\frac{d}{c}
-
\frac{1}{2\Gamma}
\ln \frac{F_{\rm th}}{F_0' (d)}
\right)
\right\rfloor,
\end{equation}
which defines the maximum allowable storage age for an entangled link of distance \(d\) and initial fidelity \(F_0'(d)\).

\subsubsection{Fidelity Qualification Probability \(p_{\rm suc}(d)\)}
For a given transmission distance \(d\), both the channel transmittance \(\eta(d)\) and the initial fidelity \(F_0'(d)\) are deterministic. Hence, the probability that a generated entangled link satisfies the fidelity threshold \(F_{\rm th}\) is
\[
p_{\rm suc}(d)
=
\begin{cases}
1, & F_0'(d)\ge F_{\rm th},\\
0, & F_0'(d)<F_{\rm th}.
\end{cases}
\]
Since \(F_0'(d)\) decreases monotonically with distance due to beam diffraction and polarization rotation, there exists a maximum transmission distance \(d_{\max}\) satisfying
\[
F_0'(d_{\max})=F_{\rm th}.
\]
Inspired by the connectivity model of unit disk graphs~\cite{de2025demonstration,nguyen2023quantum,serret2020solving}, the fidelity qualification probability can therefore be expressed as
\begin{equation}
p_{\rm suc}(d)
=
\begin{cases}
1, & d\le d_{\max},\\
0, & d>d_{\max}.
\end{cases}
\end{equation}
Substituting \(t=d/c\) into Eq.~\eqref{eq:initial_fidelity} and solving
\(F_0'(d_{\max})=F_{\rm th}\) yields
\begin{equation}
\begin{split}
&(1-2F_{\rm th})
-
\left[1-\varepsilon+6F_{\rm th}\varepsilon\right]
e^{-\frac{2R_{\rm ap}^2}{\theta^2d_{\max}^2}}
\\
&\qquad
=
-\cos\!\left(
\frac{2v_{\rm sat}d_{\max}}
{c(R_{\rm E}+h_{\rm sat})}
\right)
e^{-2\sigma_{\rm rot}^2}
\left[
1-
(1-\varepsilon)
e^{-\frac{2R_{\rm ap}^2}{\theta^2d_{\max}^2}}
\right].
\end{split}
\label{eq:max_d_trans}
\end{equation}
Equation~\eqref{eq:max_d_trans} is transcendental because \(d_{\max}\) appears simultaneously in the exponential diffraction term and the cosine polarization-rotation term. Consequently, no closed-form solution exists in the general case.
For short transmission distances (\(40\!-\!50\) km), the polarization rotation is negligible (\(e^{-2\sigma_{\rm rot}^2}\approx1\)). Neglecting this effect reduces Eq.~\eqref{eq:max_d_trans} to
\begin{equation}
d_{\max}
=
\sqrt{
\frac{-2R_{\rm ap}^2}
{\theta^2
\ln\!\left(
\dfrac{1-F_{\rm th}}
{1+\varepsilon(3F_{\rm th}-1)}
\right)}
},
\label{eq:simple_dmax}
\end{equation}
which provides a closed-form approximation for the maximum transmission distance of a one-hop LEO QISL.
For Eq.~\eqref{eq:simple_dmax} to admit a positive real solution, the denominator of the logarithm must be positive,
\begin{equation}
1+\varepsilon(3F_{\rm th}-1)>0,
\label{eq:positive_log}
\end{equation}
the logarithm itself must be negative,
\begin{equation}
\varepsilon(1-3F_{\rm th}) < F_{\rm th},
\label{eq:negative_log}
\end{equation}
and the physical constraints
\[
\varepsilon\ge0,\qquad
R_{\rm ap}>0,\qquad
\theta>0
\]
must hold. Two cases arise:

1. If \(F_{\rm th}\in[\frac13,1)\), then Eq.~\eqref{eq:positive_log} is always satisfied for \(\varepsilon\ge0\), and a positive solution for \(d_{\max}\) exists.

2. If \(F_{\rm th}\in(0,\frac13)\), Eq.~\eqref{eq:negative_log} is more restrictive, giving

\[
0
\le
\varepsilon
<
\frac{F_{\rm th}}
{1-3F_{\rm th}}.
\]

In practical applications, \(F_{\rm th} \geq \frac{1}{2}\) because it is the minimum fidelity requirement for a purification to have gain in free space. Therefore, Eq.~\eqref{eq:positive_log} is always satisfied under the free-space scenario.

\subsubsection{Link Generation Duration}
A single entanglement-generation attempt consists of three sequential phases: EPR-pair generation, quantum transmission and classical acknowledgment. The total duration is
\begin{equation*}
t_{\rm gen}
=
t_{\rm EPR}
+
t_{\rm trans}
+
t_{\rm ack},
\end{equation*}
where \(t_{\rm EPR}\) is the EPR-pair generation time, \(t_{\rm trans}\) is the one-way quantum transmission time and \(t_{\rm ack}\) is the classical acknowledgment time.
Assuming an entangled-photon source generates \(R_{\rm src}\) pairs per second,
\(
t_{\rm EPR}
=
\frac{1}{R_{\rm src}}.
\)
The quantum transmission time between two satellites separated by distance \(d\) is
\(
t_{\rm trans}
=
\frac{d}{c},
\)
where \(c\) is the speed of light. After receiving the photon, Satellite \textit{B} acknowledges successful entanglement establishment via the classical channel. The acknowledgment time comprises the RF round-trip propagation delay~\cite{mahafza2017introduction} is
\[
t_{\rm trans}^{(\rm Cls)}
=
\frac{2d}{c},
\]
and the classical processing delay is
\[
t_{\rm proc}^{(\rm Cls)}
=
t_{\rm mod}
+
t_{\rm demod}
+
t_{\rm pkt},
\]
where \(t_{\rm mod}\), \(t_{\rm demod}\) and \(t_{\rm pkt}\) denote the modulation, demodulation and packet-processing delays, respectively~\cite{kurose2019computer}. Therefore, the duration of one generation attempt is
\begin{equation*}
t_{\rm gen}(d)
=
\frac{1}{R_{\rm src}}
+
\frac{3d}{c}
+
t_{\rm proc}^{(\rm Cls)}.
\end{equation*}
The corresponding number of time slots is
\begin{equation}
C(d)
=
\left\lceil
\frac{t_{\rm gen}(d)}
{\Delta t}
\right\rceil
=
\left\lceil
\frac{1}{{\Delta t}}\left(
\dfrac{1}{R_{\rm src}}
+\dfrac{3d}{c}
+t_{\rm proc}^{(\rm Cls)}\right)
\right\rceil.
\end{equation}

\subsection{Link Utilization}

\begin{figure}[t]
  \centering
\includegraphics[width=0.475\textwidth]{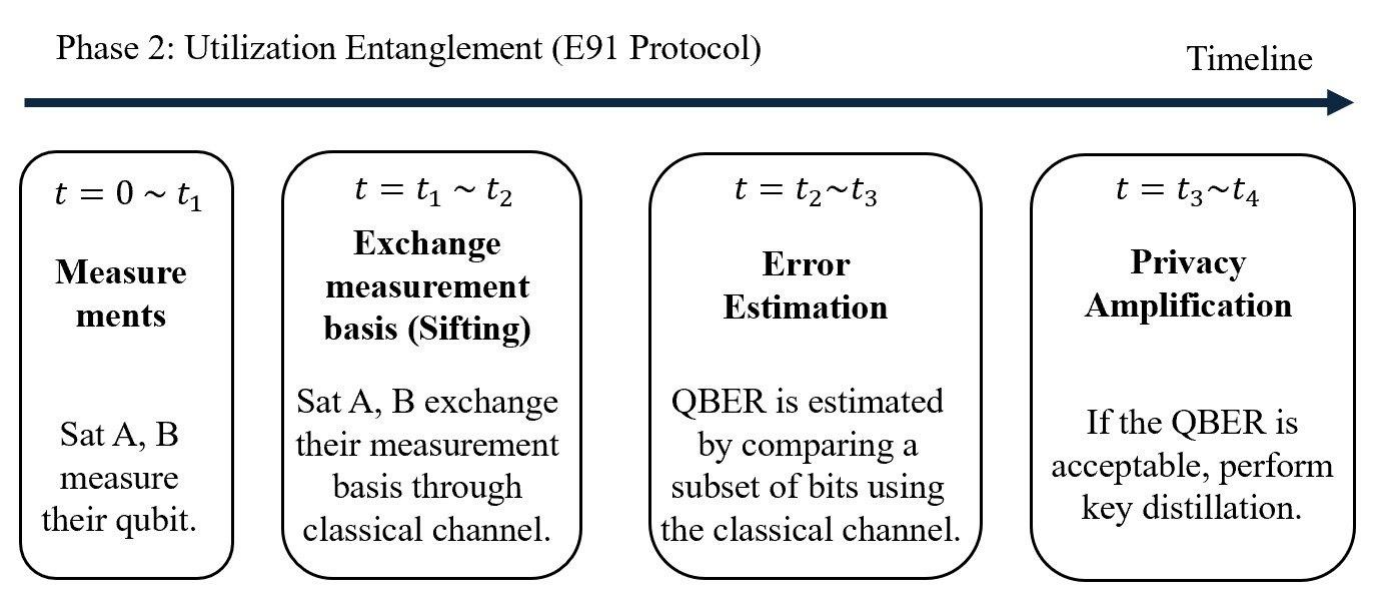}
  \caption{E91 QKD Protocol}
  \label{fig:use a link1}
\end{figure}

Once an entangled link has been established and stored in the quantum memories of satellites \textit{A} and \textit{B}, it must be consumed to complete the communication request. We consider the E91 quantum key distribution protocol~\cite{ekert1991quantum} as a representative utilization protocol because it requires multiple rounds of classical communication after quantum measurements. Fig.~\ref{fig:use a link1} illustrates the four sequential phases of the protocol for a shared Bell state \(|\Phi^+\rangle=\frac{|HH\rangle+|VV\rangle}{\sqrt{2}}\).

\subsubsection{Quantum Measurements (\(t=0\) to \(t=t_1\))}

At \(t=0\), upon receiving the classical signal to begin utilization, satellites \textit{A} and \textit{B} simultaneously measure their respective qubits. Satellite \textit{A} randomly selects a BSM basis from
\[
\mathcal{A}=\{a_1,a_2,a_3\},
\]
with orientations
\(\phi_{a_1}=0^\circ\),
\(\phi_{a_2}=45^\circ\) and
\(\phi_{a_3}=90^\circ\).
Similarly, satellite \textit{B} selects a basis from
\[
\mathcal{B}=\{b_1,b_2,b_3\},
\]
with orientations
\(\phi_{b_1}=22.5^\circ\),
\(\phi_{b_2}=67.5^\circ\) and
\(\phi_{b_3}=112.5^\circ\).
Each BSM consists of rotating the polarization analyzer to the selected basis and detecting the photon, giving a BSM duration of
\begin{equation}
t_1=t_{\rm BSM}=t_{\rm rotate}+t_{\rm detect},
\end{equation}
where \(t_{\rm rotate}\) and \(t_{\rm detect}\) denote the analyzer rotation and single-photon detection times, respectively. Each BSM produces a binary outcome \(m\in\{0,1\}\), corresponding to horizontal- or vertical-like polarization in the chosen basis. We denote the outcomes by \(m_A^{(j)}\) and \(m_B^{(k)}\) when satellites \textit{A} and \textit{B} use bases \(a_j\) and \(b_k\), respectively.

\subsubsection{Basis Sifting (\(t=t_1\) to \(t=t_2\))}

After completing the quantum measurements, satellites \textit{A} and \textit{B} simultaneously exchange their selected measurement bases over the classical RF channel. Satellite \textit{A} sends \(M_A=\{j:a_j\}\), while satellite \textit{B} sends \(M_B=\{k:b_k\}\), each requiring a one-way transmission time
\[
t_{\rm sift}^{(\rm trans)}
=
\frac{d}{c}.
\]
Upon reception, each satellite demodulates the message, processes the protocol and determines the basis compatibility. The corresponding processing time is
\begin{equation*}
t_{\rm sift}^{(\rm proc)}
=
t_{\rm demod}
+
t_{\rm pkt}
+
t_{\rm compare}.
\end{equation*}
The E91 protocol classifies the basis pairs into two categories, (i) key-generation pairs, \((a_1,b_1)\) and \((a_3,b_3)\), whose correlated outcomes are used to generate the secret key; and (ii) Bell-test pairs, \((a_1,b_2)\), \((a_2,b_1)\), \((a_2,b_2)\), \((a_2,b_3)\) and \((a_3,b_2)\), which are used to compute the Bell parameter \(S\) and detect eavesdropping. Since the two one-way transmissions occur simultaneously, the duration of this phase is
\begin{equation*}
t_2-t_1
=
t_{\rm sift}^{(\rm trans)}
+
t_{\rm sift}^{(\rm proc)}
=
\frac{d}{c}
+
t_{\rm demod}
+
t_{\rm pkt}
+
t_{\rm compare}.
\end{equation*}

\subsubsection{Error Estimation (\(t=t_2\) to \(t=t_3\))}

After basis sifting, satellite \textit{B} estimates the QBER by comparing a randomly selected subset of the sifted key with satellite \textit{A}. Let
\[
S\subset\{1,2,\ldots,n_{\rm raw}\}
\]
denote the selected test positions, where \(n_{\rm raw}\) is the number of sifted key bits. Satellite \textit{B} transmits the subset indices and corresponding measurement outcomes
\[
\{(i,m_B^{(i)}):i\in S\},
\]
to satellite \textit{A}, requiring a one-way transmission time as
\[
t_{\rm err}^{(\rm trans,1)}
=
\frac{d}{c}.
\]
Satellite \textit{A} compares the received values with its own outcomes and counts the number of disagreements as
\[
n_{\rm err}
=
\left|
\{\,i\in S:m_A^{(i)}\neq m_B^{(i)}\,\}
\right|,
\]
which requires comparison time \(t_{\rm count}\). It then returns the error count to satellite \textit{B}, requiring
\[
t_{\rm err}^{(\rm trans,2)}
=
\frac{d}{c}.
\]
Satellite \textit{B} computes the QBER,
\[
{\rm QBER}
=
\frac{n_{\rm err}}{n_{\rm test}},
\]
where \(n_{\rm test}=|S|\) is the number of test bits, which are discarded after estimation. If \({\rm QBER}>{\rm QBER}_{\rm th}\), the protocol aborts and the utilization attempt fails.
Finally, the processing time is

\begin{equation*}
t_{\rm err}^{(\rm proc)}
=
t_{\rm demod}
+
t_{\rm pkt}
+
t_{\rm count}
+
t_{\rm QBER},
\end{equation*}
giving the total duration of this phase as
\begin{equation*}
\begin{split}
t_3-t_2
&=
t_{\rm err}^{(\rm trans,1)}
+
t_{\rm err}^{(\rm proc)}
+
t_{\rm err}^{(\rm trans,2)}
+
t_{\rm demod}
+
t_{\rm pkt} \\
&=
\frac{2d}{c}
+
2(t_{\rm demod}+t_{\rm pkt})
+
t_{\rm count}
+
t_{\rm QBER}.
\end{split}
\end{equation*}

\subsubsection{Privacy Amplification (\(t=t_3\) to \(t=t_4\))}

If the estimated QBER is below the security threshold, the remaining sifted key is compressed through privacy amplification to remove any information potentially available to an eavesdropper. Let
\[
n_{\rm sift}=n_{\rm raw}-n_{\rm test}
\]
denote the remaining sifted bits. Satellite \textit{B} randomly selects a universal hash function
\(\mathfrak{h}\in\mathcal{H}\), producing a final key of length

\[
n_{\rm final}
=
n_{\rm sift}
\left(1-\mathfrak{h}(\mathrm{QBER})\right),
\]
where \(\mathfrak{h}(\mathrm{QBER})\) is the binary entropy function accounting for information leakage.
Satellite \textit{B} transmits the hash-function specification to satellite \textit{A}, requiring
\[
t_{\rm priv}^{(\rm trans,1)}
=
\frac{d}{c}.
\]
Both satellites then independently compute
\begin{equation*}
\begin{split}
\mathcal{K}_A
&=
\mathfrak{h}
\!\left(
\mathfrak{k}_A^{(1)},
\ldots,
\mathfrak{k}_A^{(n_{\rm sift})}
\right),\\
\mathcal{K}_B
&=
\mathfrak{h}
\!\left(
\mathfrak{k}_B^{(1)},
\ldots,
\mathfrak{k}_B^{(n_{\rm sift})}
\right),
\end{split}
\end{equation*}
requiring hash-computation time \(t_{\rm hash}\). Satellite \textit{A} then returns a confirmation message to satellite \textit{B}, requiring
\[
t_{\rm priv}^{(\rm trans,2)}
=
\frac{d}{c}.
\]
The corresponding processing time is

\begin{equation*}
t_{\rm priv}^{(\rm proc)}
=
t_{\rm demod}
+
t_{\rm pkt}
+
t_{\rm hash}
+
t_{\rm confirm},
\end{equation*}
where \(t_{\rm confirm}\) is the confirmation-processing time. Hence, the duration of this phase is
\begin{equation*}
\begin{split}
t_4-t_3
&=
t_{\rm priv}^{(\rm trans,1)}
+
t_{\rm hash}
+
t_{\rm priv}^{(\rm trans,2)}
+
t_{\rm priv}^{(\rm proc)}
\\
&=
\frac{2d}{c}
+
2(t_{\rm demod}+t_{\rm pkt})
+
2t_{\rm hash}
+
t_{\rm confirm}.
\end{split}
\end{equation*}
The total utilization time of the E91 protocol is therefore
\[
\begin{split}
t_{\rm util}(d)
&=
t_1
+(t_2-t_1)
+(t_3-t_2)
+(t_4-t_3)\\
&=
\frac{5d}{c}
+
t_{\rm util}^{(\rm proc)},
\end{split}
\]
where \(
t_{\rm util}^{(\rm proc)}
=
t_{\rm rotate}
+t_{\rm detect}
+5t_{\rm demod}
+5t_{\rm pkt}
+t_{\rm compare}
+t_{\rm QBER}
+2t_{\rm hash}
+t_{\rm confirm}
\)
is the total processing overhead. The corresponding number of time slots is
\begin{equation}
L(d)
=
\left\lceil
\frac{t_{\rm util}(d)}
{\Delta t}
\right\rceil
=
\left\lceil
\frac{
1 
}
{\Delta t}
\left( \dfrac{5d}{c}
+
t_{\rm util}^{(\rm proc)} \right)
\right\rceil.
\end{equation}

\subsubsection{Utilization Success Probability \(p_{\rm util}\)}

Since the link-generation success probability \(p'(d)\) already accounts for successful entanglement establishment and fidelity qualification, the utilization success probability depends only on the reliability of the classical communication, single-photon detectors and protocol synchronization.

The E91 protocol requires five one-way classical message exchanges, giving the overall classical communication success probability as
\begin{equation}
p_{\rm cls}
=
\left(p^{\rm (Cls)}_{\rm single}\right)^5,
\end{equation}
where \(p^{\rm (Cls)}_{\rm single}\) is the success probability of a single classical transmission.
Both satellites must also successfully detect their photons during the measurement phase, giving the detector success probability as
\begin{equation}
p_{\rm det}
=
p_{\rm det}^{(A)}
p_{\rm det}^{(B)},
\end{equation}
where \(p_{\rm det}^{(A)}\) and \(p_{\rm det}^{(B)}\) are the detector reliability at satellites \textit{A} and \textit{B}, respectively.
Therefore, the overall utilization success probability is

\begin{equation}
\begin{split}
p_{\rm util}
&=
p_{\rm cls}\,
p_{\rm det}\,
p_{\rm sync} \\
&=
\left(p^{\rm (Cls)}_{\rm single}\right)^5
p_{\rm det}^{(A)}
p_{\rm det}^{(B)}
p_{\rm sync},
\end{split}
\end{equation}
where \(p_{\rm sync}\) is the probability of maintaining protocol synchronization.

\section{Performance Metrics Derivation}
\label{app:PMD}

This appendix derives the performance metrics for both the pre-generation (p) and on-demand (o) strategies, including the total satisfaction rate \(\mathcal{R}_{\rm total}\), accepted request satisfaction rate \(\mathcal{R}_{\rm ac}\), average waiting time \(\mathcal{W}\), link utilization efficiency \(\mathcal{U}\) and average consumed fidelity \(\mathcal{F}\).

By Corollary~\ref{cor: Ergodicity}, both Markov chains are ergodic for \(0<\lambda<1\) and \(0<p'(d)<1\). Therefore, the ergodic theorem for finite discrete-time Markov chains~\cite{silvestrov2025coupling} guarantees that, for every state \(s\),
\begin{equation*}
\lim_{\mathcal{N}\rightarrow\infty}
\frac{1}{\mathcal{N}}
\sum_{t=1}^{\mathcal{N}}
\mathbf{1}\{S_t=s\}
=
\pi_s
\qquad
\text{a.s.},
\label{eq:ergodic}
\end{equation*}
where \(\pi_s\) is the stationary probability of state \(s\), \(\mathbf{1}\{\cdot\}\) is the indicator function and \(\mathcal{N}=\lceil T/\Delta t\rceil\) is the number of time slots in an observation window of duration \(T\). Consequently, all long-run performance metrics can be expressed in terms of the stationary distribution.

Since request arrivals follow an independent Bernoulli process with parameter \(\lambda\), the request arrival in slot \(t\) is independent of the system state \(S_t\), which is determined solely by the outcomes of previous slots.

For convenience, we define
\begin{equation*}
\gamma (d)
=
\sum_{g=1}^{G_{\max}}
[1-p'(d)]^{g-1}p'(d)
=
1-[1-p'(d)]^{G_{\max}},
\label{eq:generate_success}
\end{equation*}
which is the probability that at least one of the \(G_{\max}\) link-generation attempts succeeds, where each attempt succeeds with probability \(p'(d)\) and requires \(C(d) \cdot \Delta t\) time slots.

\subsubsection{Pre-generation Strategy}
\label{app:pre-gen_formula}

The state space of the pre-generation Markov chain is
\[
\begin{aligned}
S^{(\mathrm{p})}
&={}
\{(0,0)\}
\cup
\{(0,g):1\le g\le G_{\max}\} \\
&\cup
\{(1,i):1\le i\le K(d)-1\}
\cup
\{(2,i):1\le i\le K(d)\}.
\end{aligned}
\]
which has cardinality \( |S^{(\mathrm{p})}|=G_{\max}+2K(d) \). 
Let \( P^{(\mathrm{p})} = \bigl[p_{ij}^{(\mathrm{p})}\bigr] \)
denote the one-step transition probability matrix. Its nonzero entries are given by Eqs.~\eqref{eq: pre-gen trans prob from (0,0)}, \eqref{eq: pre-gen trans prob from (0,g)}, \eqref{eq: pre-gen trans prob from (1,i)}, \eqref{eq: pre-gen trans prob from (1,K-1)} and \eqref{eq: pre-gen trans prob from (2,i)}, which are summarized in Table~\ref{tab:pre-gen}.

\begin{table}[t]
\centering
\caption{$\mathcal{T}^{(p)}$}
\label{tab:pre-gen}
\begin{tabular}{lll}
\toprule
$S_i$ & $S_j$ & $p_{S_i,S_j}$ \\
\midrule
$(0,0)$ & $(0,0)$ & $(1-\lambda)(1-p'(d))$ \\
$(0,0)$ & $(1,1)$ & $(1-\lambda)p'(d)$ \\
$(0,0)$ & $(0,1)$ & $\lambda$ \\
$(1,i),\ 1 \le i \le K(d)-2$ & $(1,i{+}1)$ & $1-\lambda$ \\
$(1,i),\ 1 \le i \le K(d)-2$ & $(2,i{+}1)$ & $\lambda$ \\
$(1,K(d)-1)$ & $(0,0)$ & $1-\lambda$ \\
$(1,K(d)-1)$ & $(2,K(d))$ & $\lambda$ \\
$(0,g),\ g < G_{\max}$ & $(2,1)\,/\,(0,g+1)$ & $p'(d)\,/\,1-p'(d)$ \\
$(0,G_{\max})$ & $(2,1)\,/\,(0,0)$ & $p'(d)\,/\,1-p'(d)$ \\
$(2,i),\ 1 \le i \le K(d)$ & $(0,0)$ & $1$ \\
\bottomrule
\end{tabular}
\end{table}

The stationary distribution of the pre-generation chain satisfies
\[
\boldsymbol{\pi}^{(\mathrm{p})}
=
\boldsymbol{\pi}^{(\mathrm{p})}P^{(\mathrm{p})}.
\]
Its state probabilities can be obtained recursively from state \((0,0)\), since each generating state has a unique predecessor. In particular, state \((0,1)\) is reached from \((0,0)\) upon a request arrival, giving
\[
\pi_{(0,1)}^{(\mathrm{p})} (\lambda,d)
=
\lambda\pi_{(0,0)}^{(\mathrm{p})} (\lambda,d).
\]
For \(1\le g<G_{\max}\), state \((0,g+1)\) is reached from \((0,g)\) following a failed generation attempt, so
\[
\pi_{(0,g+1)}^{(\mathrm{p})} (\lambda,d)
=
[1-p'(d)] \pi_{(0,g)}^{(\mathrm{p})} (\lambda,d).
\]
Hence,

\begin{equation}
\pi_{(0,g)}^{(\mathrm{p})} (\lambda,d)
=
\lambda[1-p'(d)]^{g-1}
\pi_{(0,0)}^{(\mathrm{p})} (\lambda,d),
1\le g\le G_{\max} .
\label{eq: pre-gen_pi-gen}
\end{equation}

For the storage states, \((1,1)\) is reached from \((0,0)\) when no request arrives and proactive generation succeeds, giving
\[
\pi_{(1,1)}^{(\mathrm{p})} (\lambda,d)
=
(1-\lambda)p'(d) \pi_{(0,0)}^{(\mathrm{p})} (\lambda,d).
\]
For \(1\le i<K(d)-1\), state \((1,i+1)\) is reached from \((1,i)\) when no request arrives, so
\[
\pi_{(1,i+1)}^{(\mathrm{p})} (\lambda,d)
=
(1-\lambda)\pi_{(1,i)}^{(\mathrm{p})} (\lambda,d).
\]
Hence,
\begin{equation}
\pi_{(1,i)}^{(\mathrm{p})}  (\lambda,d)
=
p'(d) (1-\lambda)^i
\pi_{(0,0)}^{(\mathrm{p})} (\lambda,d),
1\le i\le K(d)-1.
\label{eq: pre-gen_pi-store}
\end{equation}

For the utilization states, \((2,1)\) is reached from any generating state upon a successful generation attempt, giving
\[
\pi_{(2,1)}^{(\mathrm{p})} (\lambda,d)
=
p'(d) \sum_{g=1}^{G_{\max}}
\pi_{(0,g)}^{(\mathrm{p})} (\lambda,d).
\]
For \(1\le i\le K (d) -1\), state \((2,i+1)\) is reached from \((1,i)\) when a request arrives, so
\[
\pi_{(2,i+1)}^{(\mathrm{p})} (\lambda,d)
=
\lambda\pi_{(1,i)}^{(\mathrm{p})} (\lambda,d).
\]
Using Eq.~\eqref{eq: pre-gen_pi-gen} and
\(
\sum_{g=1}^{G_{\max}}[1-p'(d)]^{g-1}
=\gamma(d)/p'(d),
\)
we obtain
\begin{equation}
\pi_{(2,1)}^{(\mathrm{p})} (\lambda,d)
=
\lambda\gamma (d)
\pi_{(0,0)}^{(\mathrm{p})} (\lambda,d).
\label{eq: pre-gen_pi-util1}
\end{equation}
Moreover,
\begin{equation}
\pi_{(2,i+1)}^{(\mathrm{p})} (\lambda,d)
=
\lambda p'(d) (1-\lambda)^i
\pi_{(0,0)}^{(\mathrm{p})} (\lambda,d),
1\le i\le K(d)-1.
\label{eq: pre-gen_pi-util}
\end{equation}

The recursions above satisfy the balance equations for all states except \((0,0)\). Since the equations in
\(
\boldsymbol{\pi}^{(\mathrm{p})}
=
\boldsymbol{\pi}^{(\mathrm{p})}P^{(\mathrm{p})}
\)
are linearly dependent for a row-stochastic transition matrix, it remains only to verify the balance equation at \((0,0)\).

The inflows to \((0,0)\) arise from: (i) its self-loop, (ii) failure of the final generation attempt from \((0,G_{\max})\), (iii) expiration of the oldest stored link from \((1,K(d)-1)\) and (iv) completion of utilization from all states \((2,i)\). Hence,
\begin{equation*}
\begin{split}
\pi_{(0,0)}^{(\mathrm{p})} (\lambda,d)
={}&
(1-\lambda)[1-p'(d)]
\pi_{(0,0)}^{(\mathrm{p})} (\lambda,d) \\
&+
[1-p'(d)]
\pi_{(0,G_{\max})}^{(\mathrm{p})} (\lambda,d)
\\
&+
(1-\lambda)
\pi_{(1,K(d)-1)}^{(\mathrm{p})} (\lambda,d)
+
\sum_{i=1}^{K(d)}
\pi_{(2,i)}^{(\mathrm{p})} (\lambda,d).
\end{split}
\end{equation*}

Substituting Eqs.~\eqref{eq: pre-gen_pi-gen},
\eqref{eq: pre-gen_pi-store},
\eqref{eq: pre-gen_pi-util1}, and
\eqref{eq: pre-gen_pi-util} gives
\begin{equation*}
\begin{split}
\pi_{(0,0)}^{(\mathrm{p})} (\lambda,d)
={}&
\Bigl[
(1-\lambda)[1-p'(d)]
+
\lambda [1-p'(d)]^{G_{\max}}\\
&+
p'(d) (1-\lambda)^{K(d)}
+
\lambda \gamma(d)  \\
&+
p'(d)\bigl((1-\lambda)-(1-\lambda)^{K(d)}\bigr)
\Bigr]
\pi_{(0,0)}^{(\mathrm{p})} (\lambda,d).
\end{split}
\end{equation*}
Using \(
\lambda [1-p'(d)]^{G_{\max}}+\lambda \gamma(d)=\lambda,
\)
and canceling the \(p'(d)(1-\lambda)^{K(d)}\) terms, the coefficient reduces to
\(
(1-\lambda)[1-p'(d)] + \lambda + p'(d) (1-\lambda)=1.
\)
Thus, the balance equation at \((0,0)\) is also satisfied, and Eqs.~\eqref{eq: pre-gen_pi-gen}--\eqref{eq: pre-gen_pi-util} satisfy all stationary balance equations.

We define the aggregate stationary probabilities of the generation, storage and utilization phases as
\[
\begin{split}
\pi_{\rm gen}^{(\mathrm{p})} (\lambda,d)
&:=
\sum_{g=1}^{G_{\max}}
\pi_{(0,g)}^{(\mathrm{p})} (\lambda,d),\\
\pi_{\rm st}^{(\mathrm{p})} (\lambda,d)
&:=
\sum_{i=1}^{K-1}
\pi_{(1,i)}^{(\mathrm{p})} (\lambda,d),\\
\pi_{\rm util}^{(\mathrm{p})} (\lambda,d)
&:=
\sum_{i=1}^{K}
\pi_{(2,i)}^{(\mathrm{p})} (\lambda,d).
\end{split}
\]
Using Eqs.~\eqref{eq: pre-gen_pi-gen},
\eqref{eq: pre-gen_pi-store},
\eqref{eq: pre-gen_pi-util1}, and
\eqref{eq: pre-gen_pi-util}, these become
\begin{equation}
\begin{aligned}
\pi_{\rm gen}^{(\mathrm{p})} (\lambda,d)
&=
\lambda\pi_{(0,0)}^{(\mathrm{p})} (\lambda,d)
\sum_{g=1}^{G_{\max}}
[1-p'(d)]^{g-1} \\
&=
\frac{\lambda \gamma(d)}{p'(d)}
\pi_{(0,0)}^{(\mathrm{p})}(\lambda,d),
\\[2mm]
\pi_{\rm st}^{(\mathrm{p})} (\lambda,d)
&=
p'(d) \pi_{(0,0)}^{(\mathrm{p})} (\lambda,d)
\sum_{i=1}^{K(d)-1}
(1-\lambda)^i
\\
&=
\frac{
p'(d)(1-\lambda)
\left[1-(1-\lambda)^{K(d)-1}\right]
}{
\lambda
}
\pi_{(0,0)}^{(\mathrm{p})} (\lambda,d),
\\[2mm]
\pi_{\rm util}^{(\mathrm{p})} (\lambda,d)
&=
\pi_{(2,1)}^{(\mathrm{p})} (\lambda,d)
+
\lambda
\sum_{i=1}^{K(d)-1}
\pi_{(1,i)}^{(\mathrm{p})} (\lambda,d)
\\
&=
\lambda
\left(
\gamma(d) \pi_{(0,0)}^{(\mathrm{p})}(\lambda,d)
+
\pi_{\rm st}^{(\mathrm{p})} (\lambda,d)
\right).
\end{aligned}
\label{eq:aggregate_stationary_probabilities}
\end{equation}

Substituting the aggregate probabilities into the normalization condition \(
\pi_{(0,0)}^{(\mathrm{p})} (\lambda,d)
+
\pi_{\rm gen}^{(\mathrm{p})} (\lambda,d)
+
\pi_{\rm st}^{(\mathrm{p})}(\lambda,d)
+
\pi_{\rm util}^{(\mathrm{p})}(\lambda,d)
=
1
\)
yields
\[
\left[
1
+
\frac{\lambda \gamma(d)}{p'(d)}
+
\lambda \gamma(d)
\right]
\pi_{(0,0)}^{(\mathrm{p})} (\lambda,d)
+
(1+\lambda)
\pi_{\rm st}^{(\mathrm{p})} (\lambda,d)
=
1.
\]
Substituting the expression for \(\pi_{\rm st}^{(\mathrm{p})}(\lambda,d)\) gives
\begin{equation}
\begin{split}
&\pi_{(0,0)}^{(\mathrm{p})}(\lambda,d)\\
&=
\left[
1
+
\frac{\lambda\gamma(d)[1+p'(d)]}{p'(d)}
+
\frac{
p'(d)(1-\lambda^2)
\left[1-(1-\lambda)^{K(d)-1}\right]
}{
\lambda
}
\right]^{-1}
\end{split}
\label{eq:stable_states_initial}
\end{equation}
The remaining stationary probabilities then follow directly from
Eqs.~\eqref{eq: pre-gen_pi-gen}--\eqref{eq:aggregate_stationary_probabilities}.
When \(K(d)=1\), the minimum storage age equals the cutoff time, making proactive storage infeasible. Consequently,
\(\pi_{\rm st}^{(\mathrm{p})} (\lambda,d) =0\) from Eq.~\eqref{eq:aggregate_stationary_probabilities} and the pre-generation strategy reduces to the on-demand strategy.

\paragraph{Request Satisfaction Rate \(\mathcal{R}^{(\mathrm{p})}(\lambda,d)\)}

We consider two request satisfaction rates, (1) \(\mathcal{R}_{\rm ac}^{(\mathrm{p})}(\lambda,d)\), conditioned on accepted requests and (2) \(\mathcal{R}_{\rm total}^{(\mathrm{p})}(\lambda,d)\), which also accounts for requests rejected while the system is locked.

A request arriving in slot \(t\) is accepted only if
\[
S_t\in
\{(0,0)\}
\cup
\{(1,i):1\le i\le K(d)-1\};
\]
requests arriving during generation or utilization are rejected. Thus, \(\mathcal{R}_{\rm ac}^{(\mathrm{p})}(\lambda,d)\) measures the fraction of accepted requests that are ultimately satisfied, whereas \(\mathcal{R}_{\rm total}^{(\mathrm{p})}(\lambda,d)\) measures the fraction of all arriving requests that are satisfied.

Since the request arrival in slot \(t\) is independent of \(S_t\), the long-run expected numbers of accepted and satisfied requests per slot are
\begin{equation*}
\begin{aligned}
N_{\rm ac}^{(\mathrm{p})}(\lambda,d)
&=
\lambda
\left[
\pi_{(0,0)}^{(\mathrm{p})}(\lambda,d)
+
\pi_{\rm st}^{(\mathrm{p})}(\lambda,d)
\right],\\
N_{\rm sat}^{(\mathrm{p})}(\lambda,d)
&=
\lambda
\left[
\gamma(d) \pi_{(0,0)}^{(\mathrm{p})}(\lambda,d)
+
\pi_{\rm st}^{(\mathrm{p})}(\lambda,d)
\right]
p_{\rm util}.
\end{aligned}
\end{equation*}
A request accepted in state \((0,0)\) is satisfied only if at least one of the \(G_{\max}\) generation attempts succeeds, with probability \(\gamma(d)\), followed by successful utilization with probability \(p_{\rm util}\). In contrast, a request accepted in a storage state requires only successful utilization. Hence,
\begin{equation}
\begin{split}
&\mathcal{R}_\text{ac}^\text{(p)}(\lambda,d) =
\frac{N_{\rm sat}^{(\mathrm{p})}(\lambda,d)}
{N_{\rm ac}^{(\mathrm{p})}(\lambda,d)} = p_{\mathrm{util}} \, \frac{ \pi_{\mathrm{st}}^\text{(p)}(\lambda,d) + \gamma \pi_{(0,0)}^\text{(p)}(\lambda,d) } { \pi_{\mathrm{st}}^\text{(p)}(\lambda,d) + \pi_{(0,0)}^\text{(p)}(\lambda,d) } \\
&= p_{\mathrm{util}} \, \frac{\gamma(d) + \frac{p'(d) (1-\lambda) \left[ 1-(1-\lambda)^{K(d)-1} \right]}{\lambda}}{1 + \frac{p'(d) (1-\lambda) \left[ 1-(1-\lambda)^{K(d)-1} \right]}{\lambda}} \\
&= p_{\mathrm{util}} \, \left[ 1 + \frac{\lambda [\gamma(d) - 1]}{\lambda + \iota(\lambda,d)}     \right]
\end{split}
\end{equation}
where 
\begin{equation*}
    \iota(\lambda,d) = p'(d) \left[ 1-\lambda - (1-\lambda)^{K(d)} \right]
\end{equation*}
\(\pi_{\mathrm{st}}^\text{(p)}(\lambda,d)\) denotes the stationary probability that a stored entangled link is available and \(\pi_{(0,0)}^\text{(p)}(\lambda,d)\) denotes the stationary probability that the system is in the state \((0,0)\).

Unlike \(\mathcal{R}_{\rm ac}^{(\mathrm{p})}(\lambda,d)\), the metric
\(\mathcal{R}_{\rm total}^{(\mathrm{p})}(\lambda,d)\) includes requests arriving while the system is locked, which are rejected and therefore counted as unsatisfied. Using the normalization condition, let
\[
\begin{split}
    \pi_{\rm lk}^{(\mathrm{p})}(\lambda,d)
&:=
\pi_{\rm gen}^{(\mathrm{p})}(\lambda,d)
+
\pi_{\rm util}^{(\mathrm{p})}(\lambda,d)\\
&=
1-\pi_{(0,0)}^{(\mathrm{p})}(\lambda,d)
-\pi_{\rm st}^{(\mathrm{p})}(\lambda,d)
\end{split}
\]
denotes the stationary probability that the system is locked.
Since request arrivals are independent of the system state, the long-run expected number of arriving requests per slot is
\[
N_{\rm total}^{(\mathrm{p})} (\lambda,d)
=
\lambda
\sum_{s\in S^{(\mathrm{p})}}
\pi_s^{(\mathrm{p})}(\lambda,d)
=
\lambda.
\]
Hence,
\begin{equation}
\begin{split}
&\mathcal{R}^{\text{(p)}}_\text{total}(\lambda,d) =
\frac{N_{\rm sat}^{(\mathrm{p})}(\lambda,d)}
{N_{\rm total}^{(\mathrm{p})}(\lambda,d)}= p_\text{util} \left[\gamma(d) \pi^{\text{(p)}}_{(0,0)}(\lambda,d) + \pi^{\text{(p)}}_\text{st}(\lambda,d) \right] \\
&= p_\text{util} \left[ \gamma(d) + \frac{p'(d) (1-\lambda) \left[ 1-(1-\lambda)^{K(d)-1} \right]}{\lambda} \right] 
\times\\ &\left[ 1 + \frac{\lambda\gamma(d)[1+p'(d)]}{p'(d)} + \frac{ p'(d)(1-\lambda^2)  \left[1-(1-\lambda)^{K(d)-1}\right] }{ \lambda } \right]^{-1} \\
& = \frac{p'(d) p_\text{util} [\lambda\gamma(d) + \iota(\lambda,d)]}{p'(d) \lambda + \lambda^2 \gamma(d) + \lambda^2 \gamma(d) p'(d) + p'(d) \iota(\lambda,d) (1+\lambda) }
\end{split}
\end{equation}

\paragraph{Average Waiting Time \(\mathcal{W}^{(\mathrm{p})}(\lambda,d)\)}

The average waiting time is the expected delay from request acceptance to successful satisfaction, conditioned on the request being satisfied. A satisfied request originates either from a storage state or from the initial state \((0,0)\).

If the request arrives in a storage state \(S_t=(1,i)\), a pre-generated link is immediately available. The system enters utilization state \((2,i+1)\) in slot \(t+1\), followed by \(L(d)\) utilization slots. Hence, the number of time slots be stored in quantum memories is
\[
\mathcal{W}_{\rm st}^{(\mathrm{p})}(d)
=
1+L(d).
\]
Conditioned on successful satisfaction, the probability that the request originated from a storage state is
\begin{equation*}
\Pr(\mathrm{st}\mid\mathrm{sat})(\lambda,d)
=
\frac{
\pi_{\rm st}^{(\mathrm{p})}(\lambda,d)
}{
\pi_{\rm st}^{(\mathrm{p})}(\lambda,d)
+
\gamma\pi_{(0,0)}^{(\mathrm{p})}(\lambda,d)
}.
\end{equation*}

If the request arrives in state \((0,0)\), the first link-generation attempt begins immediately. Let \(\mathcal{G}\), \(1\le\mathcal{G}\le G_{\max}\), denote the number of attempts required until the first successful generation, conditioned on generation succeeding within \(G_{\max}\) attempts.

Each generation attempt occupies \(C(d)\) time slots. Upon success, the link is stored for one time slot, yielding age \(\Delta t\) and fidelity \(F(1)\), before entering the \(L(d)\)-slot utilization phase. Hence, if the first successful generation occurs on the \(\mathcal{G}\)-th attempt, the number of time slots for waiting link generation is
\begin{equation*}
\mathcal{W}_{\rm gn}^{(\mathrm{p})}(\mathcal{G})(d)
=
\mathcal{G} C(d)+L(d)+2.
\end{equation*}
Conditioned on successful generation within \(G_{\max}\) attempts, the probability that the first success occurs on the \(g\)-th attempt is
\begin{equation*}
\Pr(\mathcal{G}=g\mid\mathrm{gen})(d)
=
\frac{[1-p'(d)]^{g-1}p'(d)}{\gamma(d)},\
1\le g\le G_{\max},
\label{eq:generate_success_at_g}
\end{equation*}
giving the conditional mean
\begin{equation}
\bar g (d)
:=
\mathbb{E}[\mathcal{G}\mid\mathrm{gen}](d)
=
\frac{1}{p'(d)}
-
\frac{G_{\max}(1-\gamma(d))}{\gamma(d)}.
\label{eq:expected_generation_attempts}
\end{equation}
Conditioned on successful satisfaction, the probability that the request originated from state \((0,0)\) is
\begin{equation*}
\Pr\!\bigl((0,0)\mid\mathrm{sat}\bigr)(\lambda,d)
=
\frac{\gamma(d) \pi_{(0,0)}^{(\mathrm{p})}(\lambda,d)}
{\pi_{\rm st}^{(\mathrm{p})}(\lambda,d)
+\gamma\pi_{(0,0)}^{(\mathrm{p})}(\lambda,d)}.
\end{equation*}
Therefore,
\begin{equation}
\begin{split}
\mathcal{W}^{\text{(p)}} (\lambda,d)&=
\Pr(\mathrm{st}\mid\mathrm{sat})(\lambda,d)\,
\mathcal{W}_{\rm st}^{(\mathrm{p})}(\lambda,d)\\
&+
\Pr\!\bigl((0,0)\mid\mathrm{sat}\bigr)(\lambda,d)\,
\mathcal{W}_{\rm gn}^{(\mathrm{p})}(\bar g(d))(\lambda,d)
\\
&= 1 + L(d) + \frac{ \gamma(d) \pi_{(0,0)}^{\text{(p)}}(\lambda,d) }{ \pi_{\mathrm{st}}^{\text{(p)}}(\lambda,d) + \gamma(d) \pi_{(0,0)}^{\text{(p)}}(\lambda,d) } \\
& \qquad \Biggl[ 1+\frac{C(d)}{p'(d)} 
- \frac{ C(d) G_{\max}(1-\gamma(d)) }{ \gamma(d) } \Biggr] \\
& = 1 + L(d) + \frac{\gamma(d)}{\gamma(d) + \frac{p'(d)(1-\lambda)\left[ 1 - \left( 1-\lambda \right)^{K(d)-1} \right]}{\lambda}} \\
& \qquad \Biggl[ 1+\frac{C(d)}{p'(d)} - \frac{ C(d) G_{\max}[1-\gamma(d)] }{ \gamma(d) } \Biggr] \\
& = 2 + L(d) + \frac{\lambda \kappa(d) - \iota(\lambda,d)}{\lambda \gamma(d) + \iota(\lambda,d) }
\end{split}
\end{equation}
where 
\begin{equation*}
    \kappa(d) = \frac{C(d) [\gamma(d) - [1-\gamma(d)]p'(d)G_{\max}]}{p'(d)}.
\end{equation*}

\paragraph{Link Utilization Efficiency \(\mathcal{U}^{\text{(p)}}(\lambda,d)\)}

\(\mathcal{U}\) is the long-run fraction of successfully generated links that are consumed before they expire. A utilization failure still consumes the link, so \(\mathcal{U}\) does not contain \(p_\text{util}\). In steady state, the number of links created depends on two conditions are:
\begin{equation*}
\begin{split}
    N_\text{st-created}^{\text{(p)}}(\lambda,d) &= (1-\lambda) p'(d) \pi_{(0,0)}^{\text{(p)}}(\lambda,d),\\
    &\text{(stored links entering } (1,1)) \\
    N_\text{gn-created}^{\text{(p)}}(\lambda,d) &= p'(d) \sum_{g=1}^{G_\text{max}} \pi_{(0,g)}^{\text{(p)}} (\lambda,d) = \lambda \gamma(d) \pi_{(0,0)}^{\text{(p)}} (\lambda,d),  \\
    &\text{(on-demand links entering } (2,1))
\end{split}
\end{equation*}
Every on-demand link is consumed immediately, stored links are consumed at the rate at which requests get a stored link, so
\begin{equation*}
\begin{split}
    N_\text{st-consumed}^{\text{(p)}}(\lambda,d) &= \lambda \pi_\text{st}^{\text{(p)}}(\lambda,d) \\
    N_\text{gn-consumed}^{\text{(p)}}(\lambda,d) &= N_\text{gn-created}^{\text{(p)}}(\lambda,d) = \lambda \gamma(d) \pi_{(0,0)}^{\text{(p)}}(\lambda,d)
\end{split}
\end{equation*}
Therefore,
\begin{equation}
\begin{split}
    \mathcal{U}^{\text{(p)}}(\lambda,d) &= \frac{N_\text{st-consumed}^{\text{(p)}}(\lambda,d) + N_\text{gn-consumed}^{\text{(p)}}(\lambda,d)}{N_\text{st-created}^{\text{(p)}}(\lambda,d) + N_\text{gn-created}^{\text{(p)}}(\lambda,d)} \\
    &= \frac{\lambda \pi_\text{st}^{\text{(p)}}(\lambda,d) + \lambda \gamma(d) \pi_{(0,0)}^{\text{(p)}}(\lambda,d)}{(1-\lambda) p'(d) \pi_{(0,0)}^{\text{(p)}}(\lambda,d) + \lambda \gamma(d) \pi_{(0,0)}^{\text{(p)}}(\lambda,d)}\\
    &=\frac{(1-\lambda) p'(d) \left[ 1-(1-\lambda)^{K(d)-1} \right] + \lambda \gamma(d)}{(1-\lambda) p'(d) + \lambda \gamma(d)}\\
    &= \frac{\iota(\lambda,d) + \lambda \gamma(d)}{p'(d) - \lambda p'(d) + \lambda \gamma(d)}
\end{split}
\end{equation}

\paragraph{Average Link Fidelity \(\mathcal{F}^{\text{(p)}}(\lambda,d)\)}

The consumed link fidelity is measured at the instant a link enters utilization, since quantum measurement collapses the quantum states and the subsequent classical coordination causes no further decoherence, therefore, \(p_\text{util}\) does not exist in \(\mathcal{F}\). With \(F(d,i) = F'_0(d) \cdot e^{-2\Gamma i \Delta t}\) (as Eq.~\eqref{eq: storing fidelity}), the link has two possibilities to be consumed:
\begin{itemize}
    \item Enters to state \((2,1)\): come from on-demand generation at state \((0,0)\) with fidelity \(F(d,1)\), the expected number of links at this condition is \(N_\text{ut-link}^{\text{(p)}} (\lambda,d,1) = \lambda \gamma(d) \pi_{(0,0)}^{\text{(p)}}(\lambda,d) \);
    \item Enters to state \((2,i+1)\): come from stored state \((1,i)\) with fidelity \(F(d,i+1), 1 \le i \le K(d)-1\), the expected number of links at this condition is \(N_\text{ut-link}^{\text{(p)}} (\lambda,d,i+1) = \lambda \pi_{(1,i)}^{\text{(p)}}(\lambda,d)\).
\end{itemize}
Therefore,
\begin{equation}
\begin{split}
    &\mathcal{F}^{\text{(p)}}(\lambda,d) \\
    &= \frac{F(d,1) N^{\text{(p)}}_\text{ut-link}(\lambda,d,1) + \sum_{i=1}^{K(d)-1} F(d,i+1) N^{\text{(p)}}_\text{ut-link}(\lambda,d,i+1)}{N^{\text{(p)}}_\text{ut-link}(\lambda,d,1) + \sum_{i=1}^{K(d)-1} N^{\text{(p)}}_\text{ut-link}(\lambda,d,i+1)} \\
    &= \frac{ \gamma(d) F(d,1)\pi_{(0,0)}^{\text{(p)}}(\lambda,d) + \displaystyle\sum_{i=1}^{K(d)-1} F(d,i+1)\pi_{(1,i)}^{\text{(p)}}(\lambda,d) } { \gamma(d) \pi_{(0,0)}^{\text{(p)}}(\lambda,d) + \displaystyle\sum_{i=1}^{K(d)-1} \pi_{(1,i)}^{\text{(p)}} (\lambda,d)} \\
    &= F'_0(d) e^{-2\Gamma \Delta t} \frac{\gamma(d) + \frac{p'(d)(1-\lambda)e^{-2\Gamma \Delta t} \left[ 1-\left( (1-\lambda)e^{-2\Gamma \Delta t} \right)^{K(d)-1}   \right]}{1-(1-\lambda)e^{-2\Gamma \Delta t}}}{\gamma(d) + \frac{p'(d)(1-\lambda)\left[ 1 - (1-\lambda)^{K(d)-1}  \right]}{\lambda}} \\
    & = \frac{\nu(\lambda) \lambda F'_0(d) [\gamma(d) (1-\nu(\lambda) )+ p'(d)(\nu(\lambda) - \nu(\lambda)^{K(d)})]}{(1-\lambda)[1-\nu(\lambda)][\gamma(d) \lambda + \iota(\lambda)]}.
\end{split}
\end{equation}
where 
\begin{equation*}
    \nu(\lambda) = (1-\lambda)e^{-2\Gamma \Delta t}.
\end{equation*}

\subsubsection{On-demand Strategy}

The state space of on-demand Markov chain can be written as \(S^{\text{(o)}} = \{ (0,0) \} \cup \{ (0,g) | 1 \le g \le G_\text{max} \} \cup \{ (2,g)| 1 \le g \le G_\text{max}\}\) of dimension \(|S^{\text{(o)}}| = 2 G_\text{max} +1\). The nonzero entries of the one-step transition matrix \(P^{\text{(o)}} = \left[ p_{ij}^{\text{(o)}} \right]\) (Eqs.\eqref{eq: on-de trans prob from (0,0)}\eqref{eq: on-de trans prob from (0,g)} and \eqref{eq: on-de trans prob from (2,g)}) are listed in Tab~\ref{tab:on-demand}.

\begin{table}[t]
\centering
\caption{$\mathcal{T}^{(o)}$}
\label{tab:on-demand}
\begin{tabular}{lll}
\toprule
\textbf{from} & \textbf{to} & \textbf{prob.} \\
\midrule
$(0,0)$ & $(0,0)$ & $1-\lambda$ \\
$(0,0)$ & $(0,1)$ & $\lambda$ \\
$(0,g),\ g < G_{\max}$ & $(2,g)\,/\,(0,g{+}1)$ & $p'(d)\,/\,1-p'(d)$ \\
$(0,G_{\max})$ & $(2,G_{\max})\,/\,(0,0)$ & $p'(d)\,/\,1-p'(d)$ \\
$(2,g)$ & $(0,0)$ & $1$ \\
\bottomrule
\end{tabular}
\end{table}

The stationary distribution \(\pi^{\text{(o)}} = \pi^{\text{(o)}} P^{\text{(o)}}\) is obtained by the same forward recursion as:
\begin{equation}
\begin{split}
    \pi^{\text{(o)}}_{(0,g)}(\lambda,d) &= \lambda [1-p'(d)]^{g-1} \pi^{\text{(o)}}_{(0,0)}(\lambda,d) \\
    \pi^{\text{(o)}}_{(2,g)}(\lambda,d) &= p'(d) \pi^{\text{(o)}}_{(0,g)}(\lambda,d)\\
    &= \lambda p'(d) [1-p'(d)]^{g-1} \pi^{\text{(o)}}_{(0,0)}(\lambda,d)
    \label{eq: on-de_pi-gen&util}
\end{split}
\end{equation}
for \(1 \le g \le G_\text{max}\). Then, we need to verify the balance equation at \((0,0)\), which has three inflows, the self-loop, the final failed generation attempt from \((0,G_\text{max})\) and the completion of utilization from every \((2,g), 1 \le g \le G_\text{max}\). So, \(\pi^{\text{(o)}}_{(0,0)}(\lambda,d) = (1-\lambda) \pi^{\text{(o)}}_{(0,0)}(\lambda,d) + (1-p'(d)) \pi^{\text{(o)}}_{(0,G_\text{max})}(\lambda,d) + \sum_{g=1}^{G_\text{max}} \pi^{\text{(o)}}_{(2,g)}(\lambda,d) = [(1-\lambda) + \lambda (1-\gamma(d))+ \lambda \gamma(d)] \pi^{\text{(o)}}_{(0,0)}(\lambda,d) \). Summing Eq.~\eqref{eq: on-de_pi-gen&util} over each operational phase gives the aggregate stationary probabilities as:
\begin{equation*}
\begin{split}
    \pi_\text{gn}^{\text{(o)}}(\lambda,d) &:= \sum_{g=1}^{G_\text{max}} \pi^{\text{(o)}}_{(0,g)}(\lambda,d) = \lambda \pi^{\text{(o)}}_{(0,0)} \sum_{g=1}^{G_\text{max}} [1-p'(d)]^{g-1}\\
    &= \frac{\lambda \gamma(d)}{p'(d)} \pi^{\text{(o)}}_{(0,0)}(\lambda,d) \\
    \pi_\text{ut}^{\text{(o)}}(\lambda,d) &:= \sum_{g=1}^{G_\text{max}} \pi^{\text{(o)}}_{(2,g)}(\lambda,d) = p'(d) \sum_{g=1}^{G_\text{max}} \pi^{\text{(o)}}_{(0,g)}(\lambda,d)\\
    &= \lambda \gamma(d) \pi^{\text{(o)}}_{(0,0)}(\lambda,d)
\end{split}
\end{equation*}
Substituting them into the normalization condition as \(\pi^{\text{(o)}}_{(0,0)}(\lambda,d) + \sum_g \pi^{\text{(o)}}_{(0,g)}(\lambda,d) + \sum_g \pi^{\text{(o)}}_{(2,g)}(\lambda,d) = 1\), gives us:
\begin{equation*}
    \pi^{\text{(o)}}_{(0,0)}(\lambda,d) = \left[ 1 + \frac{\lambda \gamma(d) [1+p'(d)]}{p'(d)}\right]^{-1}
\end{equation*}

\paragraph{Request Satisfaction Rate \(\mathcal{R}^{\text{(o)}}\)}

We derive two kinds of \(\mathcal{R}^{\text{(o)}}\) as pre-generation strategy. 

\begin{itemize}
    \item \(\mathcal{R}_\text{ac}^{\text{(o)}}(d)\): Requests are accepted only in state \((0,0)\), so \(N_\text{ac-request}^{\text{(o)}}(d) = \lambda \pi^{\text{(o)}}_{(0,0)}(d)\).
\end{itemize}
Every accepted request undergoes the identical generation and utilization cycle and is satisfied with probability \(\gamma(d) p_\text{util}\), independently of the arrival time slot, so \(N_\text{sat-request}^{\text{(o)}}(\lambda,d) = \lambda \gamma(d) p_\text{util} \pi^{\text{(o)}}_{(0,0)}(\lambda,d)\). Therefore,
\begin{equation}
    \mathcal{R}_\text{ac}^{\text{(o)}}(d) = \frac{N_\text{sat-request}^{\text{(o)}}(\lambda,d)}{N_\text{ac-request}^{\text{(o)}}(\lambda,d)} = \gamma(d) p_\text{util}
\end{equation}

\begin{itemize}
    \item \(\mathcal{R}_\text{total}^{\text{(o)}}(\lambda,d)\): The number of requests arriving per time slot is \(N_\text{total-request}^{\text{(o)}}(\lambda) = \lambda \sum_{s\in S^{(o)}}\pi^{\text{(o)}}_s(\lambda,d) = \lambda\).
\end{itemize}
Therefore,
\begin{equation}
\begin{split}
    \mathcal{R}_\text{total}&^{\text{(o)}}(\lambda,d) = \frac{N_\text{sat-request}^{\text{(o)}}(\lambda,d)}{N_\text{total-request}^{\text{(o)}}(\lambda)} = \gamma(d) p_\text{util} \pi^{\text{(o)}}_{(0,0)}(\lambda,d) \\
    &= \frac{\gamma(d) p_\text{util}}{1 + \frac{\lambda \gamma(d) [1+p'(d)]}{p'(d)}} = \frac{\gamma(d) p'(d) p_\text{util}}{p'(d) + \gamma(d)\lambda + \gamma(d)\lambda p'(d)}.
\end{split}
\end{equation}
We observe that \(\lambda \uparrow, \mathcal{R}_\text{total}^{\text{(o)}}(\lambda,d) \downarrow\), because a larger arrival rate increases the fraction of time the system is locked for link generation and utilization.

\paragraph{Average Waiting Time \(\mathcal{W}^{\text{(o)}}(d)\)}

Every accepted request follows the same generation and utilization route. \(1\) time slot from acceptance to the first attempt, \(\mathrm{g}\) attempts of \(C(d)\) time slots each, \(1\) time slot for the minimum storage and \(L(d)\) time slots of utilization. Therefore, the number of waiting time slots is
\begin{equation}
\begin{split}
    \mathcal{W}^{\text{(o)}}(d) &= \mathrm{g}C(d) + L(d) + 2 \\
    &= C(d) + L(d) + 2 + \frac{C(d)[1-p'(d)]}{p'(d)}\\
    &\ \ \ \ - \frac{C(d) G_{\max} [1 - \gamma(d)]}{\gamma(d)} \\
    & = 2 + L(d) + \frac{\kappa(d)}{\gamma(d)}.
\end{split}
\end{equation}

\end{document}